\documentclass[twocolumn,prb,superscriptaddress]{revtex4}
\pdfoutput=1
\usepackage{bm}
\usepackage{amsmath,amssymb}
\usepackage{graphicx}
\usepackage{amsthm}
\DeclareMathAlphabet\mathbfcal{OMS}{cmsy}{b}{n}
\begin{document}
%%%%%%%%%%%%%%%%%%%%%%%%%%%%%%%%%%%%%%%%%%%%%%%%%%%%%%%%%%%%%%%%%%%%%

\title{Charge and heat transport of soft nanosystems in the presence of time-dependent perturbations}

\author{Alberto Nocera}
\affiliation{Department of Physics, Northeastern University, Boston, MA 02115, USA}
\author{C. A. Perroni}
\affiliation{CNR-SPIN and Department of Physics ``Ettore Pancini", Universita' degli Studi di Napoli Federico II, Complesso Universitario Monte Sant'Angelo, Via
Cintia, I-80126 Napoli, Italy}
\author{V. Marigliano Ramaglia}
\affiliation{CNR-SPIN and Department of Physics ``Ettore Pancini", Universita' degli Studi di Napoli Federico II, Complesso Universitario Monte Sant'Angelo, Via
Cintia, I-80126 Napoli, Italy}
\author{V. Cataudella}
\affiliation{CNR-SPIN and Department of Physics ``Ettore Pancini", Universita' degli Studi di Napoli Federico II, Complesso Universitario Monte Sant'Angelo, Via
Cintia, I-80126 Napoli, Italy}

%%%%%%%%%%%%%%%%%%%%%%%%%%%%%%%%%%%%%%%%%%%%%%%%%%%%%%%%%%%%%%%%%%%%%
\begin{abstract}
\textbf{Background}:~Soft nanosystems are electronic nanodevices, such as suspended carbon nanotubes or molecular junctions, whose transport properties are modulated by soft internal degrees of freedom, for example slow vibrational modes. Effects of the electron-vibration coupling on the
charge and heat transport of soft nanoscopic systems are theoretically
investigated in the presence of time-dependent perturbations, such
as a forcing antenna or pumping terms between the leads
and the nanosystem.  A well established approach valid for
non-equilibrium adiabatic regimes is generalized to the case where
external time-dependent perturbations are present. Then, 
a number of relevant applications of the method are reviewed for
systems composed by a quantum dot (or molecule) described by a single electronic level coupled to a vibrational mode.\\
\textbf{Results}:~Before introducing time-dependent
perturbations, the range of validity of the adiabatic approach is discussed showing that a very good agreement with the results of
an exact quantum calculation is obtained in the limit of low level
occupation. Then, we show that the interplay between the low frequency vibrational
modes and the electronic degrees of freedom affects the
thermoelectric properties within the linear response regime
finding out that the phonon thermal conductance
provides an important contribution to the figure of merit at
room temperature. Our work has been stimulated by recent experimental results on
carbon nanotube electromechanical devices working in the
semiclassical regime (resonator frequencies in the MHz range compared
to an electronic hopping frequency of the order of tens of GHz)
with extremely high quality factors. The nonlinear vibrational
regime induced by the external antenna in such systems has been discussed within
the non-perturbative adiabatic approach reproducing quantitatively
the characteristic asymmetric shape of the current-frequency
curves. Within the same set-up, we have proved that the antenna
is able to pump sufficient charge close to the mechanical
resonance making single-parameter adiabatic charge pumping
feasible in carbon nanotube resonators. The pumping mechanism
that we observe is different from that acting in the two parameter pumping and, instead,  it is
based on an important dynamic adjustment of the mechanical motion
of the nanotube to the external drive in the weakly non-linear
regime.  Finally, stochastic forces induced by quantum and thermal fluctuations due
to the electron charging of the quantum dot are shown to affect in a significant way a
Thouless charge pump realized with an elastically deformable
quantum dot. In this case, the pumping mechanism is also shown to be magnified 
when the frequency of the external drive is resonant with the proper frequency of the deformable quantum dot. In this regime, the pumping current is not strongly reduced by the temperature, giving a measurable effect.\\
\textbf{Conclusion}:~Aim of this review has been to discuss common features of different soft nanosystems under external drive.
The most interesting effects induced by time-dependent perturbations are
obtained when the external forcing is nearly resonant with the slow
vibrational modes. Indeed, not only the external forcing can
enhance the electronic response, but it also induces nonlinear
regimes where the interplay between electronic and vibrational
degrees of freedom plays a major role.
\end{abstract}
%%%%%%%%%%%%%%%%%%%%%%%%%%%%%%%%%%%%%%%%%%%%%%%%%%%%%%%%%%%%%%%%%%%%%

%\keywords{Time dependent perturbations; thermoelectric properties; electronic transport theory; nanoelectromechanical systems; electronic charge quantum pumping}
%%%%%%%%%%%%%%%%%%%%%%%%%%%%%%%%%%%%%%%%%%%%%%%%%%%%%%%%%%%%%%%%%%%%%

\maketitle
\section{Introduction}

In some nanoelectronic devices, internal soft degrees of freedom, such as slow vibrational modes, cannot be neglected  since they actively modulate the transport properties. Indeed, 
the electron-vibration coupling significantly affects the charge and heat transport of nanoscopic devices such as molecules connected to external leads
\cite{Delerue2004,Cuevas2010,Galperin2007,Zimbovskaya2011}, and nanoelectromechanical systems\cite{Craighead2000,LaHaye2004,Blencowe2004,Ekinci2005}. 

Due to the small dimensions of the molecular bridge, the hopping of an electron
from the lead onto the molecule can significantly alter its nuclear configuration.
As a main consequence, intriguing non-linear phenomena, such as hysteresis, switching, and
negative differential conductance have been observed in molecular junctions.
In conducting molecules, either the center of mass oscillations \cite{Park2000}, or thermally
induced acoustic phonons\cite{Qin} can be the source of coupling
between electronic and vibrational degrees of freedom. 

Nanoelectromechanical systems (NEMS) are devices similar to molecular junctions. Typically, they consist of a nanobeam
resonator which is coupled to an electronic quantum dot junction. 
Famous examples of NEMS are suspended carbon nanotube (CNT) resonators, which are anchored to two metallic leads under
bias voltage. In this case, the quantum dot is embedded in the CNT itself. 
In Refs.\cite{Huttel2009,Steele2009} the motion of the CNT is 
actuated by a nearby antenna, which means that, when the external antenna 
frequency matches the natural frequency of the CNT beam, one can measure the CNT oscillation 
frequency from the electronic current response of the device. This is possible due to the
extremely high quality factors ($Q>10^{5}$) observed when the resonator
frequencies fall in MHz range compared with an electronic hopping frequency
from the leads 
of the order of tens of GHz. Recently, it has been found that 
phenomena such as switching, 
hysteresis as well as multistability can be observed in NEMS\cite{Atalaya2011}.
NEMS have been proposed as high
sensitive position and mass sensors.
\cite{Knobel2008,Naik2009,Ekinci2004u,Lassagne2008,Kacem2009,hiebert2012,Panchal2012}

Recently, research at the nanoscale has focused not only on charge 
but also on heat transport\cite{DiVentra2011,finch2009,reddy2007,paulsson2003}. 
In particular, thermopower and thermal 
conductances have been measured and theoretically calculated in molecular
junctions\cite{Romero2002,Koch2004,Segal2005,Kraviec2007,Remaggi2013}. 
The role of vibrational degrees of freedom and their coupling with
electrons have a fundamental importance on heat transport and dissipation.
Moreover,  the effect of time perturbations not only on the charge 
dynamics but also on the energy transport is becoming a new field of 
study where both electronic and vibrational degrees of freedom are
involved\cite{Hanggi2015,Michelini2011,Dubi2012,Juergens2013,Arrachea2007,Nitzan2006}.
 
It has been shown that periodic time dependent perturbations can lead 
to pumping effects depending on the frequency of the external drive. 
In this context, different setups have been studied. 
Recent experiments have shown the possibility of 
realizing single parameter charge pumping\cite{Kaest2008,Kaest2008a,Kaest2009,Fuji2008} 
on devices similar to those described above. However, the characteristic 
frequencies of the external drive are much larger than the 
electronic tunneling rate. 
In fact, in these conditions, an effective phase-shift can be produced 
due to 
electron-electron interactions, which is intrinsically generated by 
non-adiabatic 
blockade of tunneling\citep{Cavaliere2009}.
However, it has been pointed out that higher frequencies are
necessary to observe pumping currents  of the same order of magnitude 
of those observed
when a two-parameter pumping mechanism is
 present\cite{Vavi2001,Mosk2002,Torres2005,Torres2011,Agar2007}. 
Another particularly interesting experiment
has been carried out in Ref.\cite{Ganzhorn2012}, where charge 
conductance has been 
obtained at zero bias voltage, by applying a small power to 
the antenna at a frequency close to that of the internal slow resonator. 
Therefore, it is of great importance to address theoretically the 
single parameter charge pumping in 
the regime where the driving frequencies are smaller than the electronic
tunneling rates of the device and close to the frequency of the internal
vibrational mode. Indeed, in the absence of an internal degree of freedom, 
it has been theoretically demonstrated that single parameter 
charge pumping through a quantum dot in the this regime 
is poor\cite{Brouwer1998} even if the electronic correlations 
are important\cite{Splettstoesser2005,Splettstoesser2006}.

The relative magnitude of the characteristic vibrational frequency  
of the molecule or nanobeam and the hopping rate of electrons 
from the leads represents an important quantity in understanding the physics of molecular junctions and NEMS. 
  
In this review, we analyze the adiabatic regime, realized when the internal vibrational modes have frequencies smaller than the hopping rate.
 Within this regime, one can observe phenomena such 
as switching, multistability and hysteresis in molecular junctions or NEMS, and study the 
physics of NEMS subjected to periodic perturbations such those described above. 
When the transient dynamics is considered, and the uniqueness and the approach to the steady state is concerned, analytical and numerical approaches have given controversial results because of the 
length of the time scales required to reach a steady state\cite{Riwar2009,Tahir2010,Albrecht2012,Biggio2014,Rabani2013}, if any.  
A strong debate has developed in the literature regarding the existence of
multistability in these systems\cite{Albrecht2012,Biggio2014,Rabani2013} by studying models similar to those investigated in this review.
Although we do not address this issue here, in our approach multistability is not found. 

The electron-vibration coupling within the Anderson-Holstein model has received much theoretical attention both in the non-adiabatic\cite{Millis2005,Koch2005,Koch2006,Braig2003,Piovano2011,Cavaliere2010,Ziani2011} and in the adiabatic regime. In this latter case, which is the focus of our review, it has been studied in a fully out-of-equilibrium response regime with different theoretical
tools, ranging from rate equation\cite{Blanter2004,BlanterErr2005,Armour2004,Clerck2005,Boubsi2008} to non equilibrium Green function formalism
\cite{Millis2005,Galperin2005,Splettstoesser2005,Brandes2010,Pistolesi2008}. 

The adiabatic approach, which has been applied by some of the present authors in other contexts\cite{Cataudella2011}, is based on
the time scale separation between the slow dot vibrational degrees
of freedom and the fast electronic time scales involved in the thermal or
charge transport. In Refs.\cite{Brandes2010,Mozyrsky2006,Dundas2012}, 
the case of a single vibrational mode within the Anderson-Holstein model has been 
studied with the Feynman-Vernon action functional formalism in the adiabatic regime. 
We stress that, within the adiabatic approach, the  coupling between electrons and vibrational degrees of freedom can be arbitrarily large. On the other hand, the approach followed in the paper 
\cite{Dundas2012} is valid for electronic and vibrational time scales of the same order of magnitude, but it is fundamentally correct only in the regime of weak electron-vibration and vibration-vibration coupling  (negligible anharmonicity). Focus of this review will be on non-perturbative electron-vibration and vibration-vibration coupling regimes.

We point out that the adiabatic approach discussed in this 
review is somewhat different from methods based on the Ehrenfest 
dynamics \cite{Marx2000,Stefanucci2014}. Indeed, from the point 
of view of the electronic system, the adiabatic approach and the 
Ehrenfest dynamics are at the same level of approximation: 
the electron dynamics is treated considering the vibrational 
degrees of freedom classical and infinitely slow. However, 
from the point of view of the vibrational dynamics, the  
Ehrenfest approach is poorer than adiabatic approach. 
Actually, the Ehrenfest approach is similar to a mean-field 
approximation where not the correct force but the spatial 
derivative of the mean-value of the electronic Hamiltonian is derived.
 Moreover, in the  Ehrenfest approach, no dissipative and fluctuating 
 terms are evaluated for the vibrational dynamics. These terms 
 in the adiabatic approach are extremely important:  
 they satisfy the fluctuation-dissipation relation at 
 the equilibrium, and they include the semiclassical 
 corrections fundamental to treat the out-of-equilibrium 
 regime. Of course, the Ehrenfest dynamics can be combined  
 in a simpler way with ab-initio calculations\cite{li2005} 
 of the electronic and/or vibrational systems. But this is 
 not the focus of our review, which is based on model 
 Hamiltonians for both electronic and vibrational degrees of freedom.

In the absence of an external periodic perturbation and in the 
limit of small vibrational frequency with respect to electronic hopping, a generalized
Langevin equation for the displacement coordinate of the vibrational mode can be derived.
In Refs.\cite{Nocera2011}, we have shown that the same Langevin equation can be obtained 
if one performs a semiclassical approximation on the model Hamiltonian and carries out 
an adiabatic approximation \emph{directly} in the electronic Green's function 
on the Keldysh contour. The semiclassical approach naturally includes the effect of a noise 
term which stems form the quantum charge fluctuations induced by the fast time 
scales of the electronic system. The friction and the noise strengths depend by 
construction on the displacement of the oscillator from the equilibrium position.
Our semiclassical approximation, even if less general than the influence 
functional approach, allows us to disentangle exactly the quantum effects 
of the dynamics of the oscillator in the 
Langevin equation and is valid for an arbitrary
strength of electron-vibration coupling\cite{Nocera2011,Nocera2012}. 

In this review, the adiabatic approach for the local vibrational
degrees of freedom in soft nanoscopic systems has been generalized
to the case where external time-dependent periodic perturbations, such as
the effects of a forcing antenna and pumping terms between the
leads and the nanoscopic system, are present. 
In the absence of temporal perturbations, 
we have also included the presence of phononic baths, which is important when
thermal transport through the nanosystem is addressed. Even if the
approach can be applied to  multilevel electronic systems with an
arbitrary number of vibrational degrees of freedom, in this
review, we have mostly discussed the results corresponding to the
prototype system composed of a quantum dot or a molecule described by a 
single electronic level coupled to a single vibrational mode.

Before introducing temporal perturbations, we have
thoroughly studied the range of validity of the approach, focusing
first on the case of zero bias voltage at any temperature, then on
the case of zero temperature at finite bias voltages. 
The thermoelectric properties have then been analyzed within the
linear response regime focusing on the phonon thermal contribution
$G_K^{ph}$ to the figure of merit $ZT$  at room temperature. 
Parameters appropriate for junctions based on $C_{60}$ molecules
connected between different metallic leads have been considered
for the thermoelectric transport. 
We have finally generalized the treatment of the heat transport to
the case where also electron-electron interaction on the dot is
present (within the Coulomb blockade regime) and generalized the
adiabatic approach to this case. 

Then, we have analyzed the properties of the single dot in the
presence of time-dependent periodic perturbations, and in particular of an
external forcing antenna. We have included the
effect of the forcing antenna in our adiabatic scheme showing that
the resulting Langevin equation for the vibrational mode is
modified by a periodic forcing term. Moreover, the generalized force term, 
the friction and the noise strengths become functions that depend on the oscillator displacement and
acquire an explicit periodic dependence on time.

We have treated distinct systems in our unified approach. In particular we have studied
the electronic transport properties at finite bias of a NEMS device consisting 
of a vibrating suspended CNT actuated by an external antenna.  

In this setup, we show that a single parameter
charge pumping is possible. In particular, when the frequency of the antenna is close to that
of the oscillating nanotube, the amplification of the mechanical response 
generates an intrinsic imbalance in the current's response within a single 
pumping cycle, giving a net non-zero contribution. 
Interestingly, we have found theoretically that, in the 
non-linear regime, the pumping current has a non-zero 
response also to harmonics higher than one and their behaviour 
has been compared with the features of the first harmonic.

Finally, we have investigated the behaviour of a two-parameter quantum dot device
(Thouless pump) in the presence of an internal vibrational degree of freedom. 
The characteristics of the pumping current have been studied as a function of 
the electron-vibration coupling, temperature, and driving frequency.
We have assumed that both the frequency of the driving forces and the intrinsic 
oscillation frequency of the quantum dot vibration are adiabatic. We have found 
theoretically that the pumping current can be amplified by 
the internal vibration of the quantum dot. As found above, this is possible only when 
the frequency of the driving forces are close to resonance with the vibration of 
the dot. We have shown that the amplification is robust against temperature, leading to
the prediction that, different from expectation, slowly oscillating quantum dots could
have pumping effects measurable up to high temperatures.

The review is organized as follows. In sec.~2, the
general model is presented. In sec.~3 the adiabatic approach is
discussed. In sec.~4,  the range of validity of the
adiabatic approach is analyzed and, in sec.~5, the method is applied to the
case of thermoelectric transport. In sec.~6, 
the results of the adiabatic approach in the Coulomb
blockade regime are investigated. In sec.~7, the
effects of time dependent perturbations are thoroughly studied.

%%%%%%%%%%%%%%%%%%%%%%%%%%%%%%%%%%%%%%%%%%%%%%%%%%%%%%%%%%%%%%%%%%%%%

\section{Model}
In this section, we present a general Hamiltonian for a multilevel
quantum dot or molecule including the local coupling to vibrational degrees of
freedom and connected to two leads in the presence of a finite
bias voltage and temperature gradient. Contrary to previous
reviews on the subject\cite{Bode2012}, we include the effect of
time-dependent perturbations, such as an external antenna and
pumping terms between the nanosystem and the leads.

The total Hamiltonian of the system is
\begin{equation}\label{Htot}
\hat{\cal H}(t)=\hat{H}_{el}(t)+ \hat H_{osc}(t) +{\hat H}_{int}.
\end{equation}
where $\hat{H}_{el}(t)$ is the electronic Hamiltonian, $\hat H_{osc}(t)$ describes the vibrational degrees of freedom in both the leads and the dot, and
${\hat H}_{int}$ describes the interaction between the electronic and vibrational modes on the dot.

The electronic Hamiltonian $\hat{H}_{el}(t)$ is given by
\begin{equation}\label{Hel}
\hat{H}_{el}(t)=\hat{H}_{leads}+ \hat H_{dot}(t) +{\hat
H}_{leads-dot}(t),
\end{equation}
where the different terms are introduced in the following. The dot Hamiltonian is
\begin{equation}\label{Hdot}
\hat{H}_{dot}(t)= \sum_{m,l,\sigma}{\hat
c^{\dag}_{m,\sigma}}\varepsilon^{m,l}_{\sigma}(t){\hat
c_{l,\sigma}}+U \sum_{m,l}{\hat n^{\dag}_{m,\uparrow}}{\hat
n_{l,\downarrow}}
\end{equation}
where $c_{m,\sigma}$ ($c_{m,\sigma}^{\dag}$) is the standard
electron annihilation (creation) operator for electrons on the dot
levels with spin $\sigma=\uparrow,\downarrow$, where indices $m,l$ can assume positive integer values with a maximum $M$ indicating the total number of electronic levels in the quantum dot. The matrix
$\varepsilon^{m,l}_{\sigma}(t)$ is assumed diagonal in spin space,
while $U$ represents the Coulomb repulsion between the electronic
levels. We assume that only the diagonal part of the matrix
$\varepsilon^{m,l}_{\sigma}(t)$ is nonzero and contains an assigned time
dependence coming from the effects induced by an external antenna
$\varepsilon^{m,m}_{\sigma}(t)=\big(\varepsilon_{m,\sigma}+V_{G}+V_{ext}\cos(\omega_{ext}
t)\big)$, where $V_{ext}$ is the amplitude of the external antenna
potential, $\omega_{ext}$ is the driving frequency, $V_{G}$ is the
static gate potential, and
$\varepsilon_{m,\sigma}$ are the bare energies of the quantum dot
levels. It is important to notice here that in this review we will
consider two possible ways in which the electromagnetic field
generated by an external antenna couples to the dot degrees of freedom:
one is described in Eq.~(\ref{Hdot}), where the external field
excites directly the electronic dot levels (see section 7B 
and Ref.\cite{Perroni2013}); The second case will be
introduced in the Hamiltonian (\ref{Hosc}) and involves the
coupling of the field with the mechanical displacement of the dot
(see section 7A and Ref.\cite{Nocera2012}).

The Hamiltonian of the leads is given by
\begin{equation}\label{Hleads}
\hat{H}_{leads}=
\sum_{k,\alpha,\sigma}\varepsilon_{k_{\alpha}}{\hat
c^{\dag}_{k_{\alpha},\sigma}}{\hat c_{k_{\alpha},\sigma}},
\end{equation}
where the operators ${\hat c^{\dag}_{k_{\alpha},\sigma}} ({\hat
c}_{k_{\alpha},\sigma})$ create (annihilate) electrons with
momentum $k$, spin $\sigma$, and energy
$\varepsilon_{k_{\alpha}}=E_{k_{\alpha}}-\mu_{\alpha}$ in the left
($\alpha=L$) or right ($\alpha=R$) leads. The difference of the
electronic chemical potentials in the leads provides the bias
voltage $V_{bias}$ applied to the junction: $\mu_{L}=\mu+e
V_{bias}/2$, $\mu_{R}=\mu-e V_{bias}/2$, with $\mu$ average
chemical potential. The left and right leads will be considered as
thermostats in equilibrium at the temperatures $T_L=T+\Delta T/2$
and $T_R=T-\Delta T/2$, with $T$ average temperature. Therefore,
the left and right electron leads are characterized by the free
Fermi distribution functions $f_{L}(\omega)$ and $f_{R}(\omega)$,
respectively.

The coupling between the dot and the leads is described by
\begin{equation}\label{Hdot-leads}
\hat{H}_{dot-leads}(t)=
\sum_{k,\alpha,m,\sigma}(V^{m}_{k_{\alpha}}(t){\hat
c^{\dag}_{k_{\alpha},\sigma}}{\hat c_{m,\sigma}}+ h.c.),
\end{equation}
where the tunneling amplitude between the molecular dot and a
state $k$ in the lead $\alpha$ has in general a time dependent
amplitude $V^{m}_{k_{\alpha}}(t)$. 
For the sake of simplicity, we will suppose that the density of
states $\rho_{k,\alpha}$ for the leads is flat within the
wide-band approximation: $ \rho_{k,\alpha} \mapsto \rho_{\alpha}$,
$V_{k_{\alpha}}^{m}(t) \mapsto V_{\alpha}^{m} u^{m}_{\alpha} (t)
$, with $u^{m}_{\alpha} (t)$ periodic functions describing the
strength of the pumping external parameters. Therefore, the time
dependent full hybridization width matrix of the molecular
orbitals is $\hbar \Gamma^{m,n}(t,t')=\sum_{\alpha } \hbar
\Gamma^{m,n}_{\alpha} (t,t')=\sum_{\alpha } \hbar
\Gamma^{m,n}_{\alpha} u^{m}_{\alpha}(t) u^{n}_{\alpha}(t')$, with
$\hbar$ Planck constant and the tunneling rate
$\Gamma^{m,n}_{\alpha}=2\pi\rho_{\alpha}V^{m*}_{\alpha}V^{n}_{\alpha}/\hbar$.
In this review, we consider the generic asymmetric configuration:
${\bm\Gamma}_{L}(t,t')\neq {\bm\Gamma}_{R}(t,t')$,
where bold letters indicate matrices. 

The vibrational degrees of freedom in the system are described by
the Hamiltonian
\begin{equation}\label{Hosc}
\begin{aligned}
&\hat{H}_{osc}(t)= \sum_{s}\frac{\hat{p}^{2}_{s}}{
2m_{s}}+V(\hat{x}_{1},..,\hat{x}_{N})+\sum_{q,\alpha} \hbar \omega_{q_{\alpha}}{\hat
a^{\dag}_{q_{\alpha}}}{\hat a_{q_{\alpha}}}\\
&-\sum_{s}{\hat x}_{s}U_{ext}cos({\omega^{'}_{ext}t}) + \sum_{q,\alpha,s}
\left( C_{q_{\alpha}}{\hat a_{q_{\alpha}}}+h.c. \right) {\hat
x_{s}},
\end{aligned}
\end{equation}
where $m_{s}$ is the effective mass associated with the
$s$-th vibrational mode of the nanosystem,
$V(\hat{x}_{1},..,\hat{x}_{N})=\frac{1}{2}\sum_{s}k_{s}{\hat x}_{s}^{2}$ is the
harmonic potential (with $k_{s}$ the spring constants, and the
oscillator frequencies is $\omega^{s}_{0}=\sqrt{k_{s}/m_{s}}$),
$U_{ext}$ is the external antenna force,
$\omega^{'}_{ext}$ is the driving frequency and $\hat{x}_{s}(t)$ is the
displacement field of the vibrational modes of the quantum dot.

In Eq.~(\ref{Hosc}), the operators ${\hat a^{\dag}_{q,\alpha}}
({\hat a}_{q,\alpha})$ create (annihilate) phonons with momentum
$q$ and frequency $\omega_{q,\alpha}$ in the lead $\alpha$. The
left and right phonon leads will be considered as thermostats in
equilibrium at the temperatures $T_L$ and $T_R$, respectively,
which we assume to be the same as those of the electron leads. In
the following we will include also the presence of the phonon bath in
the leads when we derive the equations relevant for the adiabatic
approach. Their effect will be considered in sec.~5
where the thermoelectric properties of a molecular junction will
be analyzed. In Eq.~(\ref{Hosc}), the coupling between the
displacement $\hat{x}$ and a phonon $q$ in the lead $\alpha$ is given by
the elastic constant $C_{q,\alpha}$. In order to characterize this
interaction, one introduces the spectral density $J(\omega)$:
\begin{equation}
J(\omega)=\frac{\pi}{2} \sum_{q,\alpha} \frac{ C_{q,\alpha}^2 } {
M  \omega_{q,\alpha} }  \delta(\omega-\omega_{q,\alpha})= m \omega
\tilde{\gamma}(\omega), \label{spectral}
\end{equation}
with $M$ mass of the lead atoms and $\tilde{\gamma}(\omega)$
frequency dependent memory-friction kernel of the
oscillator\cite{Weiss2008}. In the regime  $\omega^{s}_0 << \omega_D$ for all the modes,
$\tilde{\gamma}(\omega)$ can be approximated  as real and
independent of frequency, providing  the damping rate 
$\tilde{\gamma}(\omega)  \simeq \gamma$. \cite{Weiss2008} If not specified, we consider the symmetric
configuration: $\gamma_L=\gamma_R=\gamma/2$.

In this review, we assume that the electronic and vibrational
degrees of freedom in the metallic leads are not
interacting \cite{Cuevas2010,Haug2008}, in the sense that the electron-phonon coupling active in the leads gives rise to effects on the nanoscale  
which are negligible when compared with those due to the interaction between intra-dot or intra-molecular electronic and vibrational
degrees of freedom. Therefore, the electron-vibration coupling is assumed effective only on the
quantum dot. This coupling is assumed linear in the vibrational displacements and proportional to the molecule electron occupations
\begin{equation}\label{Hint}
\hat{H}_{int}= \sum_{s,l}\lambda_{s,l}\hat{x}_{s}\hat{n}_{l},
\end{equation}
where $s=(1,..,N)$ with $N$ being the total number of vibrational modes, $l=(1,..,M)$, $\hat{x}_{s}$ is the
displacement operator of the $s$ vibrational mode,
$\hat{n}_{l}=\sum_{\sigma}c^{\dag}_{l,\sigma}c_{l,\sigma}$ is the
electronic occupation operator, and $\lambda_{s,l}$ is a matrix
representing the electron-vibrational coupling.

\begin{figure*}
\centering
\includegraphics[scale=0.25,width=16cm]{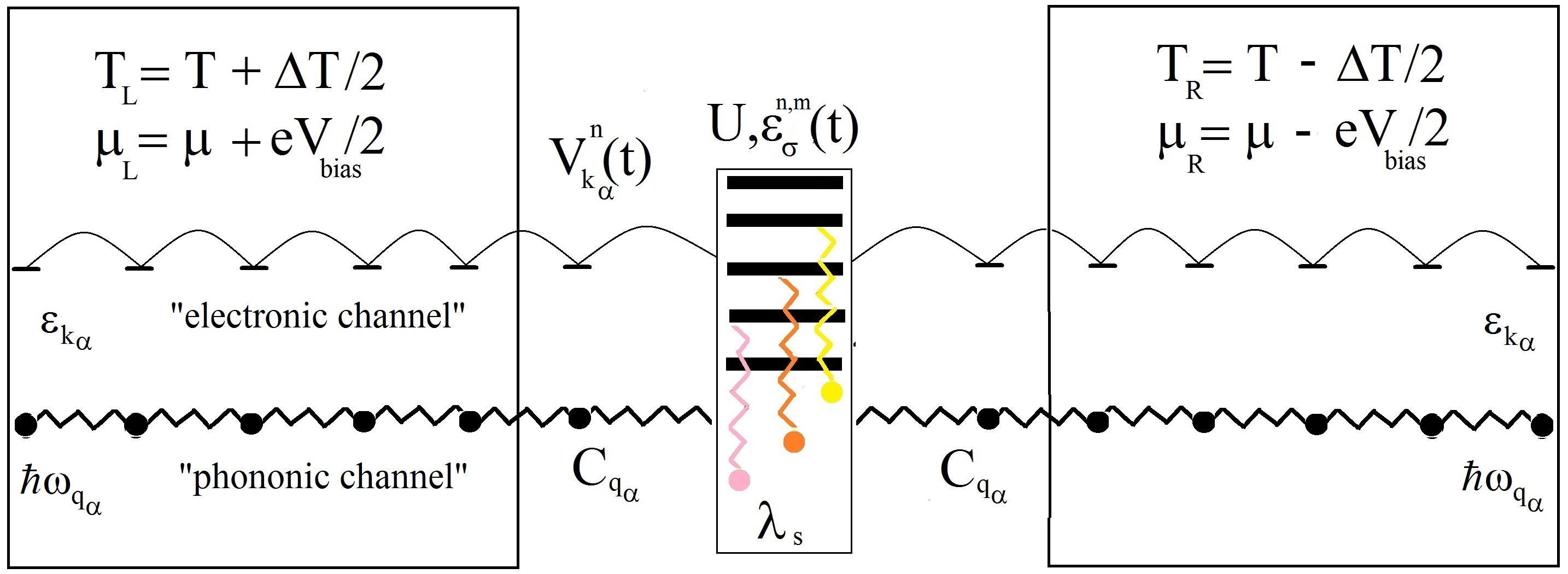}
\caption{(Color online) Scheme of the device studied in this work.
The Left Lead (LL) and the Right Lead (RL) are kept at different
chemical potentials $\mu_{L}=\mu+eV_{bias}/2$,
$\mu_{R}=\mu-eV_{bias}/2$, and different temperatures
$T_{L}=T+\Delta T/2$, $T_{R}=T-\Delta T/2$. The pumping signals
are applied using the gates $V^{n}_{k_{\alpha=L/R}}(t)$, while the
back gate (Gate) induces a $V_G$ shift to the quantum dot energy
levels $\epsilon^{m,n}_{\sigma}(t)$ in the presence of a local Coulomb
repulsion $U$ and coupling $\lambda_{s}$ with $N$ vibrational
modes. The electronic channel in the leads is indicated by $\varepsilon_{k_{\alpha}}$, while the phonon one by the energies 
$\hbar\omega_{q_{\alpha}}$. }\label{Fig1}
\end{figure*}

\section{Adiabatic approach}

In this review, we consider the electronic system coupled to
very slow vibrational modes and temporal perturbations:
$\omega^{s}_{0} \ll \Gamma^{m,n}_{\alpha}$, $d
\varepsilon^{m,n}_{\sigma}(t)/dt \ll \hbar(\Gamma^{m,n}_{\alpha})^{2}$,
$d \Gamma^{m,n}_{\alpha}/dt \ll (\Gamma^{m,n}_{\alpha})^{2}$, for each $s$,$\alpha$ 
and all
pairs of $(m,n)$. In this limit, we can treat the mechanical
degrees of freedom as classical, acting as  slow classical fields
on the fast electronic dynamics.  Therefore, the electronic
dynamics is equivalent to a time dependent multi-level problem
with energy matrix
$\varepsilon^{m,l}_{\sigma}(t)\rightarrow\varepsilon^{m,l}_{\sigma}(t)+\sum_{s}\lambda_{s}x_{s}\delta_{l,m}$,
where $x_{s}$ are now classical displacements.

We point out that an extensive presentation of the adiabatic approximation for vibrational
degrees of freedom in a quantum dot has already been presented in Ref.\cite{Bode2012}, 
but here we extend that approach to the case where a thermal gradient, phonon leads
degrees of freedom and time-dependent perturbations, such as external antenna and 
pumping terms, are present. The Langevin equation for the vibrational modes of the
quantum dot (or molecule), including all the mentioned extensions, can be cast as follows
\begin{equation}\label{Langevin}
m_{s}\ddot{x}_{s}+k_{s}x_{s}=F_{s}(t)+U_{ext}cos({\omega^{'}_{ext}t})+\xi_{s}(t),
\end{equation}
where the generalized force
\begin{equation}\label{fortot}
F_{s}(t)=F^{el}_{s}(t)+ F_L^{ph}(t)+F_R^{ph}(t)
\end{equation}
contains the contribution $F^{el}_{s}$ due to the effect of all
electronic degrees of freedom, and $F_{\alpha}^{ph}(t)$ is the
force due to the coupling to the $\alpha$ lead phonon degrees of
freedom. It can be easily shown that, in the regime investigated
in this review, in Eq.~(\ref{fortot}), one has
$F_{\alpha}^{ph}(t)=-m \gamma_{\alpha} v$, with
$\gamma_\alpha=\gamma/2$.

The electronic force, $F^{el}(t)=tr[i\lambda_{s}\textbf{G}^{<}(t,t)]$
(the trace \`{}\`{}tr\'{}\'{} is taken over the dot levels), is
defined in terms of the lesser dot matrix Green's function $\textbf{G}^{<}(t,t')$, while the fluctuating forces $\xi_{s}(t)$ will be discussed later in
this section.

For the sake of simplicity, we do not include explicitly the effect of the Coulomb
repulsion on the quantum dot Hamiltonian in deriving equations
encoding the adiabatic approximation. In Sec. 5, we
will show that, in the particular case of a single level quantum dot
with large repulsion $U$, the adiabatic approach works exactly as
in the non-interacting case with the "caveat" of treating each
Green's function pole as a non interacting
level\cite{Perroni2014a}.

Our notation is such that $\textbf{G}$ denotes full Green's
functions, while ${\mathbfcal G}$ denotes the strictly adiabatic
(or frozen) Green's functions that are evaluated for a fixed value
of ${\textbf X}\equiv\{x_{s}\}$ and $t$. Starting from the Dyson equation
\begin{equation}
\begin{aligned}
&{\bm G}^{R}(t,t')= {\bm G}^{R}_{0}(t,t')+\int dt_1\int dt_2 {\bm G}^{R}_{0}(t,t_1)\times\\
&\times{\bm \Sigma}^{R,leads}(t_1,t_2){\bm G}^{R}(t_2,t'),
\end{aligned}
\end{equation}
where $ {\bm G}^{R}_{0}(t,t')$ is the retarded Green's function in the absence of coupling to the leads,
it is straightforward to show that the adiabatic expansion (to first order in $d \Gamma^{m,n}_{\alpha}/dt$, $d
\varepsilon^{m,l}_{\sigma}(t)/dt$ and $\dot{x}_{s}$) for the retarded Green's function  ${\bm G}^{R}$ is given by
\begin{equation}\label{adiabaticGR}
\begin{aligned}
&{\bm G}^{R}\simeq {\mathbfcal
G}^{R}+\frac{i}{2}\big(\partial_{E}{\mathbfcal G}^{R}(\sum_{s}
\lambda_{s} \dot{x}_{s}+\partial_{t}{\bm
\Sigma}^{R,leads}){\mathbfcal G}^{R}\\
&-{\mathbfcal G}^{R}(\sum_s
\lambda_{s} \dot{x}_{s}+\partial_{t}{\bm
\Sigma}^{R,leads})\partial_{E}{\mathbfcal G}^{R}\big).
\end{aligned}
\end{equation}
Above, ${\mathbfcal G}^{R}(E,{\bm X},t)$ is the strictly adiabatic (frozen) retarded Green's function including the coupling with the leads
\begin{equation}
{\mathbfcal G}^{R}(E,{\bm X},t)=[E-{\bm \varepsilon}({\bm
X},t)-{\bm \Sigma}^{R,leads}(t)]^{-1}, 
\end{equation}
where ${\bm \varepsilon}({\bm X}, t)$ represents the matrix $\varepsilon^{m,l}_{\sigma}(t)+\sum_{s}\lambda_{s}x_{s}\delta_{l,m}$ and ${\bm \Sigma}^{R,leads}=\sum_{\alpha}{\bm \Sigma}^{R,leads}_{\alpha}$ is the total self-energy due to the coupling between the dot and the leads. The self-energies ${\bm\Sigma}^{R,A,<,>}_{\alpha}$ are defined as
\begin{equation}
{\bm\Sigma}^{R,A,<,>}_{\alpha,m,n}(t_1,t_2)=\sum_{k,\alpha=R,L}V^{*m}_{k_{\alpha}}(t_1)g^{R,A,<,>}_{k,\alpha}(t_1,t_2)V^{n}_{k_{\alpha}}(t_2).
\end{equation}
Notice that the imaginary part of ${\bm \Sigma}^{R,leads}_{\alpha}$ is proportional to $\hbar\bm{\Gamma}_{\alpha}(t_1,t_2)$.
Employing the above equations, for the lesser Green's function ${\bm G}^{<}$ one gets 
\begin{equation}\label{Glesadia}
\begin{aligned}
&{\bm G}^{<}\simeq{\mathbfcal
G}^{<}+\frac{i}{2}\Big[\partial_{E}{\mathbfcal G}^{<}\big(\sum_{s}
\lambda_{s}\dot{x}_{s}+\partial_{t}{\bm\Sigma}^{A,leads}\big){\mathbfcal
G}^{A}\\
&-{\mathbfcal G}^{R}\big(\sum_s \lambda_{s}
\dot{x}_{s}+\partial_{t}{\bm\Sigma}^{R,leads}\big)\partial_{E}{\mathbfcal
G}^{<}+\partial_{E}{\mathbfcal G}^{R}\big(\sum_{s}
\lambda_{s}\dot{x}_{s}\\
&+\partial_{t}{\bm\Sigma}^{R,leads}\big){\mathbfcal
G}^{<}-{\mathbfcal G}^{<}(\sum_{s} \lambda_{s}
\dot{x}_{s}+\partial_{t}{\bm\Sigma}^{A,leads})\partial_{E}
{\mathbfcal G}^{A}\Big],
\end{aligned}
\end{equation}
with ${\mathbfcal G}^{<} = {\mathbfcal
G}^{R}{\bm\Sigma}^{<}{\mathbfcal G}^{A}$.

The electron-vibration induced forces in the zero-th order
adiabatic limit, ${\bm G}^{<}\simeq {\mathbfcal G}^{<}$,
are given by
\begin{equation}\label{force0}
F_{s}^{el(0)}({\bm X},t)=-k_{s}x_{s}-\int \frac{dE}{2\pi i}
tr[\lambda_{s}{\mathbfcal G}^{<}]
\end{equation}
The leading order correction to the lesser Green's function ${\bm G}^{<}$ gives two
contributions to the electron-vibration induced forces: a term
proportional to the vibrational velocity and a pumping velocity
term acting as a driving
\begin{equation}\label{force1}
F_{s}^{el(1)}({\bm X},t)=-\sum_{s'}\theta_{s,s'}({\bm
X},t)\dot{x}_{s'}+B_{s}({\bm X},t).
\end{equation}
The first term determines the tensor ${\bm \theta}$ (obtained
after integration by parts)
\begin{equation}
\theta_{s,s'}({\bm X},t)=\int \frac{dE}{2\pi} tr\big({\mathbfcal
G}^{<} \lambda_{s}\partial_{E}{\mathbfcal G}^{R}\lambda_{s'}-
{\mathbfcal G}^{<} \lambda_{s'}\partial_{E}{\mathbfcal
G}^{A}\lambda_{s}\big),\label{thetaeq}
\end{equation}
while the second the vector $B$
\begin{equation}
\begin{aligned}
&B_{s}({\bm X},t)=\int \frac{dE}{2\pi} tr\big({\mathbfcal G}^{<}
\lambda_{s}\partial_{E}{\mathbfcal
G}^{R}\partial_{t}{\bm\Sigma}^{R,leads}\\
&- {\mathbfcal G}^{<}\partial_{t}{\bm\Sigma}^{A,leads}\partial_{E}{\mathbfcal
G}^{A}\lambda_{s}\big).
\end{aligned}
\end{equation}
The tensor ${\bm \theta}$ can be split into symmetric and
anti-symmetric contributions\citep{VonOppen2012}, ${\bm \theta}={\bm \theta}^{sym} +
{\bm \theta}^{a}$, which define a dissipative term ${\bm
\theta}^{sym}$ and an orbital, effective magnetic field ${\bm
\theta}^{a}$ in the space of the vibrational modes. The latter
interpretation is based on the fact that the corresponding force
takes a Lorentz-like form. Using ${\mathbfcal G}^{R}-{\mathbfcal
G}^{A}={\mathbfcal G}^{>}-{\mathbfcal G}^{<}$ and noting that
$2\int dE {\mathbfcal G}^{<}\partial_{E}{\mathbfcal G}^{<}=\int
dE \partial_{E}({\mathbfcal G}^{<})^2=0$, we obtain the explicit
expressions
\begin{equation}\label{gamma}
\begin{aligned}
\theta_{s,s'}^{sym}({\bm X},t)&=\int \frac{dE}{2\pi}
tr\big\{\lambda_{s}{\mathbfcal G}^{<}\lambda_{s'}\partial_{E}{\mathbfcal G}^{>}\big\}_{sym},\\
\theta_{s,s'}^{a}({\bm X},t)&=-\int \frac{dE}{2\pi}
tr\Big\{\lambda_{s}{\mathbfcal
G}^{<}\lambda_{s'}\partial_{E}({\mathbfcal G}^{R}+{\mathbfcal
G}^{A})\Big\}_{a}
\end{aligned}
\end{equation}
Here we have introduced the notation
$\{C_{s,s'}\}_{sym,a}=\frac{1}{2}\{C_{s,s'}\pm C_{s',s}\}_{sym,a}$ for
symmetric and anti-symmetric parts of an arbitrary matrix ${\bm
C}$.

We can now discuss the stochastic forces $\xi_{s}(t)$ in
Eq.~(\ref{Langevin}) in the adiabatic approximation. In the
presence of coupling with the phonon leads, the fluctuating forces
are composed of three independent terms
\begin{equation}\label{flucforces}
\xi_{s}(t) = \xi^{el}_{s}(t) + \xi_L^{ph}(t) + \xi_R^{ph}(t),
\end{equation}
where $\xi^{el}_{s}(t)$ is provided by quantum electronic
fluctuations while $\xi_{\alpha}^{ph}(t)$ is due to the $\alpha$
phonon lead. The noise term
$\xi^{el}_{s}(t)$ expresses the effects of the quantum electronic density 
fluctuations ${\hat F}=\lambda_{s}{\hat c}^{\dag}_{m}{\hat
c}_{m}$ on the oscillator motion. By analogy with fluid dynamics, it can be viewed 
as originating from the thermal and non-equilibrium fluctuations of the 
electronic ``fluid" in which the vibrating quantum dot can be considered as immersed. 
In the absence of electron-electron interactions, the Wick theorem allows to write the noise correlator as
\begin{equation}
\langle \xi^{el}_{s}(t)\xi^{el}_{s'}(t')\rangle=
tr\{\lambda_{s}\bm{G}^{>}(t,t')\lambda_{s'}\bm{G}^{<}(t',t)\},
\end{equation}
where ${\bm G}^{>}(t,t')$ is the greater Green's function with
matrix elements
\begin{equation}
G^{>}_{m,l}(t,t')=-i\langle
c_{m,\sigma}(t)c^{\dag}_{l,\sigma}(t')\rangle.
\end{equation}
At this stage, it represents a colored noise term in the Langevin
equations (\ref{Langevin}) that depends non-locally on the
dynamics of the vibrational modes and it is complicated to treat
numerically. In the adiabatic approximation, one first substitutes
the full Green's function ${\bm G}$ by the adiabatic zero-order
Green's function ${\mathbfcal G}$ and then observes that the
electronic fluctuations act on short time scales only. For this
reason the total forces $\xi_{s}(t)$ are locally correlated in
time and one has to only retain the low frequency limit of their
stochastic variance. One thus obtains a multiplicative white noise
term
\begin{equation}\label{noise}
\begin{aligned}
\langle \xi^{el}_{s}(t)\xi^{el}_{s'}(t')\rangle&\simeq
tr\{\lambda_{s}{\mathbfcal G}^{>}({\bm
X},t)\lambda_{s'}{\mathbfcal G}^{<}({\bm X},t)\}\\&=D({\bm
X},t)\delta(t-t'),
\end{aligned}
\end{equation}
where
\begin{equation}
D_{s,s'}({\bm X},t)=\int \frac{dE}{2\pi}
tr\big\{\lambda_{s}{\mathbfcal G}^{<}\lambda_{s'}{\mathbfcal
G}^{>}\big\}_{sym}.
\end{equation}
The fluctuating forces coming from the fluctuations of the phonon
leads $\xi_{\alpha}^{ph}(t)$ have the following property
\begin{equation}\label{langevin1b}
\langle \xi_{\alpha}^{ph}(t) \rangle=0,\;\;\;\; \langle
\xi_{\alpha}^{ph}(t) \xi_{\alpha}^{ph}(t') \rangle= 2  k_{B}
T_{\alpha} m \gamma_{\alpha}  \delta(t-t'). \nonumber
\end{equation}
Combining the three terms, one gets the total fluctuating forces
$\xi_{s}(t)$ such that
\begin{equation}
\langle \xi_{s}(t) \rangle=0,\;\;\;\; \langle \xi_{s}(t)
\xi_{s'}(t') \rangle= D_{eff}^{s,s'}({\bm X},t) \delta(t-t'),\label{Lange}
\end{equation}
where the effective position dependent noise term
$D_{eff}^{s,s'}({\bm X},t)$ is
\begin{equation}\label{Dxt}
D_{eff}^{s,s'}({\bm X},t)=D_{s,s'}({\bm X},t)+ k_{B} (T_L+T_R) m \gamma.
\end{equation}

Once the forces Eqs.~(\ref{force0})-(\ref{force1}) and the noise terms (\ref{noise}) 
are calculated, Eq.~(\ref{Langevin}) represents a set of non-linear Langevin equations
in the unknown dynamics $x_{s}(t)$ which are explicitly dependent on time. 
Even for the simple case where only one vibrational degree of freedom is present, 
the stochastic differential equation should be solved numerically. 
In Ref.\cite{Nocera2011}, the present authors have applied a Runge-Kutta algorithm 
adapted to the stochastic nature of the equation. When an explicit time dependence is present, 
such as in the case where 
the charge pumping is studied\cite{Perroni2013,Perroni2014b}, the periodic 
nature of the temporal perturbations has allowed to extend the algorithm straightforwardly.
Indeed, by sampling the occurrences of ${\bm X},\dot{\bm X}$ in the phase space at times $t$ separated the characteristic period of the perturbations,
one can calculate the oscillator distribution
functions $P({\bm X},{\bm V},t)$ (where ${\bm V}=\dot{\bm X}$) during a single period,
and, therefore, all the properties of the vibrational modes. Using this
function, one can determine the time evolution of an electronic or
vibrational observable $O({\bm X},{\bm V},t)$ as:
\begin{equation}\label{average}
\langle O\rangle(t)=\int \int d {\bm X} d {\bm V} P({\bm X},{\bm
V},t) O({\bm X},{\bm V},t).
\end{equation}

The electronic observables, such as charge and heat currents, can be evaluated
exploiting the slowness of the vibrational degrees of freedom.

\subsection{Electronic charge and heat currents}

In this subsection, we discuss the adiabatic expansion for the current passing
through the quantum dot. An expression valid in the absence of time dependent perturbations it has already been provided in Ref.\cite{VonOppen2012}, but here we re-present that derivation in the presence of time dependent perturbations.
The definition of the current through lead $\alpha$ is given by
\begin{equation}\label{currentdef}
I_{\alpha}(t)=-e \langle {\dot
N}_{\alpha}\rangle=\sum_{n,k\in\alpha}V_{k_{\alpha},n}(t)\langle
c^{\dag}_{k_{\alpha}}(t)c_{n}(t)\rangle+h.c.
\end{equation}
where $N_{\alpha}=\sum_{r\in\alpha}c^{\dag}_{r}c_{r}$. Using the
definitions of self-energy and Green's functions\cite{Haug2008}
the current can be expressed in terms of the Green's function of
the dot and the self-energy of the dot-leads coupling
\begin{equation}\label{currentdef1}
\begin{aligned}
I_{\alpha}(t)&=e\int dt' tr\big\{ {\bm
G}^{R}(t,t'){\bm\Sigma}^{<}_{\alpha}(t',t)\\&+{\bm
G}^{<}(t,t'){\bm\Sigma}^{A}_{\alpha}(t',t)\big\}+h.c.\;.
\end{aligned}
\end{equation}
We can now apply the adiabatic expansion to the above expression
employing the formulas for the Green's functions ${\bm G}^{R}$
(\ref{adiabaticGR}) and ${\bm G}^{<}$ (\ref{Glesadia})
\begin{equation}\label{current1}
\begin{aligned}
&I_{\alpha}(t)=e\int \frac{dE}{2\pi} tr\Big\{ {\mathbfcal
G}^{R}{\bm\Sigma}^{<}_{\alpha}+{\mathbfcal
G}^{<}{\bm\Sigma}^{A}_{\alpha}-\frac{i}{2}\big(\partial_{t}{\mathbfcal
G}^{R}\partial_{E}{\bm\Sigma}^{<}_{\alpha}\\
&+\partial_{t}{\mathbfcal
G}^{<}\partial_{E}{\bm\Sigma}^{A}_{\alpha}\big)+\frac{i}{2}\big(\partial_{E}{\mathbfcal
G}^{R}\partial_{t}{\bm\Sigma}^{<}_{\alpha}+\partial_{E}{\mathbfcal
G}^{<}\partial_{t}{\bm\Sigma}^{A}_{\alpha}\big)\Big\}+h.c..
\end{aligned}
\end{equation}
We split the current into an adiabatic contribution $I_{0}$, 
two terms proportional to the velocity $\dot{\bm X}$ and the time
derivative of the pumping parameters $u_{\alpha}(t)$ (describing
the lead-dot self-energy ${\bm\Sigma}(t)$):
$I_{\alpha}=I^{0}_{\alpha}+I^{1}_{\alpha}+I^{2}_{\alpha}$. The
zeroth order adiabatic contribution is given by
\begin{equation}
I^{0}_{\alpha}({\bm X},t)=e\int \frac{dE}{2\pi} tr\Big\{
\big({\mathbfcal G}^{R}-{\mathbfcal
G}^{A}\big){\bm\Sigma}^{<}_{\alpha}+{\mathbfcal
G}^{<}\big({\bm\Sigma}^{A}_{\alpha}-{\bm\Sigma}^{R}_{\alpha}\big)\Big\},
\end{equation}
where we have collected the strictly adiabatic terms from
Eqs.(\ref{adiabaticGR}) and (\ref{Glesadia}). Now we turn to the
first order corrections, restricting our considerations to the wide-band limit. 
The contribution to the current (\ref{current1}) which is linear in the velocity of the
vibrational modes reads (after integration by parts)
\begin{equation}\label{I1}
\begin{aligned}
&I^{1}_{\alpha}({\bm X},t)=e\int \frac{dE}{2\pi} i\sum_{s'}{\dot
x}_{s'}tr\Big\{ \big(\partial_{E}{\mathbfcal
G}^{R}\big)\lambda_{s'}{\mathbfcal G}^{R}{\bm\Sigma}^{<}_{\alpha}
\\&+\Big[ \big(\partial_{E}{\mathbfcal
G}^{<}\big)\lambda_{s'}{\mathbfcal G}^{A}
-{\mathbfcal G}^{A}\lambda_{s'}\big(\partial_{E}{\mathbfcal
G}^{<}\big)\Big]{\bm\Sigma}^{A}_{\alpha}\Big\}+h.c.,
\end{aligned}
\end{equation}
while the term $I^{2}_{\alpha}(t)$ coming from the pumping
perturbations is
\begin{equation}\label{I2}
\begin{aligned}
&I^{2}_{\alpha}({\bm X},t)=-e\int \frac{dE}{2\pi i}tr\Big\{
\big(\partial_{E}{\mathbfcal G}^{R}\big)\partial_{t}{\bm
\Sigma}^{R,leads}{\mathbfcal G}^{R}{\bm\Sigma}^{<}_{\alpha}\\&+\Big[
\big(\partial_{E}{\mathbfcal G}^{<}\big)\partial_{t}{\bm
\Sigma}^{R,leads}{\mathbfcal G}^{A}-{\mathbfcal G}^{A}\partial_{t}{\bm
\Sigma}^{R,leads}\times \\&\times\big(\partial_{E}{\mathbfcal
G}^{<}\big)\Big]{\bm\Sigma}^{A}_{\alpha}\Big\}+h.c..
\end{aligned}
\end{equation}

Analogously to Eq.~(\ref{currentdef1}), the electronic energy current
passing through the dot from the lead $\alpha=R,L$ is defined as
$J^{el}_{\alpha}=-\langle{\dot H}_{lead_{\alpha}}\rangle$. In the
case where the chemical potentials of the leads are not
time-dependent, one can easily show that all the expressions
derived for the electronic current $I_{\alpha}$ and its adiabatic
expansion are formally identical to those valid for the energy
current $J^{el}_{\alpha}$, with the only "caveat" that one has to
substitute the self-energies ${\bm\Sigma}^{R,A,<,>}_{\alpha}$ with
the functions ${\bm\Xi}^{R,A,<,>}_{\alpha}$ defined as
\begin{equation}
\begin{aligned}
&\Xi^{R,A,<,>}_{m,n}(t_1,t_2)=\sum_{k,\alpha=R,L}V^{m *}_{k_{\alpha}}(t_1)\epsilon_{k_{\alpha}}g^{R,A,<,>}_{k,\alpha}(t_1,t_2)\times \\
&\times V^{n}_{k_{\alpha}}(t_2).
\end{aligned}
\end{equation}

In analogy with the terms in the deterministic and fluctuating
forces of the Langevin equations (\ref{Langevin}), the total
energy current $J$ involving the oscillator is composed of three
terms \cite{Dundas2012}:
\begin{equation}
J =\sum_s \left(
J^{el}_{\lambda,s}+J_{L,s}^{ph}+J_{R,s}^{ph}\right),
\end{equation}
where $J^{el}_{\lambda,s}$ originates from the electron level and
depends on the electron-vibration coupling. It can be also
expressed as
\begin{equation}
J^{el}_{\lambda,s}=\langle v_s \left[ \xi^{el}_s(t)-
\theta_{\lambda,s}({\bm X},) v_s \right] \rangle,
\end{equation}
while $J_{\alpha_s}^{ph}$ comes from the $\alpha$ phonon lead
\begin{equation}
J_{\alpha_s}^{ph}=\langle v_s \left[ \xi_{\alpha}^{ph}(t)- m
\gamma_{\alpha} v_s  \right] \rangle,
\end{equation}
where $v_s =\dot{x}_s$.
These quantities have to be evaluated along the dynamics. Once the
stationary state is reached, the energy conservation requires that
the total energy current $J$ vanishes.

In the next section we discuss the validity of the adiabatic approximation, that is based on the separation between the slow vibrational and fast electronic timescales.

\section{Range of validity of adiabatic approach: single level molecule}
\label{sec:singsitoT0}

In this section, we investigate the range of validity of the
adiabatic approach together with the semiclassical treatment of
the vibrational degrees of freedom. To this aim, we consider, as in the rest of the paper, the
simple case where the nanoscopic system is represented by a
molecule modeled as a single electronic level ($M=1$) locally
interacting with a single vibrational mode ($N=1$, see Fig. \ref{Fig1}).
This means that the focus is on a molecular (or quantum dot) electronic orbital
which is sufficiently separated in energy from other orbitals. We
have in mind, for instance, the $C_{60}$ molecule when the LUMO energy differs
from the HOMO energy for more than $1$ eV. Even when the
degeneracy of the LUMO is removed by the contact with Ag-leads, the
splitting gives rise to levels which are separated by an energy of
the order of $0.5$ eV \cite{Lu2003}. Furthermore, the energy of the molecular
orbital can be tuned by varying the gate voltage $V_{G}$ and
unless otherwise stated we do not consider time-dependent
perturbations.

We will focus on the center of mass mode as the only relevant
vibrational mode for the molecule, which is expected to have the
lowest frequency for large molecules. In fact, in $C_{60}$
molecules, experimental results provide compelling evidence for a
coupling between electron dynamics and the center of mass motion
where $\hbar \omega_0$ has been estimated to be of the order of
$5$ meV \cite{Park2000}. Furthermore, as reported in experimental
measurements \cite{Park2000}, the effects of the electron-vibration interaction 
are not negligible in junctions with $C_{60}$ molecules. Within these assumptions, in
Eq.~(\ref{Htot}), the interaction Hamiltonian ${\hat H}_{int}$
reduces to the same interaction term of the single impurity
Anderson-Holstein model \cite{Cuevas2010} and the
dot-oscillator coupling sets the characteristic polaron energy and
length scales
\begin{equation}
E_{P}=\frac{\lambda^2}{2m\omega_{0}^{2}},\;\;l_{P}={\lambda/m\omega_{0}^{2}}.
\end{equation}
Equation (\ref{Langevin}) reduces in this case to a
single Langevin equation \cite{Nocera2011,Perroni2014}. Since the
main objective of this section is to discuss the range of validity
of the adiabatic approach for the electronic properties, we hereby report the expression for the displacement dependent electronic
spectral function $A(\omega,x)$
\begin{equation}\label{Spec}
A(\omega,x)= \frac{\hbar\Gamma}{I^2(\omega,x)+(\hbar\Gamma)^2/4},
\end{equation}
with $I(\omega,x)=\hbar \omega - V_G -2E_{P} x$.  Within the adiabatic approach,
the actual electronic spectral function $A(\omega)$ is
\begin{equation}\label{Specnew}
A(\omega)=\int_{-\infty}^{+\infty}P(x)A(\omega,x),
\end{equation}
where $P(x)$ is the reduced position distribution function of the
oscillator in the absence of time dependent forces. Another
important quantity is the average kinetic energy of the oscillator
\begin{equation}
\langle E_{C}\rangle= \frac{1}{2}\textrm{m}
\langle \textrm{v}^{2} \rangle,
\label{Ekin}
\end{equation}
which is calculated through the reduced velocity distribution function $P(\textrm{v})$ of the oscillator
in the absence of time dependent forces.

\subsection{Equilibrium conditions at T$\neq0$}
In this subsection, we consider equilibrium conditions with both leads
at temperature $T$. When the gate voltage is tuned in such a way that the electronic level of the dot is far from the bias window ($|V_G| \gg 0$), a very low electronic density is on the dot. We emphasize that, in this limit,
the electronic Green functions of the model can be exactly
calculated if, as assumed in this revew, the wide band limit is
used for the leads \cite{almb78,mahan2000}. In this section we intend to compare an approach which takes fully into
account the quantum nature of the oscillator, and it is valid
also at very low temperatures ($T \le \omega_0$), with the adiabatic approach. 
We stress that, within the off-resonant regime, the oscillator dynamics is very
weakly perturbed by the effects of the electron-vibration
coupling, but it remains sensitive to the coupling to phonon
leads.

\begin{figure}
\hbox{\hspace{0.5cm}\includegraphics[width=8cm,height=9.0cm]{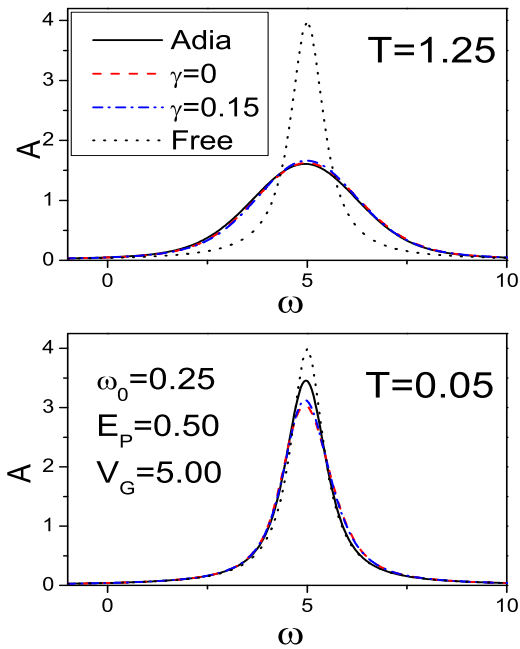}}
\caption{(Color online) Analysis of the validity of the adiabatic
approximation. Electronic spectral function as a function
of the frequency (in units of $\Gamma$) for adiabatic approach
(solid line), fully quantum low density approach (dashed line for $\gamma=0$,
dash-dotted line for $\gamma=0.15$), and free case (dotted line) at temperature $T=1.25 \hbar
\Gamma / k_{B}$ (Upper Panel) and temperature $T=0.05 \hbar \Gamma /
k_{B}$ (Lower Panel). In all the plots, $E_P=0.5 \hbar \Gamma$,
$V_G=5 \hbar \Gamma$, and $\omega_0=0.25 \Gamma$.} \label{figqua}
\end{figure}

The spectral function calculated within the exact approach in the
regime of low level occupation has been compared with that
obtained within the adiabatic approach. We focus on the electronic spectral function $A(\omega)$ since in the linear response regime (low bias voltage, absence of time perturbations) all the transport coefficients can be expressed as integrals of $A(\omega)$ (see next subsections and the remaining sections of the present paper). It is clear that a good agreement between the adiabatic approach and the exact approach in the low density regime  can be used as a reliable test of the validity of the adiabatic scheme. The comparison between the two approaches represents a new and interesting part of the review and it will allow us to assess that quantities calculated within the adiabatic approach are very reliable even in the regime of higher density.

In the upper panel of Fig. \ref{figqua}, we consider the spectral functions at
$T=1.25 \hbar \Gamma / k_{B}$, which is close to room temperature
for $\hbar\Gamma\sim20meV$. Moreover, we consider the off-resonant
regime $V_G=5  \hbar \Gamma$. The spectral weight up to $0$ (position of the
chemical potential) indicates that the level occupation $n$ is
small (less than $0.1$). The agreement between the spectral
functions calculated within the two approaches is excellent. The
peak positions for both approaches are at $\omega=V_G$ and the
widths of the curves match perfectly. The role of $\gamma$ is not
relevant, since, in any case, it is much smaller than $\Gamma$.
Obviously, with decreasing the temperature, the two approaches
tend to differ. In the lower panel of Fig.\ref{figqua}, we have
considered a very low temperature ($T=0.05 \hbar \Gamma / k_{B}$)
where a worse agreement is expected. We point out that the
agreement between the two approaches is still good. Actually, the
exact approach at low molecule occupation slightly favors a small
transfer of spectral weight at high frequency. In any case, the strong
similarities in the spectral function will point out to analogous
behaviours of electron transport properties within the two
approaches. In conclusion, we can say that the semiclassical adiabatic approach describes correctly the system down to quite low temperatures. 

In the linear response regime, without time perturbations, the spectral function allows to derive all the Green functions using the fluctuation-dissipation theorem. Clearly, out-of equilibrium (finite bias and/or time perturbations) not only the imaginary part of the retarded Green function is important, but, as discussed in the previous section, also other Green functions are relevant. In this review, we will analyze the effects of non-equilibrium Green functions directly on the out-of-equilibrium response functions (for example, currents and pumping charges). It is well known that, in the out-of-equilibrium regime, the self-consistent adiabatic approach becomes progressively  more exact in comparison with the equilibrium and linear response regime. For example, in the next subsection, we will show that the bias voltage introduces an additional effective temperature for the vibrational degrees of freedom, therefore the semiclassical approach represents more and more the physical situation.

\subsection{Non-equilibrium conditions at T=0}

\begin{figure}
\hbox{\hspace{-1.85cm}\includegraphics[width=12cm,height=10.0cm,angle=0]{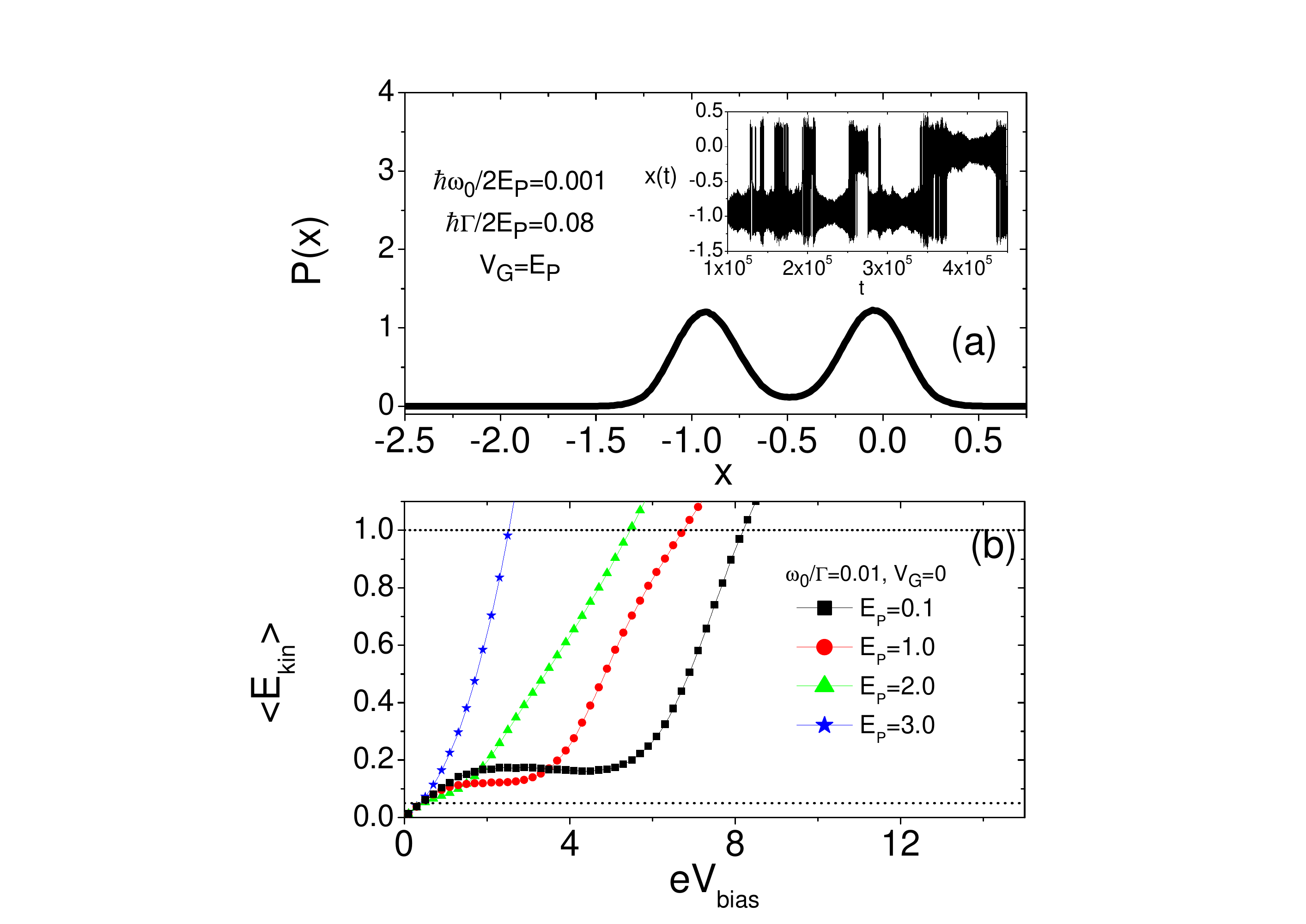}}
\caption{(Color online). Analysis of the validity of the adiabatic
approximation. Panel (a): Dimensionless position
distribution probability for $eV_{Bias}/2E_{P}=0.1$. 
Inset of panel (a): solution of the Langevin equation Eq.~(\protect\ref{Lange}) for the same values of parameters as in the main plot. Panel (b): Average kinetic
energy $\langle E_{kin}\rangle$ as function of the bias voltage for different interaction strengths $E_{P}$. Adapted from Ref.~\cite{Nocera2011}.}\label{Fig2a}
\end{figure}

\begin{figure}
\hbox{\hspace{-1.3cm}\includegraphics[width=12cm,height=9.0cm,angle=0]{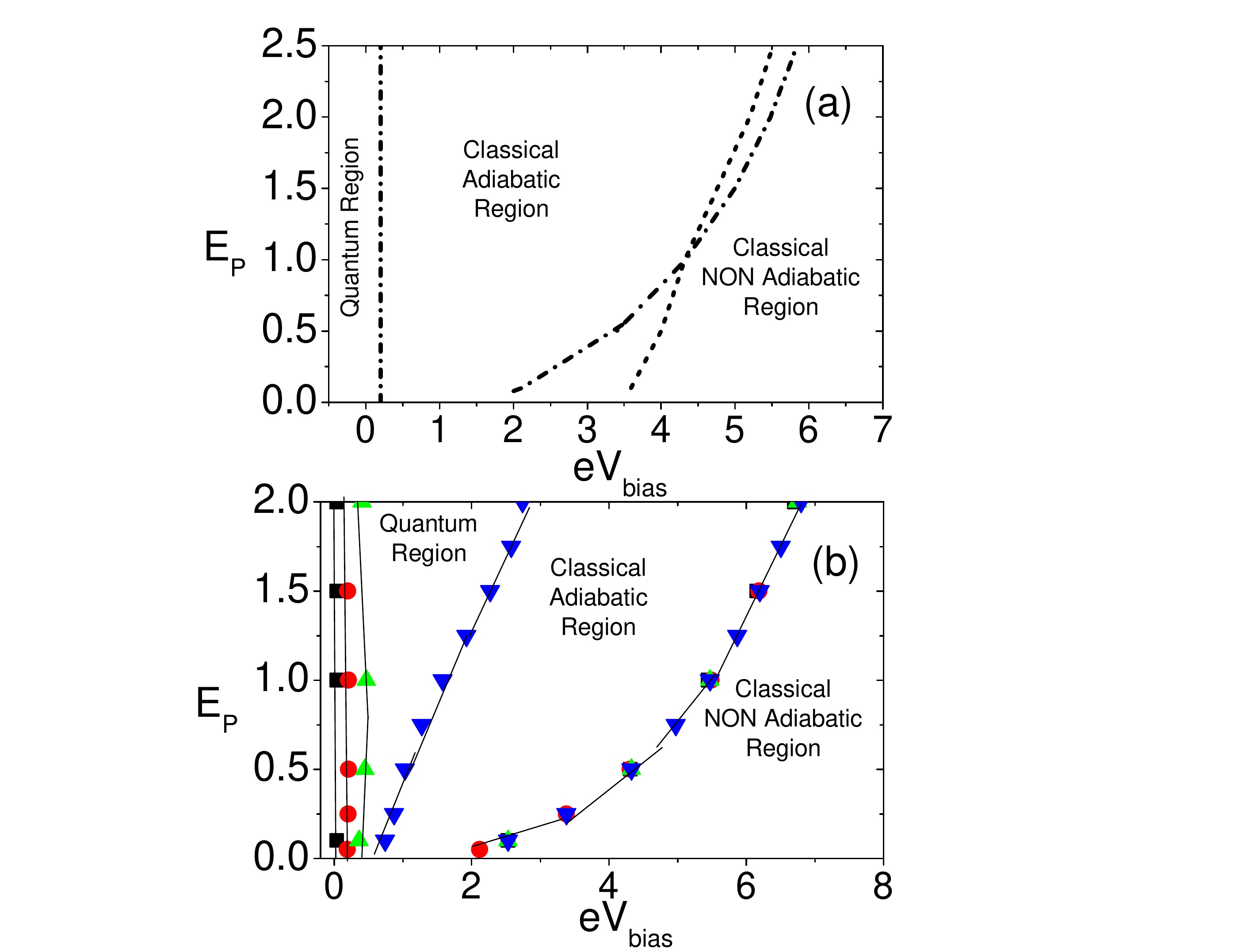}}
\caption{(Color online). Phase diagrams expressing the validity of the adiabatic
approximation. Panel (a): Phase diagram at
fixed adiabatic ratio $\omega_0/\Gamma=0.05$. The dashed
(black) line indicates the QR-CAR crossover for $V_G=0$ and $V_G=1$. The dotted and dashed dot-
ted lines indicate the CAR-CNAR crossover for $V_G=0$ and $V_G=1$, respectively. Panel (b): Phase
diagram at fixed gate voltage $V_G=0$ (asymmetric static potential) for different 
adiabatic ratios
$\omega_{0}/\Gamma = 0.01-0.05-0.1-0.25$. Adapted from Ref.~\cite{Nocera2011}.}\label{Fig2b}
\end{figure}

In this subsection, we consider $\gamma=0$ (no coupling to the lead phonon degrees of freedom) and no temperature gradient between the electronic leads which have been considered at zero temperature $T_{L}=T_{R}=T=0$. A careful analysis of this regime has been provided by the present authors in Ref.\cite{Nocera2011}. Here, we review the main results of that analysis for the sake of completeness. The first step of that analysis consists in obtaining the displacement distribution probability, $P(x)$, associated to the dynamics of the nanosystem vibrational mode. It results from the solution of a single Langevin equation with no forcing term, see Eq.~(\ref{Langevin}) and Ref.\cite{Nocera2011}. In panel (a) of Fig.\ref{Fig2a}, the $P(x)$ is shown in the strong coupling regime $E_{P}>>\hbar\Gamma$ at equilibrium for zero temperature. We highlight here that the bi-modality is provided by the interaction between the electronic and vibrational degrees of freedom which modifies the generalized force in Eq.~(\ref{fortot}).  The actual dynamics of the position of the oscillator $x(t)$ is plotted in the inset of the same panel in Fig.\ref{Fig2a} and shows the 'jumps' between the two symmetric potential wells. The kinetic energy of the oscillator provides a useful quantity for checking the validity of the adiabatic approximation. In panel (b) of Fig.\ref{Fig2a}, it is plotted as a function of the bias voltage applied to the junction. As one increases the electron-oscillator interaction, one can see that a non-monotonic region at intermediate bias voltages appears. Interestingly, for small bias voltages, all the curves show a linear behaviour (not shown), meaning that in this regime the non-equilibrium electronic bath provides an effective temperature proportional to the bias $eV_{bias}$. As already stated in Ref.\cite{Nocera2011}, by using the kinetic energy of the oscillator one can build a phase diagram of the model defining the range of validity of the adiabatic approximation. By comparing the average kinetic energy with the vibrational energy $\hbar\omega_0$ and the characteristic electronic energy $\hbar\Gamma$, one can define a Quantum Region (QR) ($\langle E_K\rangle<\hbar\omega_0/2$), a classical adiabatic region (CAR) ($\hbar\omega_0/2<\langle E_K\rangle<\hbar\Gamma$), and finally a classical non-adiabatic region in the phase diagram (CNAR) ($\langle E_K\rangle>\hbar\Gamma$). The analysis is reported in panels (a-b) of Fig.\ref{Fig2b}). We expect that the adiabatic approximation is reliable in the intermediate classical adiabatic region that occupies the majority of the area shown in panel (a) of Fig.\ref{Fig2b}, where the phase diagram in the plane ($V_G,eV_{bias}$) is shown. The quantum region is confined in a striped area on the left of the plot corresponding to small values of the bias voltages, as physicallly expected. The boundary between the CAR and the CNAR regions is slightly dependent on the gate voltage, showing that in the regime of low dot density occupation the validity of the approach is stronger. The plot reported in panel (b) of Fig.\ref{Fig2b} shows the phase-diagram in the same plane as in panel (a) for different values of the vibrational energy $\hbar\omega_0$, showing that the boundary QR-CAR moves to larger values of the bias voltage while the CAR-CNAR boundary is almost unaffected. As a consistency check of the adopted approach we observe that, if one increases the characteristic energy of the vibrational mode, the area of validity of the semiclassical adiabatic approximation shrinks.

\section{Charge and heat transport} \label{thermo}

In this section we analyze the electronic transport properties of our quantum dot (or molecule) in the
regime of validity of the adiabatic approach. In particular, we will focus on electronic properties as electrical and thermal conductivities  resulting from the
average over the dynamical fluctuations of the oscillator motion.
To this aim, we report the zero-th order adiabatic expression for the electronic current $I^{(0)}$
(averaged over the distribution probability of the oscillator $P(x,\textrm{v})$)
\begin{equation}
I^{(0)} = \frac{e}{\hbar} \int_{-\infty}^{+\infty} \frac{ d
(\hbar \omega)}{8 \pi}  \hbar\Gamma[f_{L}(\omega)-f_{R}(\omega)]
A(\omega),\label{currs}
\end{equation}
and the conductance $G$
\begin{equation}
G=\left( \frac{e^2}{\hbar} \right) \left( \frac{\hbar
\Gamma}{4} \right) \int_{-\infty}^{+\infty} \frac{ d (\hbar
\omega)}{2 \pi} A(\omega) \left[ -\frac{\partial
f(\omega)}{\partial (\hbar \omega)} \right],\label{condus}
\end{equation}
where $A(\omega)$ is the spectral function defined in Eq.~(\ref{Specnew}),  with $f(\omega)=1/(\exp{[\beta (\hbar \omega-\mu)]}+1)$ the free
Fermi distribution corresponding to the chemical potential $\mu$ and the temperature $T$, and $\beta=1/k_{B} T$

\begin{figure}
\hbox{\hspace{-0.7cm}\includegraphics[width=10cm,height=8.0cm,angle=0]{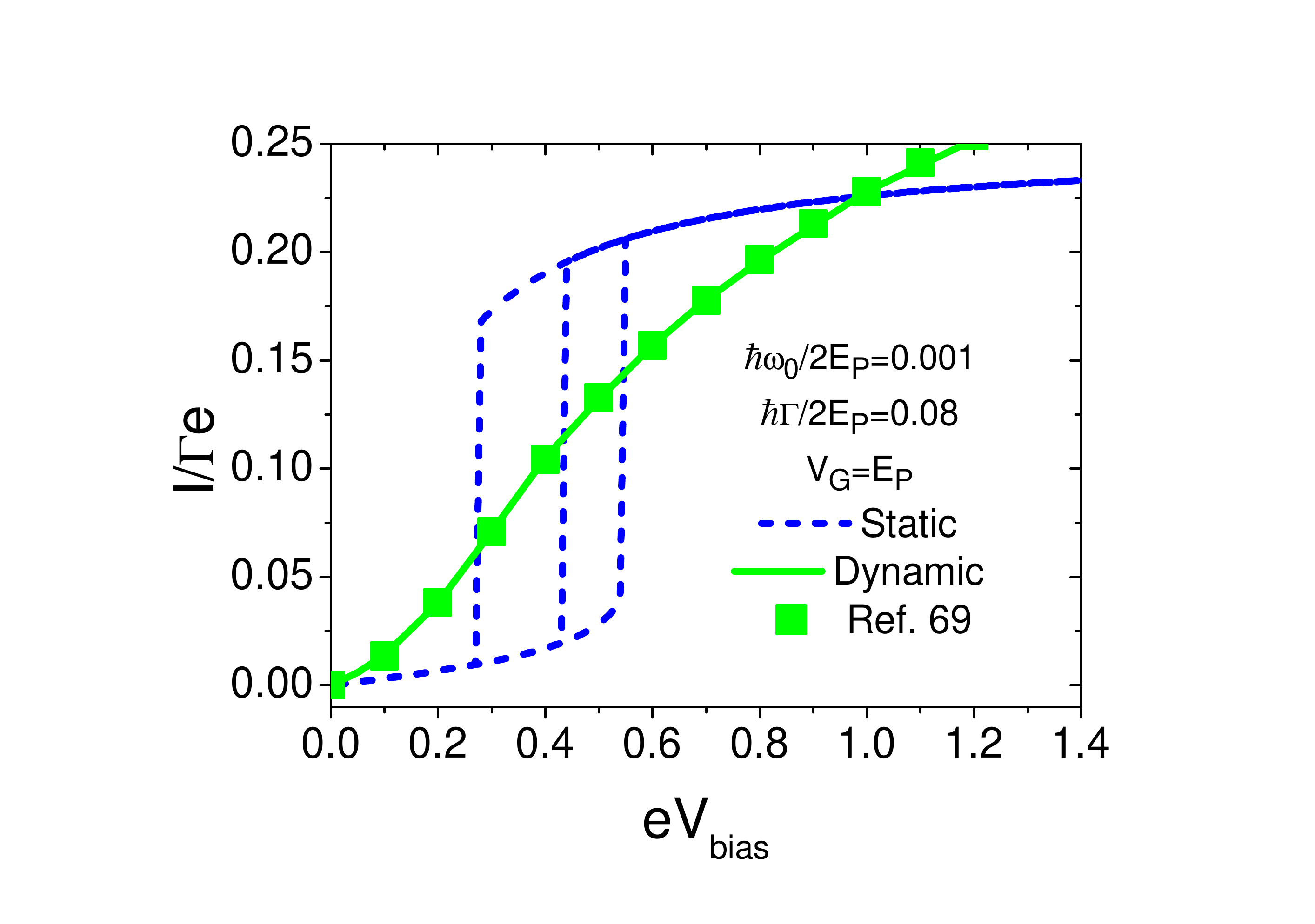}}
\caption{(Color online). Electronic transport properties within the range of validity of the adiabatic approximation. Current ($e\Gamma$ units) voltage
($eV_{bias}$ in $2 E_{P}$ units) characteristic of the device. Solid (green) curve is
drown from Ref.\cite{Pistolesi2008}, squares indicate the results of the semiclassical adiabatic approach adopted in this paper
and dashed (blue) line indicates the I-V in the infinite mass static approximation. Adapted from Ref.~\cite{Nocera2011}.}\label{Fig3}\end{figure}

We can now discuss briefly one of the main results for the current-voltage characteristic given by the adiabatic approach at zero temperature. In Fig.\ref{Fig3}, the I-$V_{bias}$ characteristic for $V_{G}=E_P=2$ is shown comparing the adiabatic approach adopted in this paper (squares), with the limit where the mass of the oscillator is considered as infinitely large (static approximation) and no dynamics is calculated (dashed line). Furthermore, we report, as a full line (green on line), data provided by an independent calculation in Ref.\cite{Pistolesi2008}, which agree completely with our calculations. As one can see, the results obtained in the adiabatic limit completely wash out the hysteresis effects and bi-stability that is obtained in the static approximation, showing that the inclusion of the slow dynamics of the oscillator structurally modifies the I-$V_{bias}$ curve. A similar scenario is obtained for the electronic conductance (not shown in Fig.\ref{Fig3}). In this case, the dynamical correction due to the adiabatic approach gives a substantial broadening of the resonance peak that is shifted by the polaronic effect (see Ref.\cite{Nocera2011}).

We devote the rest of this section to the discussion of the thermoelectrical properties of a molecular junction in the linear response regime. Our focus will be manly on 
the role of the electron-vibration coupling. To this aim, we introduce a
temperature gradient between the leads, and focus on the linear
response regime around the average chemical potential $\mu$ and
the temperature T ($\Delta T \rightarrow 0 $, $V_{bias}
\rightarrow 0$). We also introduce the possibility of presence of
interaction between the relevant vibrational mode of the
molecule and the phonons in the leads. Indeed, if this mode is
elastically coupled with a neighbor atom of the leads by a spring
with constant $k'$, one gets for the dissipative contribution of
the leads in the dynamics of the mode $\gamma \simeq 16 k'^2 / (m M
\omega_D^3)$. As in the previous section, we consider parameters
appropriate for molecular junctions based on $C_{60}$ molecules.
Taking the mass $m$ of the $C_{60}$ molecule and the atomic mass
$M$ of Ag, Au, and Pt (typical metallic leads), $\hbar \gamma$ is
of the order of $7.68$ meV, $7.74$ meV, and $2.98$ meV,
respectively. The smallest value of coupling to phonon baths is
due to the largest Debye frequency $\omega_D$ of platinum. In any
case, $\hbar \gamma \simeq 3-8$ meV for these metals, therefore
$\omega_0$ is of the same order of $\gamma$.

Within the general approach discussed in the previous sections, we can
calculate all the observable quantities relevant for studying
the thermoelectric properties. For example, the Seebeck
coefficient is given by $S=-G_S/G$, where the charge conductance $G$ has been defined in Eq.~(\ref{condus}), and 
\begin{equation}
G_S=  \left( \frac{e}{\hbar} \right) \left( \frac{\hbar \Gamma}{4
T} \right) \int_{-\infty}^{+\infty} \frac{ d (\hbar \omega)}{2
\pi} (\hbar \omega) A(\omega) \left[ -\frac{\partial
f(\omega)}{\partial (\hbar \omega)} \right], \label{conducts}
\end{equation}
with $f(\omega)$ the free Fermi distribution. Then, we will
calculate the electron thermal conductance $G_K^{el}=G_Q+T G_S S$,
with
\begin{equation}
G_Q=  \left( \frac{1}{\hbar T} \right) \left( \frac{\hbar
\Gamma}{4 } \right) \int_{-\infty}^{+\infty} \frac{ d (\hbar
\omega)}{2 \pi} (\hbar \omega)^2 A(\omega) \left[ -\frac{\partial
f(\omega)}{\partial (\hbar \omega)} \right]. \label{conductq}
\end{equation}

In order to estimate the thermal conductance, one can determine the vibrational energy
currents directly from the derivative of the oscillator energy.
\cite{Pana2012} The oscillator is directly in contact with phonon
leads, but only indirectly with electron leads due to the coupling between the oscillator and the electronic level on the molecule (see Fig.\ref{Fig1}).
The phonon thermal conductance $G_K^{ph}$ can be
calculated within the linear response
regime\cite{Wang2007a,Wang2008} around the temperature $T$ as
\begin{equation}
G_K^{ph}=\lim_{\Delta T \rightarrow 0}
\frac{(J_L^{ph}-J_R^{ph})}{2 \Delta T}.
\end{equation}
The total thermal conductance is then given by the sum $G_K$ of the electron and phonon thermal conductance:
\begin{equation}
G_K=G_K^{el}+G_K^{ph}.
\end{equation}
Therefore, one can easily evaluate the total figure of merit $ZT$ (valid in the linear response regime): 
\begin{equation}
ZT=\frac{G S^2 T}{G_K}.
\end{equation}
When the coupling of the center of mass mode to the metallic leads can be neglected
 ($\gamma=0$), $G_K=G_K^{el}$, so that $ZT =ZT^{el}$, which
characterizes the electronic thermoelectric properties. When $\gamma$ is different from zero, the contribution of the thermal phonon conductance $G_K^{ph}$ is not negligible on the determination 
of the figure of merit $ZT$. Not only the electronic transport functions but, as discussed below, also $G_K^{ph}$ are sensitive to the effects of the electron-vibration coupling which tend to reduce the value of $ZT$.

In this section, we assume $\hbar \Gamma \simeq 20$ meV as energy
unit. As a consequence, $\Gamma$ will be the frequency unit. We
will also assume $\omega_0=0.25 \Gamma$ and vary $\gamma$ from $0.15
\Gamma$ to $0.40 \Gamma$ (simulating, as discussed above, the effects of different
metallic leads). We will measure times in units of $1/\Gamma$,
temperatures in units of $\hbar \Gamma /k_{B}$ (ambient temperature
$T_A \simeq 1.25$ in these units). Finally, we fix the
average chemical potential $\mu=0$. Our analysis will mainly focus on the phonon
energy transmission and on the electronic level variation with
respect to the leads chemical potential. These variations can be
controlled, in our model, changing the gate potential $V_{G}$ from
$V_{G}=0$, when the electronic level coincide with the lead chemical potential
(resonant case), to a very different value $|V_{G}| >>0$
(off-resonant case).

\begin{figure}
\centering
\includegraphics[width=10cm,height=9.0cm,angle=0]{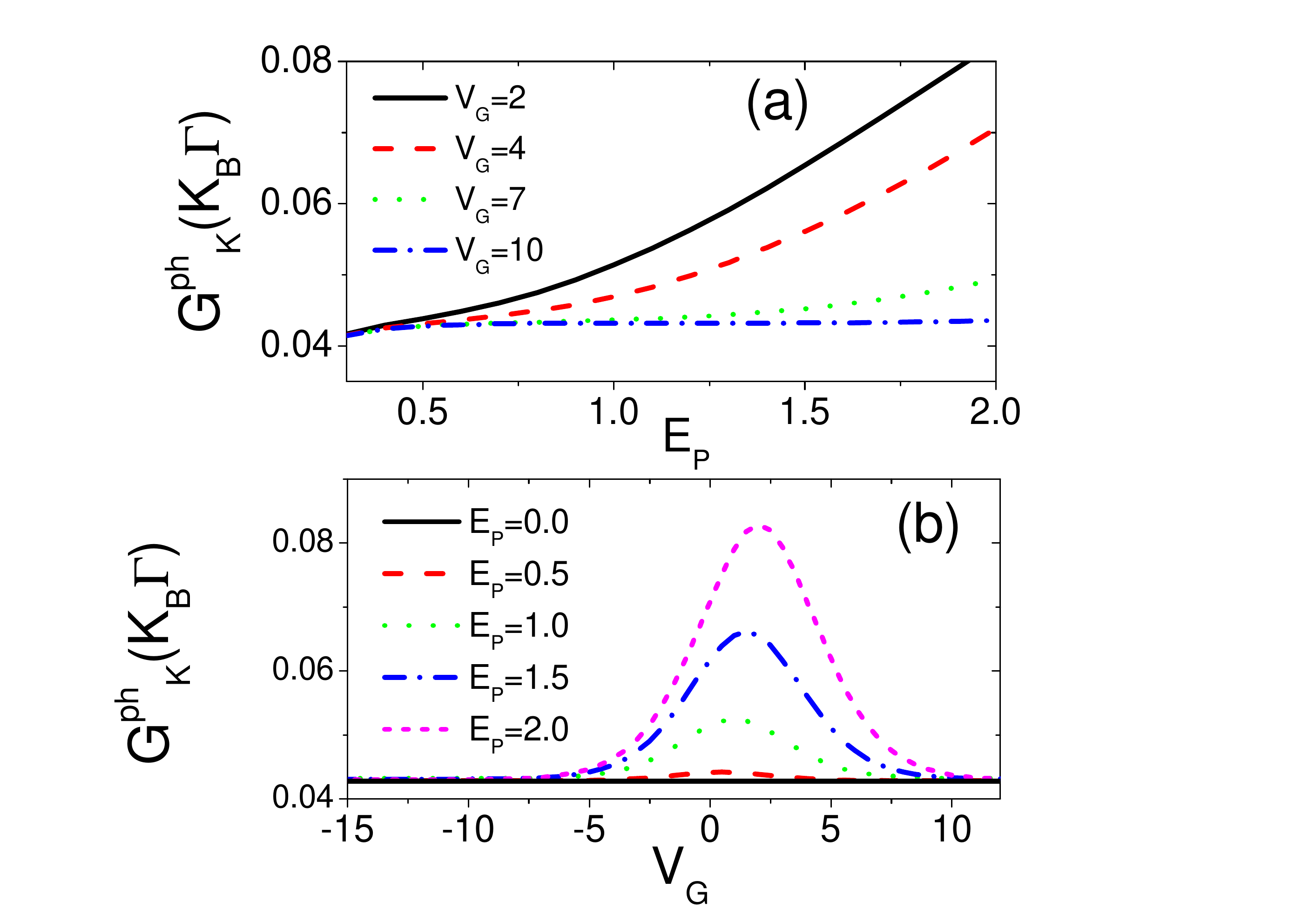}
\caption{(Color online). Phonon thermal conductance in the linear response regime, in the range of validity of the adiatic approximation. 
Panel (a): Phonon thermal conductance
$G_K^{ph}$ (in units of $k_{B} \Gamma$) as a function of
electron-vibration coupling $E_P$ (in units of $\hbar \Gamma$) for
different values of gate voltage $V_{G}$ (in units of $\hbar
\Gamma$). Panel (b): Phonon thermal conductance $G_K^{ph}$ (in
units of $k_{B} \Gamma$) as a function of gate voltage $V_{G}$ (in
units of $\hbar \Gamma$) for different values of
electron-vibration coupling $E_P$ (in units of $\hbar \Gamma$). 
In all the panels, the oscillator damping rate $\gamma=0.15
\Gamma$, $T=1.25 \hbar
\Gamma / k_{B}$ (close to room temperature), and $\omega_0=0.25
\Gamma$. Adapted from Ref.~\cite{Perroni2014}.}\label{Fig5a}
\end{figure}

In order to investigate heat exchange with the molecule, the leads phonon
spectrum is assumed to be acoustic.  For silver (atomic number
Z=47), gold (Z=79), and platinum (Z=78) leads considered in
experimental measurements, \cite{Yee2011} the Debye frequency is
such that $ \hbar \omega_D$ is of the order of $18.5$ meV, $15.1$
meV, and $20.7$ meV, respectively. \cite{Kittel2004} Therefore, $
\hbar \omega_D \simeq 15-20$ meV for these leads. In any case, as
for any large molecule, the center of mass mode is such that
$\omega_0 << \omega_D$.

In Fig. \ref{Fig5a}, we focus on the
phonon thermal conductance $G_K^{ph}$. At moderate values of the coupling between the molecular oscillator and the 
lead phonon bath ($\gamma \simeq 0.15\Gamma$), we  find that, for weak electron-vibration coupling $E_P$
(see panel (a) of Fig. \ref{Fig5a}) or in the off-resonant regime
$|V_{G}| \gg 0$ (see panel (b) of Fig. \ref{Fig5a}), $G_K^{ph}$
reaches its lowest value that is close to $0.04 k_{B} \Gamma \simeq
16$ pW/K. One obtains  this numerical value  when  electron-vibration
effects are neglected (in the off-resonant regime, the electron level density is so low that the effective electron-vibration coupling is negligible).
This value corresponds only to the contribution given by the
phonon leads. We point out that this asymptotic value of
$G_K^{ph}$ is always larger than the values of $G_K^{el}$
corresponding to $E_P=0$. Therefore, $G_K^{ph}$ plays the major
role in determining the total thermal conductance $G_K$ for weak
electron-vibration coupling. If one considers larger values of
$\gamma$ (for example $\gamma \simeq 0.4 \Gamma$), $G_K^{ph}$
plays an even more important role in $G_K$.

In panel (b) of Fig.\ref{Fig5a}, we show that $G_K^{ph}$ always
gets larger with increasing the electron-vibration coupling $E_P$.
Actually, the electron-oscillator coupling gives rise to an
additional damping rate on the vibrational dynamics whose effect is to enhance the thermal
conductivity $G_K^{ph}$. In a certain sense, due to the electron-vibration coupling, the molecular vibrational degrees of freedom are more effectively coupled to the lead phonons favoring 
the heat exchange between them. 
In the quasi-resonant regime (small $V_G$), the
increase of $G_K^{ph}$ can be also favored by the softening of 
oscillator frequency\cite{Nocera2012} due to the enhanced effects of electron-vibration coupling.  

As discussed above, the effects of the electron-vibration coupling
on the oscillator dynamics depends not only on the strength of the
coupling $E_P$, but also on the occupation of the electronic
level.  Actually, the behaviour of $G_K^{ph}$ is strongly dependent
on the value of gate voltage $V_{G}$. As shown in panel (a) of
Fig. \ref{Fig5a}, in the quasi-resonant case ($V_{G}=2$), the
increase of $G_K^{ph}$ as a function of the electron-vibration
coupling $E_P$ is marked. Actually, for $E_P=2$, the value of
$G_K^{ph}$ is doubled. On the other hand, in the off-resonant
regime of low level occupation, the dynamics of the oscillator is
poorly influenced by the electron-vibration effects, even if $E_P$
is not small. Finally, in the panel (b) of Fig. \ref{Fig5a}, we
have analyzed the behaviour of $G_K^{ph}$ as a function of the gate
voltage $V_{G}$ for different values of $E_P$. As expected,
$G_K^{ph}$ shows the largest deviations from the asymptotic value
in the quasi-resonant case. We point out that the peak value is
practically coincident with the value of $E_P$, therefore,
$G_K^{ph}$ is strongly sensitive to the renormalizations of the
electron level induced by the electron-vibration coupling. Indeed,
the role of the phonon thermal conductance $G_K^{ph}$ is important in inducing a suppression of $ZT$, which will discussed in the next section.

\section{Coulomb blockade regime}
\label{eeinter}

In molecular junctions and quantum dots, the strong Coulomb repulsion usually reduces the 
electronic charge fluctuations and suppresses the double occupation of the electronic levels. 
These phenomena are known as Coulomb blockade effects. In order to include this effect in the 
adiabatic approach  discussed in the previous sections, we generalize it to the case in which 
the electronic level can be double occupied and a strong \emph{finite} 
local Coulomb repulsion $U$ is added together with the electron-vibration interaction.

The starting point is the observation that, in the absence of electron-oscillator interaction,
and in the limit where the coupling of the dot to the leads is small $\hbar\Gamma<<U$ (first
correction in $\hbar\Gamma/U$ upon the atomic limit, see section 12.11 of Ref.~\cite{Haug2008}),
the single particle electronic spectral function on the dot is characterized by two spectral 
peaks separated by an energy interval equal to $U$. The peak at $\hbar\omega\simeq-U$ describes
the single occupied electronic level, while the peak at $\hbar\omega\simeq0$ the double occupied one.

In the previous sections, we have seen that, one of the main effects of an adiabatic 
oscillator on the spectral peak at finite temperature is to give an 
extra broadening and a shift proportional to the oscillator-oscillator coupling energy $E_P$.
We therefore expect that, in the presence of an adiabatic oscillator $\hbar\omega_0 
\ll \hbar\omega_D \simeq \hbar\Gamma \leq k_{B} T$ 
one can perturb each spectral peak of the quantum dot independently, obtaining (see details in ref.~\cite{Perroni2014a}) at the zero order of the adiabatic approach
\begin{eqnarray}
A(\omega,x) &=& [1-\rho(x)]\frac{\hbar \Gamma}{(\hbar \omega - \epsilon -\lambda x )^2+(\hbar \Gamma)^2/4}+  \nonumber \\
&& \rho(x) \frac{\hbar \Gamma}{(\hbar \omega - \epsilon -\lambda x
-U )^2+(\hbar \Gamma)^2/4}, \label{function}
\end{eqnarray}
where $\rho(x)$ is the electronic level density per spin. In our computational scheme, 
$\rho(x)$ should be self-consistently calculated for a fixed displacement $x$ 
of the oscillator through the following integral 
$\rho(x)=\int_{-\infty}^{+\infty} \frac{d (\hbar \omega)}{2 \pi i}
G^<(\omega,x)$, with the lesser Green function $
G^<(\omega,x)=\frac{i}{2} [f_L(\omega)+f_R(\omega)] A(\omega,x)$. 
The above approximation is valid if the electron-oscillator interaction is not too large, such that
$\hbar\Gamma\simeq E_P<<U$ and the peaks of the spectral function can 
be still resolved\cite{Perroni2014a}.

This section is intended to report a summary of the results presented in ref.~\cite{Perroni2014a}.
In this reference, it has been shown that, in order to study the heat transport through 
a quantum dot junction at finite $U$, the above approximation for the electronic Green's function 
is a reliable starting point
for the adiabatic approach. The effect of the phonon leads has also been 
included in the adiabatic expansion and it has been 
shown that the oscillator dynamics is described by a 
Langevin equation with the same structure of that derived in sec.~3.
Before discussing the numerical results, it is useful to analyze 
the properties of the electron-oscillator damping rate $\theta'(x)
=\theta(x)/m$, with $\theta(x)$ defined in Eq.~(\ref{thetaeq}) (explicitly calculated in
Ref. \cite{Perroni2014a}) and appearing explicitly in the Langevin equation of the oscillator.
The amplitude of the peaks of $\theta'(x)$, although increasing as a function of the 
electron-oscillator coupling $E_P$,
is always smaller than the damping contribution (the realistic value 
$\gamma=0.15$ is assumed in the second part of this section)
coming from the coupling with the phononic leads. 
Therefore, regardless of its shape, which depends also 
on the dot occupation and obviously on the Coulomb repulsion $U$, $\theta(x)$ 
constitutes a small perturbation of the spatially constant damping rate induced 
by the coupling to the phononic leads.

In the following, unless otherwise stated, we consider a Coulomb repulsion equal to $U=20$ 
(being the largest energy scale in the problem)
and a value $T=1.25$ mimicking a regime close to room temperature conditions.

\begin{figure}
\hbox{\hspace{-0.5cm}\includegraphics[width=10cm,height=9.0cm,angle=0]{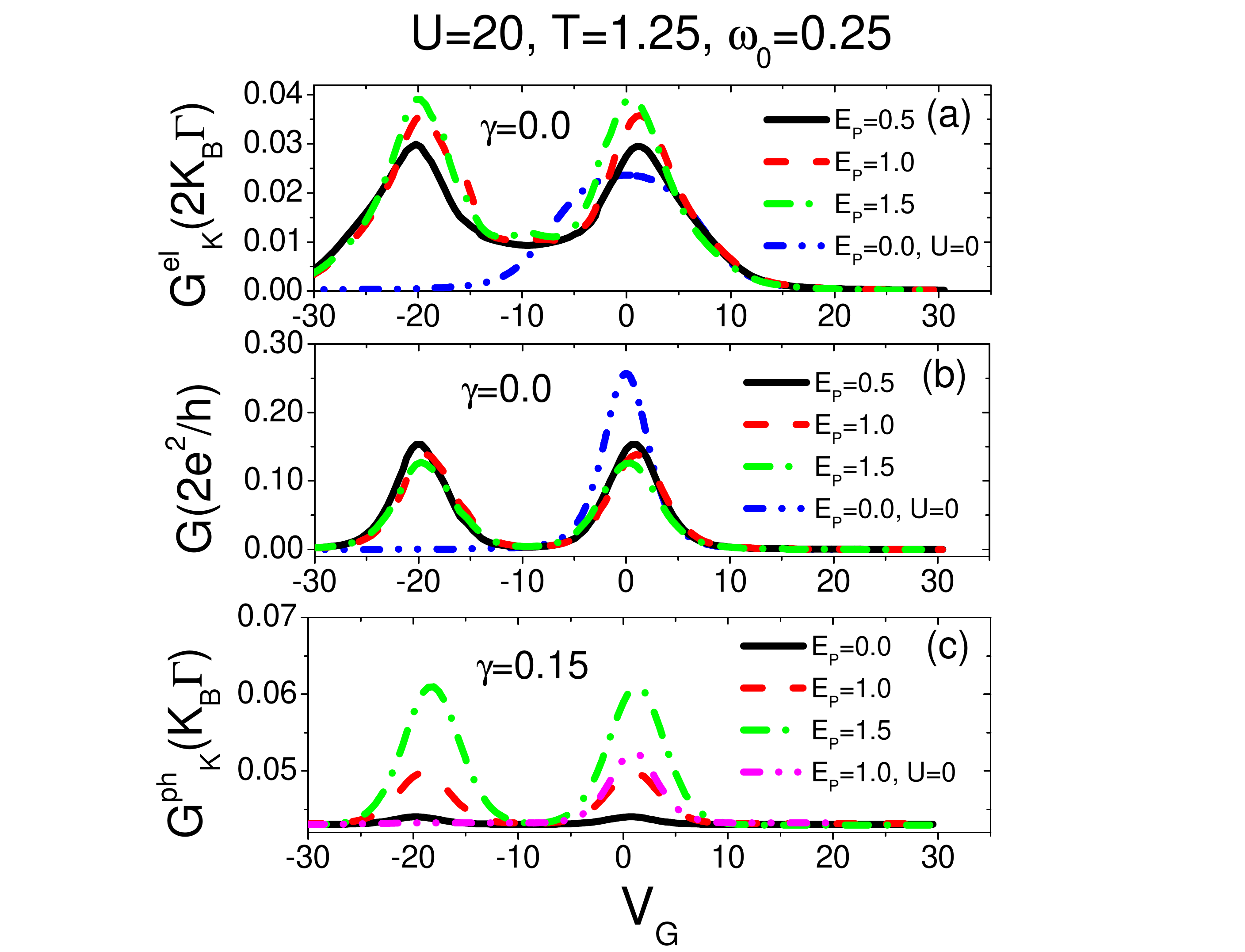}}
\caption{(Color online). Thermal transport properties in the linear response 
regime, in the range of validity of the adiabatic approximation and including 
the electron-electron interaction. 
Panel (a) Electron thermal conductance $G_K^{el}$ in
units of $2 k_{B} \Gamma$ as a function of the gate
voltage $V_{G}$ (in units of $\hbar \Gamma$) for different values
of electron-vibration coupling $E_P$ (in units of $\hbar \Gamma$).
Panel (b) Same as in panel (a) for the electron conductance G in
units of $2 e^2/h$. Panel (c) Phonon thermal conductance
$G_K^{ph}$ (in units of $k_{B} \Gamma$) with 
the oscillator damping rate $\gamma=0.15 \Gamma$. 
Adapted and reproduced with permission of \textcopyright~IOP Publishing from Ref.~\cite{Perroni2014a}. All rights reserved.}\label{Fig6a}
\end{figure}

\begin{figure}
\hbox{\hspace{-0.85cm}\includegraphics[width=11cm,height=9cm,angle=0]{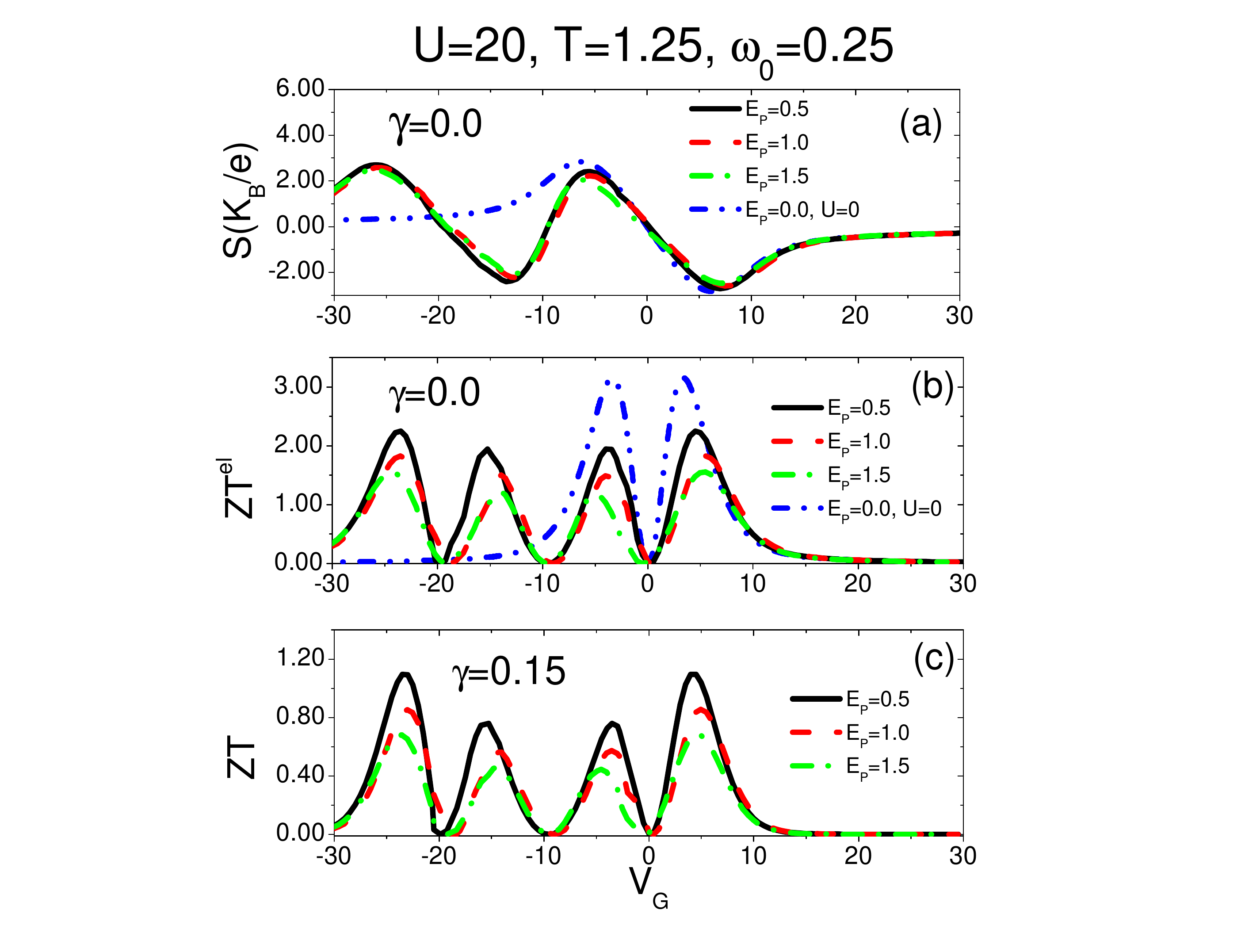}}
\caption{(Color online)Thermal transport properties in the linear response regime, 
within the range of validity of the adiatic approximation and including 
the electron-electron interaction. 
Panel (a) Seebeck coefficient $S$ in units
of $k_{B}/e$ for different values of electron-vibration $E_P$ (in
units of $\hbar \Gamma $). Panel (b): Dimensionless thermoelectric figure of
merit $ZT^{el}$ for the electronic system only. Panel (c) shows the
total thermoelectric figure of merit. 
Adapted and reproduced with permission of \textcopyright~IOP Publishing from Ref.~\cite{Perroni2014a}. All rights reserved.}\label{Fig6b}
\end{figure}

In order to better distinguish thermal electronic effects from those related to the 
direct link between the vibrational molecular modes and the lead phonon bath,  we
first discuss briefly the thermoelectric electronic properties
when $\gamma=0$. In the panels (a)-(b) of Fig. \ref{Fig6a} and in Fig.\ref{Fig6b}, 
we study electronic properties of the quantum dot junction 
as a function of the electron-vibration coupling $E_P$ and the gate voltage
$V_{G}$. The dashed-dotted-dotted (blue) line indicate results obtained for $U=0$ and
no electron-oscillator coupling $E_P=0$. In panel (c) of fig.~\ref{Fig6a} the magenta 
dashed-dotted-dotted line indicates the $G_{K}^{ph}$ for $E_P=1.0$ and $U=0.0$.

We start discussing the main characteristics of the electronic conductance $G$ (panel (b)
of Fig. \ref{Fig6a}) in linear response. Notice that, the height of the peak at $V_G\simeq 0$
of the free case is much higher than the peak at non zero electron-oscillator coupling. 
This effect is expected since the electron-electron interaction usually suppresses the 
electronic conduction. The height of the second peak at $V_{G} \simeq -20$
has the same height of the first peak and can be estimated to be of the order of 
$10^{-1}$ $e^2/h$ (($e^2/h$ is about $3.87 \times
10^{-5}$ S)). 
Away from the main peaks, $G$ has values close to the experimental data 
estimation for a $C_{60}$ junction\cite{Yee2011}.
The double peak shape of the $G$ vs. $V_G$ curve follows closely the behaviour of the $A(\omega,x)$.
Finally, observe that the main effects of the electron-oscillator interaction are a reduction 
of the amplitude of the peaks, an increase of their broadening, and a shift 
their position towards negative energy.

%The charge conductance $G$ is expected to be smaller than the free
%one due to the effects of interactions. As shown in the panel (b)
%of Fig. \ref{Fig6a}, close to room temperature, $G$ has peak values
%of the order of  $10^{-1}$ $e^2/h$ ($e^2/h$ is about $3.87 \times
%10^{-5}$ S). In particular, for $V_{G} \simeq 20$, we have
%checked that $G$ is of the order of $10^{-3}$ $e^2/h$ in agreement
%with the order of magnitude of experimental data in $C_{60}$
%\cite{Yee2011}. As expected, the conductance as a function of the
%gate voltage $V_{G}$ follows a behaviour similar to the double-peak
%structure of the spectral function as a function of the frequency.
%Therefore, $G$ has maxima for $V_{G} \simeq 0=\mu$ and $V_{G}
%\simeq -U$, and a minimum at $V_{G} \simeq -U/2$.

We now investigate the properties of the Seebeck coefficient $S$ of the junction in the 
panel (a) of Fig. \ref{Fig6b}. Notice that $S$ has a peculiar
oscillatory behaviour as a function of $V_G$, with positive peaks
and negative dips. The height of the peaks (same as that of the dips in absolute value)
is about $2 k_{B}/e$ ($k_{B}/e$ is about $86$ $\mu$eV/K). 
Notice that the Seebeck coefficient $S$ is negligible
around the position where the electronic conductance presented the main peaks, that is at 
$V_G\simeq 0$ and $V_G\simeq-20$. This property is a result of the 
strong electron-electron interaction $U$\cite{Liu2010}. Interestingly, 
for large positive values of $V_{G}$, $S$ is small and negative (n-type behaviour). 
In particular, for $V_{G}=20$, $S$ can be estimated to be $-0.45 k_{B}/e \simeq -
38.5 \mu V/K$, close in magnitude of the experimental data
provided for a $C_{60}$ junction in ref.~\cite{Yee2011}.

Another important property to mention in the comparison of $S$ and $G$
is that, away from the points $V_{G}\simeq-20$ and $V_{G}\simeq 0$, while $G$ reduces its amplitude,
$S$ increases its amplitude in absolute value. Moreover,
even in this case, 
the main effect of the electron-oscillator interaction is to reduce the amplitude of the
response function, for all the values of $V_G$ investigated. 
The shift of the peaks of $G$ and of the zeroes of $S$ is
governed by $E_P$ ($n=0.5$ for $V_{G}=E_P$). We point out that, at
fixed gate voltage, while the conductance $G$ (Fig.~\ref{Fig6a}b) shows a 
variation of less than $10\%$, the Seebeck
coefficient shows a larger sensitivity (between $10\%$ and $20\%$) to the change 
of coupling $E_P$ explored (see Fig.~\ref{Fig6b}a). This occurs for energies close to the
minimum and the maximum. For larger values of $V_{G}$, there is an
inversion in the behaviour of $S$ with increasing the coupling
$E_P$.
%It is interesting to analyze the behaviour of  
%the conductance $G$ and the Seebeck coefficient $S$ upon crossing the level resonance 
%(at $V_G=0$ in the absence of many-body interactions on the molecule). 
%These two quantities show opposite behaviours: close to the electronic resonance, 
%$G$ is peaked, $S$ vanishes, while, far from the resonance, $G$ is reduced, $S$ tends 
%to acquire large values in modulus (a maximum and a minimum corresponding to electron 
%or hole transport, respectively).
%As shown in the panels (a) of Fig. \ref{Fig6b} and (b) of Fig. \ref{Fig6a}, it is clear that
%the most relevant effect of the coupling $E_P$ on $G$ and $S$ is to shift the stationary points of the curves and reduce the
%magnitude of the response function. 

In panel (a) of Fig. \ref{Fig6a}, we also 
study the electronic thermal conductance $G_K^{el}$ as a function of $V_G$.
The principal characteristic to mention is that, by increasing the electron-oscillator coupling
$E_P$, the height of the two-peaks (again, similarly to the conductance $G$, notice 
the two peak structure\cite{Liu2010} of $G_K^{el}$)
at $V_{G}\simeq 0$ and $V_G\simeq-20$ increases. This is due to the opening 
of extra channels of conduction due the larger broadening of the spectral peaks.
Indeed, for $E_P=1$, the $G_K^{el}$ assumes
values $0.05 k_{B} \Gamma \simeq
20$ pW/K which are close to those estimated experimentally in hydrocarbon 
molecules\cite{Wang2007}(50pW/K). Notice that 
the height of the peaks ($0.01$ $k_{B} \Gamma$ where $k_{B} \Gamma$ is about $4.198
\times 10^{-10}$ W/K for $\hbar \Gamma \simeq 20$ meV) 
are smaller than  the thermal conductance quantum
$g_0(T) = \pi^2 k_{B}^{2} T/(3 h)$ at room temperature $T=1.25
\hbar \Gamma \simeq 300 $K ($g_0(T) \simeq 9.456 \times 10^{-13}
(W/K^2) T)$ \cite{Jezouin2013}.

%As reported in Fig. \ref{Fig6a}, a different behaviour is shown by the thermal electron conductance
%$G_K^{el}$. Indeed, $G_K^{el}$ can be enhanced with increasing the
%electron-oscillator coupling $E_P$. For example, at the resonance
%and at $E_P=1$, $G_K^{el}$ is of the order of $0.05 k_{B} \Gamma \simeq
%20$ pW/K for different temperatures, therefore it gets closer to
%the conductance of the order of $50$ pW/K measured in hydrocarbon
%molecules.\cite{Wang2007}  
 
%Actually, within the adiabatic
%approach, more energetic channels open with increasing the
%electron-vibration coupling since the effective level is
%renormalized by the oscillator dynamics that becomes less
%localized.  As shown in the panel (a) of Fig.\ref{Fig6a}, with varying the
%level energy $V_{G}$, the electron thermal conductance $G_K^{el}$
%shows the characteristic double peak structure due to correlation
%effects \cite{Liu2010}. The peaks values of $G_K^{el}$ are of the
%order of a few $0.01$ $k_{B} \Gamma$ ($k_{B} \Gamma$ is about $4.198
%\times 10^{-10}$ W/K for $\hbar \Gamma \simeq 20$ meV). Therefore,
%the peak values are smaller than  the thermal conductance quantum
%$g_0(T) = \pi^2 k_{B}^{2} T/(3 h)$ at the room temperature $T=1.25
%\hbar \Gamma \simeq 300 $K ($g_0(T) \simeq 9.456 \times 10^{-13}
%(W/K^2) T)$ \cite{Jezouin2013}.

It is useful to observe the similarities between the $G_K^{el}$ 
shown in Fig.~\ref{Fig6a}a and the phonon thermal conductance
$G_K^{ph}$ reported in Fig.~\ref{Fig6a}c. Indeed, 
they share a double peak shape and 
the property of increasing their amplitude with $E_P$.
Notice that the phonon thermal conductance $G_K^{ph}$ in fig.~\ref{Fig6a}c has been obtained
using $\gamma= 0.15 \Gamma$ and assumes in this case 
an amplitude comparable with $G_{K}^{el}$. The amplitude of $G_K^{ph}$
increases as a function of $\gamma$ so a larger value of this quantity would imply 
a larger contribution to the total thermal conductance $G_K$. 
Overall, at fixed gate voltage away from the peaks 
at $V_G\simeq 0$ and $V_{G}\simeq-20$, $G_K^{el}$ has an amplitude larger than $G_K^{ph}$.

%We point out that the behaviour of the electron thermal conductance
%$G_K^{el}$ shown in the panel (a) of  Fig. \ref{Fig6a} bears a
%strong resemblance with that of the phonon thermal conductance
%$G_K^{ph}$ reported in panel (c) of the same figure. Both have a double peak
%structure, and both are enhanced by the electron-vibration
%coupling. Moreover, $G_K^{el}$ acquires values larger than those
%of $G_K^{ph}$ in the energy region $-U \le V_{G} \le 0$.
%Obviously, the values of these quantities are comparable for the
%chosen value of phonon induced damping rate $\gamma= 0.15 \Gamma$.
%If one consider larger values of $\gamma$ (for example $\gamma
%\simeq 0.4 \Gamma$), then $G_K^{ph}$ would play a major role in
%the total thermal conductance $G_K$.  In any case, the values of
%$G_K^{ph}$ and $G_K^{el}$  differ for $V_{G} \gg 0$ and $V_{G}
%\ll -U$ since $G_K^{ph}$ acquires a finite asymptotic value
%(obtained even in the absence of interactions on the molecule),
%while $G_K^{el}$ goes rapidly to zero.

The results described above concerning the electronic conductance $G$, 
Seebeck coefficient $S$, and thermal conductance $G_{K}^{el}$ all combine in 
giving a pure elecronic figure of merit $ZT^{el}$. This latter quantity has been 
shown in panel (b) of Fig. \ref{Fig6b}. Notice the four peak structure with amplitude larger than $2$.
The dashed-dotted-dotted blue curve shows the results for $U=0$, confirming that the
effect of the electron-electron interaction is to reduce the amplitude and shift the
peaks of the figure of merit, even neglecting the effect of the phononic leads. 
Increasing the electron-oscillator coupling $E_P$, the behaviour of $G$, $S$ and $G_{K}^{el}$
cooperates toward a further reduction of the height of the figure of merit peaks,
which can assume a common average height smaller than $2$.
Observe finally that the position of the peaks in $ZT^{el}$ roughly coincides with 
the position of the peaks and dips of the Seebeck coefficient.

%As shown in the panel (b) of Fig. \ref{Fig6b}, we analyze the
%behaviour of the electronic thermoelectric figure of merit
%$ZT^{el}$ neglecting the contribution from $G_K^{ph}$ (coupling $\gamma=0$). The
%quantity $ZT^{el}$ shows four peaks whose values are larger than
%$1$, but smaller than the peak value around $3$ obtained in the
%absence of interactions. We stress that the peak values of
%$ZT^{el}$ at room temperature are almost coincident with maxima
%and minima of the Seebeck coefficient $S$. Actually, close to room
%temperature, the small values of the conductance $G$ are fully
%compensated by the large values of the Seebeck coefficient $S$.
%With increasing the electron-vibration coupling $E_P$, the
%reduction of $G$ and $S$ combines with the enhancement of
%$G_K^{el}$ leading to a sensible reduction of the figure of merit
%$ZT^{el}$.  For example, for $E_P=1$, the
%peak value is smaller than $2$. It is important to emphasize that a value 
%close or larger than $2$ is considered 
%relevant for actual thermoelectric applications. 
%Therefore, even if one neglects the role of phonon
%thermal conductance, the effect of electron-electron and
%electron-vibration interactions is able to induce a reduction of
%the figure of merit.

We can now study the behaviour of the thermal properties of 
the junction when also a coupling $\gamma$ different from zero 
of the vibrational mode with the phononic leads is present.

We begin discussing the phonon thermal conductance $G_{K}^{ph}$ which has been 
already briefly described above in comparison with the electronic contribution $G_K^{el}$.
As observed early in this section, the $G_{K}^{ph}$ behaviour is strongly linked to
the \emph{total} damping rate of the oscillator. In the case where $\gamma  = 0.15 \Gamma$,
the electron-oscillator contribution to the damping $\theta'(x)$ is smaller than the
phononic leads contribution, so one expects that the behaviour of $G_{K}^{ph}$ is mostly 
influenced by the electron-oscillator coupling $E_P$. Indeed, this is what is 
observed in panel (c) of Fig.\ref{Fig6a}. If $E_P$ is not too large, 
the estimated height of peaks (as always at $V_{G}\simeq 0$ and $V_{G}\simeq-20$) of the $G_{K}^{ph}$
roughly coincides with an estimation of the same quantity in the absence of electron-electron
and electron-oscillator interaction given in ref.~\cite{Perroni2014}.
Furthermore, panel (c) of fig.~\ref{Fig6a} shows that the increase of the amplitude of $G_{K}^{ph}$
strongly depends on the gate voltage, while, at $E_P=1$, the height of the main peaks is $0.05 
k_{B} \Gamma \simeq 20$ pW/K, which is close to the thermal conductance 
measured for molecules anchored to gold\cite{Wang2007,Meier2014}.
Summarizing, in the presence of a realistic 
coupling with the vibrational degrees of freedom of the leads, electron-electron interaction,
and within the limits of validity of the adiabatic approximation, 
the phonon thermal conductance depends crucially on the electron-oscillator induced 
rate of damping.

We finally explore the effect of the phonon thermal conductance on the total 
figure of merit $ZT$, which is plotted in panel (c) of fig.~\ref{Fig6b} as a function 
of the gate voltage with increasing values of the electron-oscillator coupling $E_P$.
Comparing with the results obtained for the pure electronic $ZT^{el}$, one can see that
the effect of $G_K^{ph}$ is to further reduce the amplitude 
of the figure of merit. The height of the peaks
of $ZT$ is slightly above unity for an intermediate coupling $E_P=0.5$, 
pointing to a reduction by a factor of $2$ when compared with those observed in $ZT^{el}$.
If the electron-oscillator is further increased, the $ZT$ peaks' height can reduce to values 
below the unity. We can therefore conclude this section observing that the combined effect of 
a large electron-electron interaction strength, an interaction with an adiabatic oscillator, 
and the presence of vibrational degrees of freedom in the leads,
has the final result of reducing the thermoelectric capabilities of a molecular dot device. 
Nevertheless, it has been shown that for a large set of the model parameters, the 
total figure of merit $ZT$ is still significant, pointing to possible future 
thermoelectric applications of these devices.

%Finally, in panel (c) of Fig. \ref{Fig6b}, we focus on the total figure of
%merit $ZT$ as a function of the level energy $V_{G}$ for different
%values of electron-vibration coupling $E_P$ at $U=20  \hbar
%\Gamma$ including the effects of the phonon thermal conductance
%($\gamma=0.15 \Gamma$). From the comparison with the results
%discussed above, it emerges that the phonon
%thermal conductance $G_K^{ph}$  induces an additional suppression
%of $ZT$. For the realistic value of  $E_P=0.5$ (intermediate
%coupling regime), the peak values of $ZT$ are decreased by a
%factor of $2$ in comparison with $ZT^{el}$, therefore the
%reduction of $ZT$ is not strong. Only for unrealistically large
%electron-vibration couplings ($E_P$ larger than $\hbar \Gamma$), $ZT$
%acquires peak values less than unity. Summarizing, the cooperative
%effects of phonon leads, electron-electron and electron-vibration
%interactions on the molecule are able to weaken the thermoelectric
%performance of this kind of device. However, within a realistic
%regime of parameters, the thermoelectric figure of merit $ZT$ is
%still of the order of unity, making these devices potentially useful for thermoelectric applications.

\section{Time dependent perturbations}

\subsection{Suspended CNT with external antenna
effects: electronic transport} \label{antenna}

\begin{figure*}
\centering
\includegraphics[width=13cm,height=6.0cm]{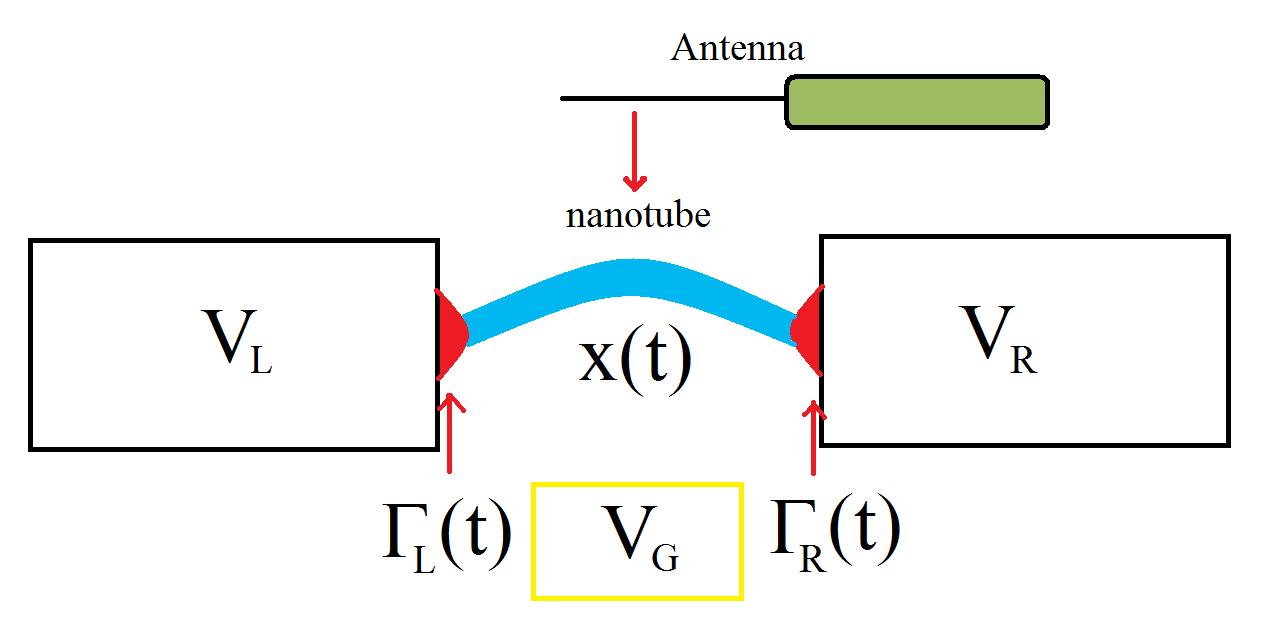}
\caption{Sketch of the device investigated in section 7. A carbon nanotube
is suspended between two metal leads to which a bias voltage
$V_L=-V_R=V_{bias}/2$ is applied. The motion of the nanotube is activated by an external antenna. The contanct giving the gate potential $V_G$ is also shown. In the case of a two parameter charge pumping, one has also a time dependent modulation of the potential barriers between the leads and the nanotube, $\Gamma_L(t)$, and $\Gamma_R(t)$.} \label{figCNT}
\end{figure*}

In this section we discuss the effect of the time-dependent perturbation induced by an external antenna on molecular devices. 
Our work has been motivated by recent transport experiments performed on suspended carbon nanotubes\cite{Steele2009,Huttel2009}, where an external temporal periodic perturbation given by a nearby antenna actuates the flexural motion of the suspended nanotube. A sketch of the device studied in this section is reported in Fig.~\ref{figCNT}.
The main  assumption is that the coupling between charge and vibrational degrees of freedom is affecting directly the mechanical displacement of the nanotube. 
In the general scheme outlined in Sec.2, the driving considered in this section can be described by (the semiclassical approximation is assumed) 
\begin{equation}\label{eq7.1}
\begin{aligned}
\hat{H}_{driving}(t)&=-xU_{ext}cos({\omega^{'}_{ext}t}),\;\;V_{ext}=0,\\
&\hbar\Gamma_{\alpha}(t)=\hbar\Gamma_{\alpha},\;\;\varepsilon(t)=V_{G},
\end{aligned}
\end{equation}
which means that in Eq.~(\ref{Hel}), the dot $\hat H_{dot}(t)$ and the ${\hat
H}_{leads-dot}(t)$ parts are assumed time independent and the driving is acting directly on the oscillator Hamiltonian in Eq.~(\ref{Hosc}), where the coupling with the vibrational degrees of freedom in the leads has also been neglected. Unless otherwise stated, in this and the next sections, we neglect the effect of a \emph{finite} Coulomb repulsion $U$. We stress that in suspended Carbon nanotubes devices, often a description in terms of a spinless electronic level is sufficient for capture the main physics since the distance between adjacent electronic levels of the dot is assumed very large.

\begin{figure}
\hbox{\hspace{-0.5cm}\includegraphics[width=9cm,height=9.0cm,angle=0]{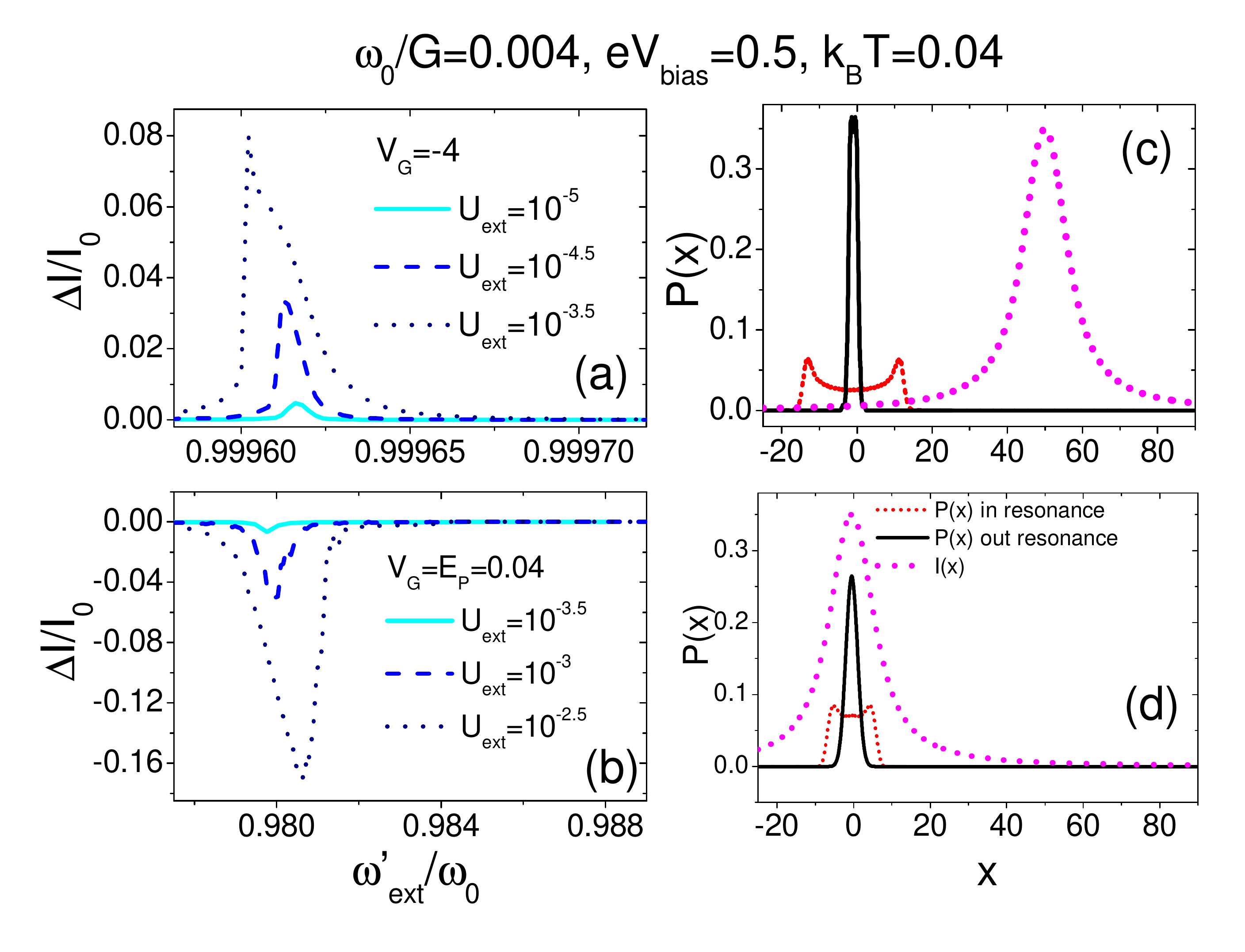}}
\caption{(Color online). Electronic transport properties in the presence of 
an external periodic time-dependent perturbation.
Panel (a) and panel (b): Normalized current change
($\Delta I/I_{0}$) as a function of the external frequency
($\omega'_{ext}/\omega_{0}$) for different antenna amplitudes. 
Panels (c) and (d) show the distribution $P(x)$ both out of
mechanical resonance and at mechanical resonance for the larger value of antenna amplitude
considered in panels (a) and (b). The short-dashed line
(magenta) represents current as function of position $I(x)$. 
Adapted from Ref.~\cite{Nocera2012}.}\label{Fig7a}
\end{figure}

\begin{figure}
\hbox{\hspace{-0cm}\includegraphics[width=9cm,height=9.0cm,angle=0]{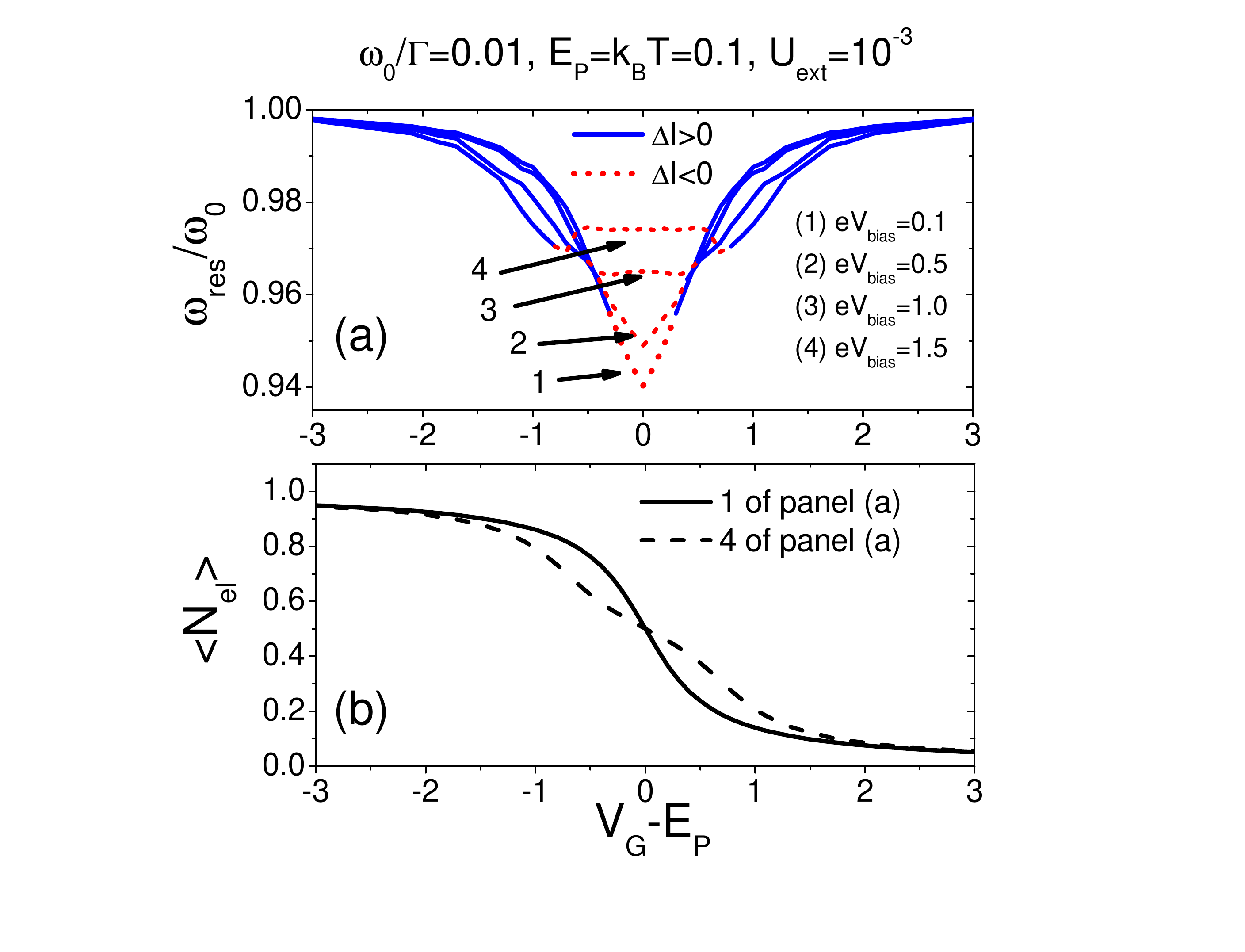}}
\caption{(Color online). Interplay between the resonator frequency and the charge density.
Panel (a):
Resonator frequency at resonance against effective gate voltage
(shifted of $E_{P}$) for different bias voltages.
Solid (blue online) and
short-dashed (red online) portions of each curve indicate
resonance frequency values with positive and negative current
change $\Delta I$, respectively. Panel (b): Electronic occupation
at resonance frequency against effective gate voltage (shifted of
$E_{P}$). Adapted from Ref.~\cite{Nocera2012}.}\label{Fig7b}
\end{figure}

As thoroughly discussed by us in Ref.\cite{Nocera2012} and in Ref.\cite{Weick2011}, the regime of parameters relevant for the experiments is very accurately described by the adiabatic approximation for the vibrational degrees of freedom. Indeed, the vibrating nanotube is oscillating at a frequency in the MHz range so that $\omega_{0}/\Gamma <<1$, where $\Gamma$ is the electronic tunneling rate. Also, a strong coupling between the electronic and vibrational degrees of freedom is realized in the experiments ($E_{P}/\hbar\omega_{0}=10$), while for the other parameters one has $\hbar\omega_{0}<<E_{P} (\sim k_{B}T) << eV_{bias} \leq \hbar\Gamma$. We observe here that the Langevin equation describing the dynamics of the CNT flexural motion acquires a forcing term
\begin{eqnarray}\label{LangevinAntenna}
m \ddot{x} &+& A(x)\dot{x}=F(x)+\xi(t)+
U_{ext}\cos(\omega^{'}_{ext}t),
\end{eqnarray}
where $U_{ext}$ and $\omega^{'}_{ext}$ represent the amplitude and the
external antenna frequency, respectively. In this section, the
coupling with the phonon leads is neglected ($\gamma=0$). In Fig.\ref{Fig7a} and Fig.\ref{Fig7b}, we report a
summary of the main results obtained by us in Ref.\cite{Nocera2012}. In this review, we
intend to focus on the particular case when one tunes the frequency of the external   antenna across the natural vibrational frequency of the oscillating nanotube. 

In panels (a) and (b) of Fig.\ref{Fig7a}, we report the electronic current change 
($\Delta I/I_{0}$, where $\Delta I=I-I_{0}$ and $I_{0}$ is the current observed in 
the absence of external antenna) as a function of the antenna frequency $\omega^{'}_{ext}$ in two different regimes of the gate voltage applied on the nanotube. In panel (a),  we show the case of low occupancy where a small current 
is flowing through the CNT (the external gate voltage with $|V_{G}|>>0$ tunes the electronis level of the CNT away from the bias window), while in panel (b) a high current regime with
$V_{G}=E_P$ is shown. As already discussed by us in Ref.\cite{Nocera2012}, 
a characteristic peak (panel (a)) and dip (panel (b)) structure is observed 
in qualitative agreement with the experimental results, including the 
particular shape of the curves reported when one increases the amplitude of the 
antenna field $U_{ext}$. 

A characteristic triangular shape is obtained in complete
agreement with experimental results\cite{Huttel2009} accompained by a shift in the
position of the resonance frequency with respect to small amplitude regime. 
This nonlinear behaviour has been understood by analyzing the properties of the 
force $F(x)$ (Eq.~(\ref{fortot})) in the equation of motion which has nonlinear terms
stemming from the electron-vibrational interaction. 
Softening, hardening and shape of the curves are 
usually related to the sign of the cubic nonlinear term in the force\cite{Nayfeh2008}. 
When the external gate voltage is tuned in such a way that the current 
flow through the device is blocked ($|V_{G}|>>E_P$) 
the sign of this term is positive, giving a net softening effect. 
When the external gate 
voltage tunes the electronic level of the quantum dot within the conduction 
window ($V_{G}=E_{P}$) and for bias values
sufficiently small, the sign of the cubic nonlinear term is
negative providing an hardening.

In the experiments described in Refs.\cite{Steele2009,Huttel2009} it is has been shown that, in conditions of mechanical resonance, one can control the oscillation frequency of the nanotube by tuning the gate voltage. Motivated by this finding, in panel (a) of Fig\ref{Fig7b} we review the results obtained by us in Ref.\cite{Nocera2012} concerning the natural frequency of the nanotube as a function of the gate voltage applied to the junction. As in the actual experiment, the position of the frequencies are detected by calculating the electronic current change (analyzed in Figure~11 in the regime of small antenna amplitude).
%as in panels (a-b) (in the small antenna amplitude regime). 
The solid part of the curves (red online) indicates that a positive current change peak was found (as in panel (a)), while the dashed part (blue online) indicates that a negative current change was found (as in panel (b)).

If one observes curve (1) of Fig.\ref{Fig7b}a, a characteristic V shape curve is found in almost quantitative agreement with experimental results in Ref.\cite{Huttel2009}. The observed renormalization of the resonance frequency can be related to
the variations of the electronic occupation as function of the gate voltage (see solid line in panel Fig.\ref{Fig7b}b). Increasing the bias voltage to values closer to $\hbar\Gamma$ or larger (line (3) and (4) of Fig. \ref{Fig7b}a), one obtains a non-trivial renormalization of the resonance frequency as function of the gate. We note that for $eV_{bias}=1.5\hbar\Gamma$ (line (4) of Fig. \ref{Fig7b}a), a fine structure represented by two very small dips appears. This feature has been experimentally observed recently in Ref.\cite{Meerwaldt2012}.

\subsection{Single parameter charge pumping}
\label{singpump}

In this review, as stated before, we will discuss some of the effects that arise when an external temporal periodic driving gets resonant with the internal frequency of the quantum nanosystem.

In this subsection, we consider a nano-system very similar to that considered previously but with a different coupling between the electric field produced by the external antenna and the quantum dot degrees of freedom. In particular, we assume that the coupling is affecting directly the electronic gate potential ($U_{ext}=0$ and $V_{ext}\neq0$ in Eq.~(\ref{Hosc})). 

The driving considered in this section can be described by
\begin{equation}\label{eq7.2}
\begin{aligned}
\hat{H}_{driving}(t)&=(V_{G}+V_{ext}\cos(\omega_{ext}
t))\sum_{\sigma}{\hat
c^{\dag}_{\sigma}}{\hat
c_{\sigma}},\\
&U_{ext}=0,\;\; \hbar\Gamma_{\alpha}(t)=\hbar\Gamma_{\alpha}
\end{aligned}
\end{equation}
which means that in Eq.~(\ref{Hel}), the ${\hat
H}_{leads-dot}(t)$ part as well as the oscillator Hamiltonian in Eq.~(\ref{Hosc}) are assumed time independent, while the coupling with the vibrational degrees of freedom in the leads has also been neglected even if in this section the electronic temperature of the leads is assumed being non-zero.

As thoroughly discussed by us in Ref.\cite{Perroni2013}, when the external antenna frequency is close to mechanical resonance with the natural frequency of the dot, one can observe a charge current flowing through the system realizing a single parameter quantum pumping. We point out that the conventional quantum pumping can only be realized with two out of phase driving parameters\cite{Thouless1983,Altshuler1999,Brouwer1998,Buitelaar2008,Wei2001}, for example left and right lead voltages or one lead and the gate voltages. 

\begin{figure}
\hbox{\hspace{0cm}\includegraphics[width=9cm,height=9.0cm,angle=0]{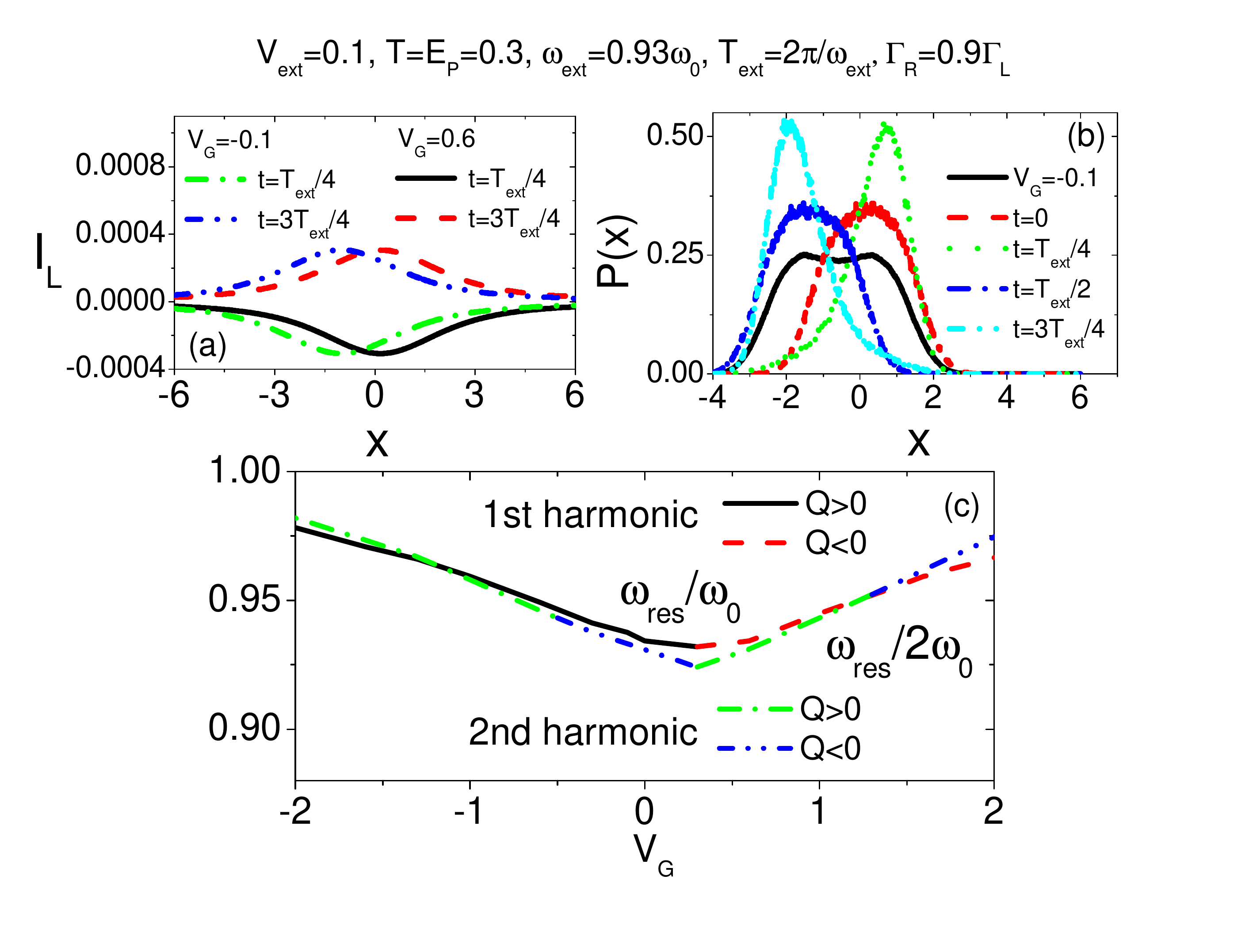}}
\caption{(Color online). Single parameter charge pumping mechanism, and oscillator frequency as a
function of the gate voltage.
Panel (a): Left current $I_L$ as a
function of oscillator position $x$ for different times $t$ and
static gate voltages $V_G$. Panel (b) Reduced position
probability distribution $P$ as a function of $x$ for different
times $t$ at fixed static voltage $V_G$. The time averaged
distribution is in black line. Panel (c): Softening of the
resonance frequency corresponding to the first and to the second
harmonic as a function of the static gate voltage $V_G$.
Adapted and reproduced with permission of EPL from Ref.~\cite{Perroni2013}.}\label{Fig8a}
\end{figure}

We hereby review\cite{Perroni2013} the main mechanism that allows the single parameter charge pumping in a system that can be realized by a suspended carbon nanotube quantum dot. In panel (a) of Fig.\ref{Fig8a}, we report the electronic current $I_L(x)$ flowing through the system as a function of the CNT vibrational displacement. In particular, we show $I_L(x)$ at a quarter and three quarter of the period $T_{ext}$
for different values of static gate $V_G$.

As one can see, close to mechanical resonance, in the first half-period it
shows a different behaviour from that in the second half-period (in
the case where $\omega_{ext}=0.93 \omega_0$). This involves that
the average on a period is different from zero, allowing to pump
charge into the nanotube.

At fixed $V_G$, $I_L(x,t)$ acquires a minimum at a quarter of period, while
a maximum at three quarter of period. Moreover, the static gate
induces a shift of the curves toward positive values for negative
$V_G$, but negative values for positive $V_G$. The shifts of $I_L$ are compared with the behaviour 
of $P(x,t)$ at the resonance. As shown in panel (b) of  Fig.\ref{Fig8a}, in addition to the new center 
of the distribution due to the coupling
$E_P$, the distribution averaged on a period (black line) is bimodal due to the resonance phenomenon. 
In the supplementary material  of Ref.\cite{Perroni2013}, we point out that the bimodal character 
is present only close to the resonance. 
For the reasons explained above, the tail of the distribution probability $P(x,t)$ is
always able to intercept a spatial region where $I_L$ is not zero giving a average non-zero charge 
pumped through the nanotube.

We point out that the pumping mechanism is due to a relevant dynamical adjustment of the oscillator 
to the single external drive and cannot be understood in terms of a phase shift between the 
external drives, such as in the two-pumping parameter mechanism, \cite{Brouwer1998} or 
between the ac gate voltage and the
parametrically excited mechanical oscillations. \cite{Gorelik2013} 
Moreover, we have considered a configuration where the inversion symmetry has been broken: the 
coupling $\Gamma_L$ to the left lead is slightly different from the coupling $\Gamma_R$ 
($\Gamma_R=0.9 \Gamma_L=0.9 \Gamma$).  A small asymmetry is sufficient to induce a single-parameter
 pumping even if the system is in the adiabatic regime.

\begin{figure}
\hbox{\hspace{-1cm}\includegraphics[width=11cm,height=9.0cm,angle=0]{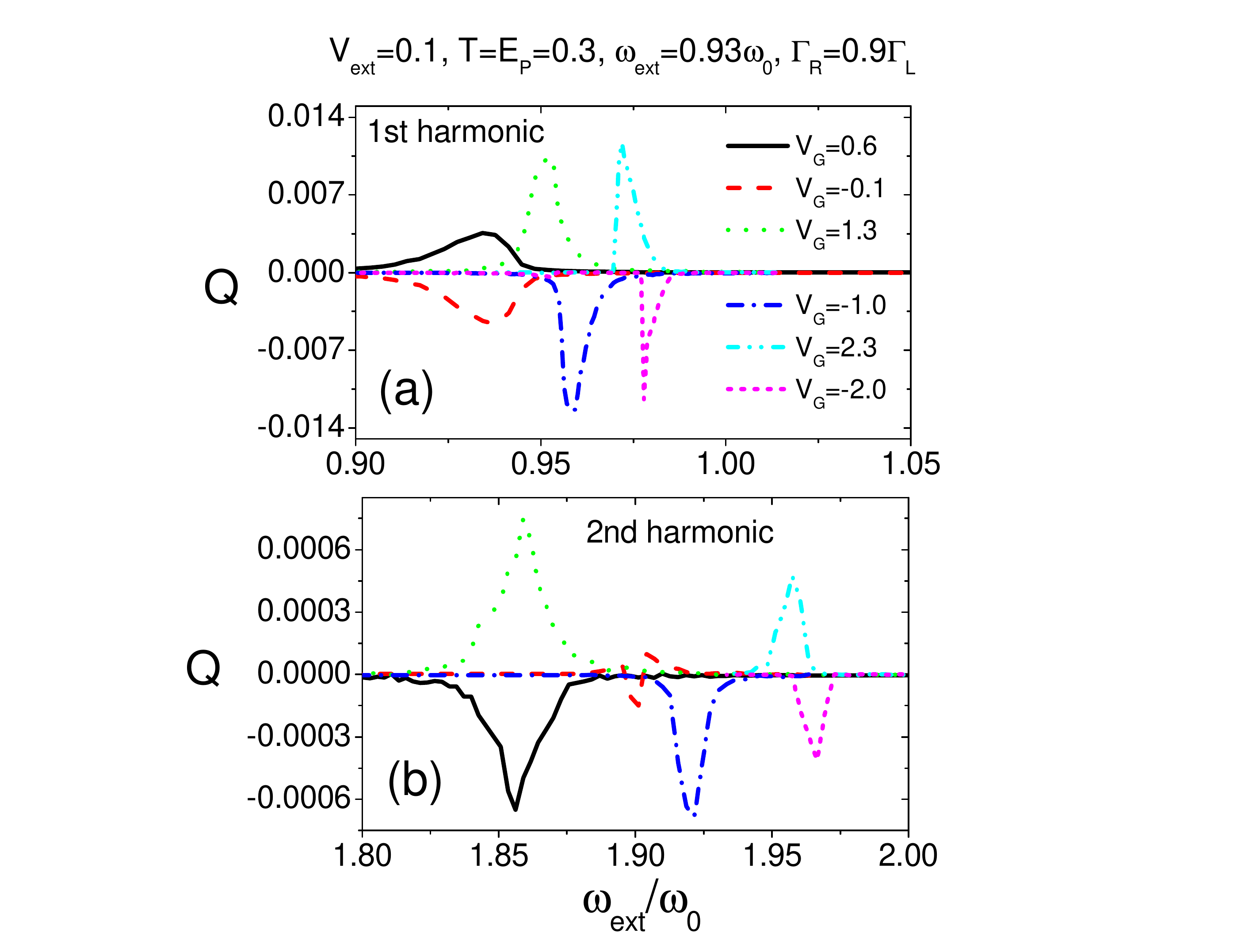}}
\caption{(Color online) Single parameter charge pumped at zero bias.
Panel (a) Charge $Q$ (in units of
$e$) as a function of the external frequency in the interval close
to $1 \times \omega_0$ with varying the static gate $V_G$. Panel (b):
Charge $Q$ (in units of $e$) as a function of the external
frequency in the interval close to $2\times\omega_0$ with varying the
static gate $V_G$. 
Adapted and reproduced with permission of EPL from Ref.~\cite{Perroni2013}.
}\label{Fig8b}
\end{figure}

In analogy to what done in out-of-equilibrium conditions in the previous subsection, we here review 
our theoretical proposal\cite{Perroni2013} of controlling the oscillation frequency of the nanotube 
by tuning the gate voltage in conditions of mechanical resonance even at zero bias where pumping 
is realized. In Fig.\ref{Fig8a}c, we show the natural frequencies of the nanotube as a function of 
the gate voltage applied to the junction. We point out that the particular coupling between 
the electronic and vibrational degrees of freedom allows us to excite also higher harmonics 
of mechanical vibration of the nanotube.

As in the experiments described in Ref.\cite{Steele2009,Huttel2009}, the position of the natural 
CNT frequencies are detected by calculating the change pumped as a function of the antenna frequency 
(as in Fig.\ref{Fig8b}) for different gate voltages. As shown in the panel (c) of Fig. \ref{Fig8a}
($V_{ext}=0.1$, slightly non-linear regime), the first harmonic
resonance has a characteristic V shape similar to that seen in the previous section for a 
different antenna-vibration coupling. We point out that the softening is
symmetrical with respect to $V_G-E_P$ even if the pumped charges
have opposite signs (see below). 
We found that the characteristics (softening, hardening) of the second harmonic 
are similar to those of 
the first harmonic. Once the frequency of the first harmonic has been individuated, 
in order to explore the behaviour of the system at higher harmonics 
it is sufficient to tune the external antenna frequency close to integer multiples of the 
proper frequency $n\times \omega_{0}$. We expect that higher harmonics, which could be 
experimentally excited by a larger antenna power, have a similar behaviour.

We notice that in panels (a) and (b) of Fig. \ref{Fig8b} the sign of the 
pumped charge depends on that of $V_G$. In
fact, this is due to the different behaviour of the currents
for positive and negative $V_G$. As a result, there is a specular
symmetry with respect to $V_G-E_P$ ($E_P=0.3$ in panels (a) and
(b) of Fig. \ref{Fig8a}). Moreover, with increasing $V_G$, 
the shape of the charge-frequency curves tends to be more triangular as a function of the frequency, meaning that the response becomes progressively non-linear with
features of the Duffing oscillator. \cite{Nocera2012,Nayfeh2008}
In Ref.\cite{Perroni2013}, we have studied the current-frequency 
response of the device by increasing the antenna power well above the linear regime. 
We observed that the pumped current at the first harmonic increases to a maximum, 
where the second harmonic response becomes appreciable. 
By increasing the antenna power further, the response at the first harmonic 
reduces while the second harmonic increases up to a maximum value as the 
first harmonic. The response is successively transferred to higher harmonics, 
and eventually the total pumped charge changes sign.
We finally discuss the response for frequencies close to second harmonic. As shown in panel (b) of 
Fig. \ref{Fig8b}, in the weakly non-linear regime ($V_{ext}=0.1$), some charge (less than ten per 
cent of that corresponding to the first harmonic) is
pumped close to those frequencies. Moreover, the frequency response shows a very complex behaviour 
with several maxima and minima that should be experimentally observed in future experiments. 

\subsection{Noise-assisted pumping}
\label{pumpthou}

In this section, we investigate another nano-system where charge pumping effects can be reinforced 
and amplified against temperature and noise when one excites the system close to the mechanical 
resonance with an external driving. In particular, we here review some of recent results obtained 
by us in Ref.\cite{Perroni2014b}.  We consider a conventional quantum pumping 
scheme\cite{Thouless1983} where the tunneling
amplitudes between the dot and the leads ($u_{\alpha} (t)$ represents the strength of
the pumping) are oscillating periodically in time due to external fields. We consider that no 
antenna effect and no bias voltage is present with $\gamma=0$ (no coupling with phonon leads). 

In the general scheme outlined in Sec.2, the driving considered in this section can be summarized in 
the following equation 
\begin{equation}\label{eq7.3}
\begin{aligned}
\hat{H}_{driving}(t)&=\hat{H}_{dot-leads}(t)=
\sum_{k,\alpha,\sigma}(V_{k_{\alpha}}(t){\hat
c^{\dag}_{k_{\alpha},\sigma}}{\hat c_{\sigma}}+ h.c.),\\
&V_{ext}=0,\;\; U_{ext}=0,
\end{aligned}
\end{equation}
which means that in Eq.~(\ref{Hel}), the dot $\hat H_{dot}(t)$ part is assumed time 
independent ($\varepsilon(t)=V_{G}$) and the driving is acting directly on the coupling
between the dot and the leads. In particular, $V_{k_{\alpha}}(t)=V_{k_{\alpha}}u_{\alpha}(t)$ 
giving $\hbar\Gamma_{\alpha}(t,t')=\hbar\Gamma_{\alpha}u_{\alpha}(t)u_{\alpha}(t')$, 
where $u_{\alpha}(t)=1+S\cos(\omega_{P}t+\phi_{\alpha})$, with $S$ amplitude of the pumping driving 
with frequency $\omega_P$ 
and phase $\phi_{\alpha}$, with phase shift $\Delta\phi=\phi_L-\phi_R$. 
Finally, the oscillator Hamiltonian is assumed time independent.

In order to apply the adiabatic approximation, we have assumed that the external time 
dependent perturbations are slowly varying in time together with the mechanical mode vibrational 
motion: $\omega_0 \ll \Gamma_0$ and $d \Gamma /dt \ll \Gamma_0^2$, where we have assumed 
that $\Gamma_L = \Gamma_R = \Gamma_0$. 

In Ref.\cite{Perroni2014b}, we have shown that the adiabatic approach
leads to a Langevin dynamics for the vibrational mode where the external fields give rise to a 
forcing term as in the case of an antenna. Indeed, the deterministic part of the force appearing in 
the Langevin equation contains also a dissipative term proportional to the velocity
\begin{equation}
F^{(1)}_{det}(x,\textrm{v},t)=-\lambda N^{(1)}(x,\textrm{v},t)=-A(x,t)v + B(x,t)
\dot{\Gamma}(t),
\end{equation}
with the coefficients $A(x,t)$ (positive definite) and $B(x,t)$
taken from Ref.\cite{Perroni2014b}. Remarkably, we have also verified that the noise strength 
associated with this force fulfills the fluctuation-dissipation theorem at each oscillator position 
and time $D(x,t)=2 k_{B} T A(x,t)$, where the position dependent noise $D(x,t)$ has been calculated 
in general in Eq.~(\ref{Dxt}). 

From the solution of the Langevin equation, one can calculate the
oscillator distribution function $P(x,v,t)$ and compute all the
observable quantities as in the previous section. As stated above, in the regime of
adiabatic pumping, one has $\omega_P \ll\Gamma_0$, and $\omega_0
\ll \Gamma_0$, so that the dimensionless ratio $ r_P= \omega_P /
\omega_0 $ is of the order of unity. The regime of weak pumping is defined by the 
condition $S\ll1$, where $S$ is proportional to the amplitude 
of the pumping terms.   
Throughout the section, we will assume $\omega_0=0.1 \Gamma_0$.

\begin{figure}
\hbox{\hspace{-0.5cm}\includegraphics[width=9cm,height=9.0cm,angle=0]{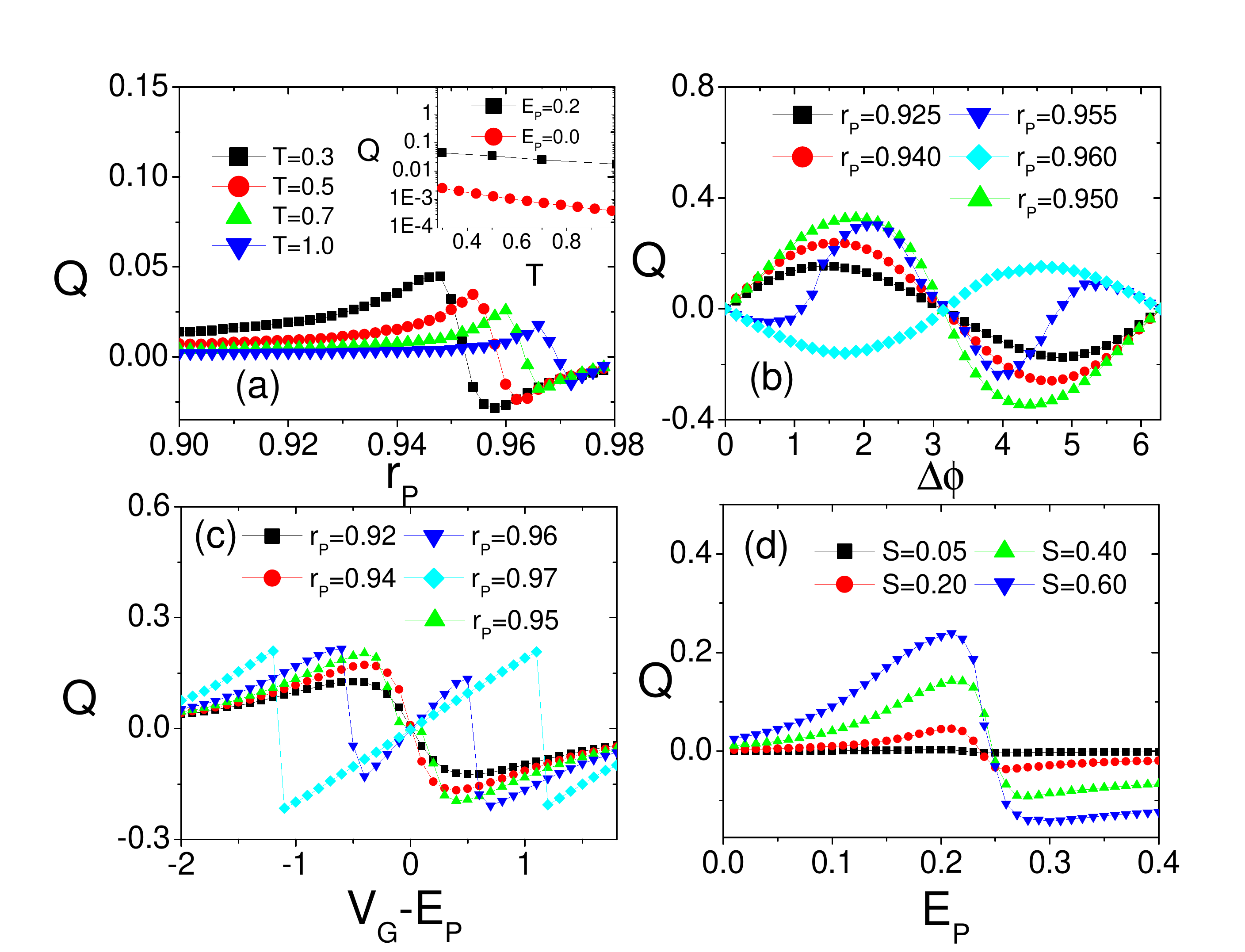}}
\caption{(Color online). Electronic charge pumped in a two parameter pumping device in the 
presence of an adiabatic vibrational degree of freedom.
Panel (a): The pumped charge $Q$ as a
function of the external frequency for different temperatures. 
Inset: The value of the charge at the maximum as a function
of the temperature is compared with the pumped charge for
$E_{P}=0$. In this panel $T=0.3$, $V_{G}=-0.1$, $S=0.20$,
$\Delta\phi=\pi/4$. Panel (b): The pumped charge $Q$ for $S=0.5$ as a
function of the phase difference $\Delta \phi$ for different
values of $r_P$. In this panel $T=0.3$, $V_{G}=-0.1$, and $E_{P}=0.2$. 
Panel (c): The
pumped charge $Q$ as a function of $V_G-E_{P}$ for different
values of $r_P$. In this plot
$T=0.3$, $V_{G}=-0.1$, $E_{P}=0.2$, $S=0.50$, and
$\Delta\phi=\pi/4$. Panel (d): The pumped charge $Q$ as a function
of the electron-oscillator coupling  $E_{P}$ for different values
of the pumping strength $S$. In this plot $T=0.3$, $V_{G}=-0.1$,
$r_{P}=0.945$, and $\Delta\phi=\pi/4$. Notice that the
pumping strength $S=0.6$ corresponds to the driving ratio
$\Gamma^{\omega}_{\alpha}/\Gamma^{0}_{\alpha}\approx1.765$, which
is very close to the maximal value of $2$.
Adapted and reproduced with permission of \textcopyright~IOP Publishing 
from Ref.~\cite{Perroni2014b}. All rights reserved.}\label{Fig9}
\end{figure}

In panel (a) of Fig. \ref{Fig9}, we plot the pumped charge $Q=T_{P} \langle{\bar I}_L\rangle$ 
as a function of the pumping frequency $\omega_P$, hence as a function of $r_P$ for 
different temperatures. We point out that $\langle{\bar I}_L\rangle$ is the average current over 
one period of the driving $\langle{\bar I}_L\rangle=(1/T_P)\int_{0}^{T_P} \langle I_L\rangle(t) dt$,
 where $\langle I_L\rangle(t)$ has been obtained averaging over $P(x,v,t)$. 

As one can see, the curves show a
characteristic peak-dip structure around the renormalized frequency
$r_{eff}=\omega_{eff}/\omega_0$, where the charge pumped through the system is zero.
This compensation effect (observed also in Ref.\cite{Citro2009}) is due to a 
non-trivial dynamical adjustment of the vibrational mode distribution probability
against the temporal variation of the current density in the phase
space\cite{Perroni2014b}. 
Note that, due to the electron-oscillator interaction, a strong softening of the bare
frequency is expected as a function of the system parameters\cite{Nocera2012,Nocera2013}. 
As shown in the inset of panel (a), this fact affects the behaviour of the pumped charge 
as a function of the temperature. Even if decreasing with temperature, the pumped charge
in resonance conditions is always larger than the same quantity in the absence of 
interaction with a vibrational degree of freedom.

Another key element to explain the mechanism of 
cooperation between the vibrational and electronic 
degrees of freedom in the pumped charge is non-linearity.
In order to amplify the non-linearity effects, in panel (b) of Fig.\ref{Fig9}
we show the pumped charge $Q$ at pumping strength $S=0.5$ as a
function of the phase difference $\Delta \phi$. Different
values of the external frequency $\omega_P$ are shown: away from the resonance regime
 ($r_P=0.950$), the response has a perfect sinusoidal shape, meaning that only the first 
harmonic is contributing. In resonance conditions, the response is distorted by the 
contribution of many harmonics.

In panel (c) of Fig. \ref{Fig9} we show the pumped charge $Q$ as
a function of $V_G$ for different values of $r_P$. We note an interesting threshold
behaviour as a function of the frequency of the pump.
Indeed, the curves correspondent to $r_P=0.92$,
$r_P=0.94$ and $r_P=0.95$, which satisfy the condition  $r_P\leq r_{eff}$, where $r_{eff}=0.95$ 
for $V_G-E_{P}\simeq 0$, show a change of sign just at $V_G-E_{P} \simeq 0$. Differently, 
for $r_P=0.96$ and $r_P=0.97$, $Q$ suddenly changes sign
sign at finite values of $V_G-E_{P}$. For larger values of $r_P$ (not shown in Fig. \ref{Fig9}), 
the pumped charge flattens, then, for $r_P$ close to unity, it tends to small negative values. 
Even in the presence of electron-oscillator interaction, when 
the gate voltage is tuned very far from the conduction 
window ($|V_G|>>E_P$) the softening of the 
oscillator frequency tends to zero in agreement with Eq.~(9) of 
Ref.\cite{Citro2009}. 
Finally, in the panel (d) of Fig.\ref{Fig9} we study the pumped charge $Q$ as
a function of the electron-oscillator coupling $E_{P}$ for
different values of the pumping strength $S$. We notice that, around the intermediate
coupling $E_p=0.2$, the pumping charge has, in absolute value, the maximum increase 
as function of the pumping strength. Also, for $S>0.05$, an interesting change of sign 
in the pumped charge at $E_P\simeq 0.25$ occurs.

\section{Conclusions}
In conclusion, we have generalized the adiabatic approach for
nanoscopic systems in the presence of slow vibrational degrees of
freedom to the case where time-dependent
perturbations are acting on the system. Focus has been on the
prototype model system consisting of a single electron level
with a slow single vibrational mode in the parameter regimes appropriate to different soft 
nanosystems, such as molecular junctions and NEMS.

In this work, we have identified the range of parameters 
where the adiabatic approach is reliable in the absence of time
dependent perturbations. We have constructed the phase
diagram (see Fig.~\ref{Fig2b}) of the model in the presence of an applied finite bias
voltage at zero temperature. The average kinetic energy of the
vibrational mode is shown to play a crucial role in the
establishing the validity of the method (see Fig.~\ref{Fig2a}b). Its meaning is related to
the effective excitation energy of the vibrational modes
dynamically induced  by a bias voltage or a temperature gradient.
When this quantity is larger than the static vibrational energy of
the modes, the adiabatic approach can be meaningfully applied to
study the charge or heat transport.

At zero bias voltage and finite temperature, a comparison with a calculation 
which is \emph{exact} in the low charge density limit on the dot has shown that 
the semiclassical adiabatic approach describes accurately the device down to 
quite low temperatures (see Fig.~\ref{figqua}).
 
We have studied the current-voltage characteristic at zero temperature 
(see Fig.~\ref{Fig3}), observing a
complete cancellation of hysteresis or finite discontinuity jumps typical of the 
static infinite mass approximation. 

For sufficiently large electron-oscillator interaction strength, 
contrary to the expectations, we find a region of parameters' space where the
kinetic energy decreases as a function of the bias voltage.
Correspondingly, a finite electronic current flow is observed in the device, 
contrary to the static limit where it was completely blocked.

We have studied the thermoelectric properties within the linear response regime at room temperature 
(see Fig.~\ref{Fig5a}). In particular, we have analyzed the role played by the phonon 
thermal contribution $G_K^{ph}$  on the thermoelectric figure of merit $ZT$ in the 
presence of electron-vibration coupling. We have found that the $G_K^{ph}$ is of 
the same order of electronic thermal conductance $G_K^{el}$ and it gets larger 
with increasing the electron-vibration coupling. Moreover, deviations from the 
Wiedemann-Franz law are progressively reduced with increasing the electron-vibration coupling. 
Therefore, the figure of merit $ZT$ depends appreciably on the behaviour of $G_K^{ph}$ and 
electron-vibration coupling. Indeed, for realistic parameters of the model, $ZT$ can be 
substantially reduced, but it can still have peaks of the order of unity with enhancements 
due to temperature increase.

We have then included the effect of a strong local repulsive
electron-electron, addressing the thermal transport in
the Coulomb blockade regime (see Figs.~\ref{Fig6a} and \ref{Fig6b}). Within the intermediate
electron-vibration coupling regime, the phonon thermal conductance $G_K^{ph}$ has a behaviour similar 
to the electron thermal conductance
$G_K^{el}$. With increasing the electron-vibration coupling, they both get larger as in the absence 
of electron-electron interaction, while the
charge conductance $G$ and the thermopower $S$ get smaller. The main result is that
the figure of merit $ZT$ depends considerably on the behaviour of
$G_K^{ph}$ and intramolecular interactions. Indeed, for realistic
parameters of the model, $ZT$ can be substantially reduced, but
its peak values can be still of the order of unity indicating that
our results can be very interesting for applications.

In the presence of time dependent perturbations, we have shown
that the vibrational modes are driven in dynamical states that can
be very well described in terms of our adiabatic approach. In particular, we
have studied a single level quantum dot realized by a suspended
carbon nanotube including, in a non-perturbative way, the effect
of the antenna actuating the nanotube motion (the sketch of the device has been 
reproduced in  Fig.~\ref{figCNT}).
For the scope of this review, we have reproposed\cite{Nocera2012} the main features of the device 
when the antenna drives the system close to the mechanical resonance with the natural 
vibrational frequency (see Figs.~\ref{Fig7a} and \ref{Fig7b}). The current-frequency 
curves have been studied, showing a very good agreement with the experimental results. 
Here the nonlinear effects are understood without adding extra nonlinear terms to the effective force
exerted on the resonator \cite{Steele2009,Weick2010,Weick2011},
but they are shown to be naturally included in our adiabatic scheme.

In the presence of the same antenna, for frequencies close to the
mechanical resonance, we have shown that it is possible to realize
single-parameter adiabatic charge pumping (see Figs.~\ref{Fig8a} and \ref{Fig8b}). 
The mechanism\cite{Perroni2013} is
different from that active in the two-parameter pumping since it
requires a dynamic adjustment of the mechanical motion of the
nanotube to the external drive. Moreover, the excitation of the
second harmonic is feasible showing a similarity of the softening
with the first harmonic.

Finally, we have studied the two parameter quantum pumping through
a molecular level coupled to a slow vibrational mode (see Fig.~\ref{Fig9}). Again, we have studied 
the device close to resonance conditions, showing that in this regime the presence of dissipation 
and noise does not destroy the pumping mechanism and, even, amplify it.
One of the main results has been the observation of reinforcement of the charge pumping as a function 
of the temperature close to resonant condition with respect to the situation where no vibrational 
motion of the dot is allowed. Furthermore, we have observed finite jumps in the charge $Q$ vs 
gate voltage curves at finite values of $V_G$, and an amplification of charge pumping by increasing 
the strength of the driving. These effects could be observable in future experiments.

In the future, our approach could be extended to the study of the thermal transport away from the 
linear regime in NEMS of molecular junctions
in the presence of time dependent perturbations\cite{Michelini2011,Hanggi2015}.
New directions in the field include also the possibility to control 
directly the vibrational degrees of freedom in order to manipulate heat
flow by use of time-varying thermal bath temperatures or various other external fields\cite{Li2012}.

Summarizing, we have discussed common features of different soft nanosystems, such as molecular 
junctions and NEMS, under external drive.
The effects induced by time-dependent perturbations are very 
marked when the external forcing is nearly resonant with the vibrational modes. 
Indeed, close to the mechanical resonance, the external temporal perturbations 
induce nonlinear regimes where the interplay between electronic and vibrational 
degrees of freedom plays a major role. We believe that our work
could represent a guide for future studies of more
realistic models of multi-level electronic systems coupled to many slow vibrational degrees of freedom, 
in the presence of time dependent perturbations, such as pumping and external forcing
antennas.

\begin{acknowledgements}
C. A. P. acknowledges partial financial support from the Progetto Premiale CNR/INFN EOS 
``Organic Electronics for Innovative Research Instrumentation". 
C. A. P. and V. C. acknowledge partial financial support from the regione Campania project L.R. 
N.5/2007  ``Role of interfaces in magnetic strongly correlated oxides: manganite heterostructures." 
\end{acknowledgements}
\bibliography{v8_review_adiabatico}

\begin{thebibliography}{109}
\expandafter\ifx\csname natexlab\endcsname\relax\def\natexlab#1{#1}\fi
\expandafter\ifx\csname bibnamefont\endcsname\relax
  \def\bibnamefont#1{#1}\fi
\expandafter\ifx\csname bibfnamefont\endcsname\relax
  \def\bibfnamefont#1{#1}\fi
\expandafter\ifx\csname citenamefont\endcsname\relax
  \def\citenamefont#1{#1}\fi
\expandafter\ifx\csname url\endcsname\relax
  \def\url#1{\texttt{#1}}\fi
\expandafter\ifx\csname urlprefix\endcsname\relax\def\urlprefix{URL }\fi
\providecommand{\bibinfo}[2]{#2}
\providecommand{\eprint}[2][]{\url{#2}}

\bibitem[{\citenamefont{Delerue and Lannoo}(2004)}]{Delerue2004}
\bibinfo{author}{\bibfnamefont{C.}~\bibnamefont{Delerue}} \bibnamefont{and}
  \bibinfo{author}{\bibfnamefont{M.}~\bibnamefont{Lannoo}},
  \emph{\bibinfo{title}{Nanostructures: theory and modelling}}
  (\bibinfo{publisher}{Springer Science \& Business Media},
  \bibinfo{year}{2004}).

\bibitem[{\citenamefont{Scheer and Cuevas}(2010)}]{Cuevas2010}
\bibinfo{author}{\bibfnamefont{E.}~\bibnamefont{Scheer}} \bibnamefont{and}
  \bibinfo{author}{\bibfnamefont{J.~C.} \bibnamefont{Cuevas}},
  \emph{\bibinfo{title}{Molecular electronics: an introduction to theory and
  experiment}}, vol.~\bibinfo{volume}{1} (\bibinfo{publisher}{World
  Scientific}, \bibinfo{year}{2010}).

\bibitem[{\citenamefont{Galperin et~al.}(2007)\citenamefont{Galperin, Ratner,
  and Nitzan}}]{Galperin2007}
\bibinfo{author}{\bibfnamefont{M.}~\bibnamefont{Galperin}},
  \bibinfo{author}{\bibfnamefont{M.~A.} \bibnamefont{Ratner}},
  \bibnamefont{and} \bibinfo{author}{\bibfnamefont{A.}~\bibnamefont{Nitzan}},
  \bibinfo{journal}{J. Phys. Condens. Matter} \textbf{\bibinfo{volume}{19}},
  \bibinfo{pages}{103201} (\bibinfo{year}{2007}).

\bibitem[{\citenamefont{Zimbovskaya and Pederson}(2011)}]{Zimbovskaya2011}
\bibinfo{author}{\bibfnamefont{N.~A.} \bibnamefont{Zimbovskaya}}
  \bibnamefont{and} \bibinfo{author}{\bibfnamefont{M.~R.}
  \bibnamefont{Pederson}}, \bibinfo{journal}{Phys. Rep.}
  \textbf{\bibinfo{volume}{509}}, \bibinfo{pages}{1 } (\bibinfo{year}{2011}),
  ISSN \bibinfo{issn}{0370-1573},
  \urlprefix\url{http://www.sciencedirect.com/science/article/pii/S0370157311002195}.

\bibitem[{\citenamefont{Craighead}(2000)}]{Craighead2000}
\bibinfo{author}{\bibfnamefont{H.~G.} \bibnamefont{Craighead}},
  \bibinfo{journal}{Science} \textbf{\bibinfo{volume}{290}},
  \bibinfo{pages}{1532} (\bibinfo{year}{2000}),
  \eprint{http://www.sciencemag.org/content/290/5496/1532.full.pdf},
  \urlprefix\url{http://www.sciencemag.org/content/290/5496/1532.abstract}.

\bibitem[{\citenamefont{LaHaye et~al.}(2004)\citenamefont{LaHaye, Buu,
  Camarota, and Schwab}}]{LaHaye2004}
\bibinfo{author}{\bibfnamefont{M.~D.} \bibnamefont{LaHaye}},
  \bibinfo{author}{\bibfnamefont{O.}~\bibnamefont{Buu}},
  \bibinfo{author}{\bibfnamefont{B.}~\bibnamefont{Camarota}}, \bibnamefont{and}
  \bibinfo{author}{\bibfnamefont{K.~C.} \bibnamefont{Schwab}},
  \bibinfo{journal}{Science} \textbf{\bibinfo{volume}{304}},
  \bibinfo{pages}{74} (\bibinfo{year}{2004}),
  \eprint{http://www.sciencemag.org/content/304/5667/74.full.pdf},
  \urlprefix\url{http://www.sciencemag.org/content/304/5667/74.abstract}.

\bibitem[{\citenamefont{Blencowe}(2004)}]{Blencowe2004}
\bibinfo{author}{\bibfnamefont{M.}~\bibnamefont{Blencowe}},
  \bibinfo{journal}{Phys. Rep.} \textbf{\bibinfo{volume}{395}},
  \bibinfo{pages}{159} (\bibinfo{year}{2004}), ISSN \bibinfo{issn}{0370-1573},
  \urlprefix\url{http://www.sciencedirect.com/science/article/pii/S0370157304000080}.

\bibitem[{\citenamefont{Ekinci and Roukes}(2005)}]{Ekinci2005}
\bibinfo{author}{\bibfnamefont{K.~L.} \bibnamefont{Ekinci}} \bibnamefont{and}
  \bibinfo{author}{\bibfnamefont{M.~L.} \bibnamefont{Roukes}},
  \bibinfo{journal}{Rev. Sci. Instrum.} \textbf{\bibinfo{volume}{76}},
  \bibinfo{eid}{061101} (\bibinfo{year}{2005}),
  \urlprefix\url{http://scitation.aip.org/content/aip/journal/rsi/76/6/10.1063/1.1927327}.

\bibitem[{\citenamefont{Park et~al.}(2000)\citenamefont{Park, Park, Lim,
  Anderson, Alivisatos, and McEuen}}]{Park2000}
\bibinfo{author}{\bibfnamefont{H.}~\bibnamefont{Park}},
  \bibinfo{author}{\bibfnamefont{J.}~\bibnamefont{Park}},
  \bibinfo{author}{\bibfnamefont{A.~K.} \bibnamefont{Lim}},
  \bibinfo{author}{\bibfnamefont{E.~H.} \bibnamefont{Anderson}},
  \bibinfo{author}{\bibfnamefont{A.~P.} \bibnamefont{Alivisatos}},
  \bibnamefont{and} \bibinfo{author}{\bibfnamefont{P.~L.}
  \bibnamefont{McEuen}}, \bibinfo{journal}{Nature}
  \textbf{\bibinfo{volume}{407}}, \bibinfo{pages}{57} (\bibinfo{year}{2000}).

\bibitem[{\citenamefont{Qin et~al.}(2001)\citenamefont{Qin, Holleitner, Eberl,
  and Blick}}]{Qin}
\bibinfo{author}{\bibfnamefont{H.}~\bibnamefont{Qin}},
  \bibinfo{author}{\bibfnamefont{A.~W.} \bibnamefont{Holleitner}},
  \bibinfo{author}{\bibfnamefont{K.}~\bibnamefont{Eberl}}, \bibnamefont{and}
  \bibinfo{author}{\bibfnamefont{R.~H.} \bibnamefont{Blick}},
  \bibinfo{journal}{Phys. Rev. B} \textbf{\bibinfo{volume}{64}},
  \bibinfo{pages}{241302} (\bibinfo{year}{2001}),
  \urlprefix\url{http://link.aps.org/doi/10.1103/PhysRevB.64.241302}.

\bibitem[{\citenamefont{H\"{u}ttel et~al.}(2009)\citenamefont{H\"{u}ttel,
  Steele, Witkamp, Poot, Kouwenhoven, and van~der Zant}}]{Huttel2009}
\bibinfo{author}{\bibfnamefont{A.~K.} \bibnamefont{H\"{u}ttel}},
  \bibinfo{author}{\bibfnamefont{G.~A.} \bibnamefont{Steele}},
  \bibinfo{author}{\bibfnamefont{B.}~\bibnamefont{Witkamp}},
  \bibinfo{author}{\bibfnamefont{M.}~\bibnamefont{Poot}},
  \bibinfo{author}{\bibfnamefont{L.~P.} \bibnamefont{Kouwenhoven}},
  \bibnamefont{and} \bibinfo{author}{\bibfnamefont{H.~S.~J.}
  \bibnamefont{van~der Zant}}, \bibinfo{journal}{Nano Lett.}
  \textbf{\bibinfo{volume}{9}}, \bibinfo{pages}{2547} (\bibinfo{year}{2009}),
  \bibinfo{note}{pMID: 19492820}.

\bibitem[{\citenamefont{Steele et~al.}(2009)\citenamefont{Steele, H\"{u}ttel,
  Witkamp, Poot, Meerwaldt, Kouwenhoven, and van~der Zant}}]{Steele2009}
\bibinfo{author}{\bibfnamefont{G.~A.} \bibnamefont{Steele}},
  \bibinfo{author}{\bibfnamefont{A.~K.} \bibnamefont{H\"{u}ttel}},
  \bibinfo{author}{\bibfnamefont{B.}~\bibnamefont{Witkamp}},
  \bibinfo{author}{\bibfnamefont{M.}~\bibnamefont{Poot}},
  \bibinfo{author}{\bibfnamefont{H.~B.} \bibnamefont{Meerwaldt}},
  \bibinfo{author}{\bibfnamefont{L.~P.} \bibnamefont{Kouwenhoven}},
  \bibnamefont{and} \bibinfo{author}{\bibfnamefont{H.~S.~J.}
  \bibnamefont{van~der Zant}}, \bibinfo{journal}{Science}
  \textbf{\bibinfo{volume}{325}}, \bibinfo{pages}{1103} (\bibinfo{year}{2009}),
  \eprint{http://www.sciencemag.org/content/325/5944/1103.full.pdf},
  \urlprefix\url{http://www.sciencemag.org/content/325/5944/1103.abstract}.

\bibitem[{\citenamefont{Atalaya et~al.}(2011)\citenamefont{Atalaya, Isacsson,
  and Dykman}}]{Atalaya2011}
\bibinfo{author}{\bibfnamefont{J.}~\bibnamefont{Atalaya}},
  \bibinfo{author}{\bibfnamefont{A.}~\bibnamefont{Isacsson}}, \bibnamefont{and}
  \bibinfo{author}{\bibfnamefont{M.~I.} \bibnamefont{Dykman}},
  \bibinfo{journal}{Phys. Rev. Lett.} \textbf{\bibinfo{volume}{106}},
  \bibinfo{pages}{227202} (\bibinfo{year}{2011}),
  \urlprefix\url{http://link.aps.org/doi/10.1103/PhysRevLett.106.227202}.

\bibitem[{\citenamefont{Knobel}(2008)}]{Knobel2008}
\bibinfo{author}{\bibfnamefont{R.~G.} \bibnamefont{Knobel}},
  \bibinfo{journal}{Nat. Nanotechnol.} \textbf{\bibinfo{volume}{3}},
  \bibinfo{pages}{525} (\bibinfo{year}{2008}).

\bibitem[{\citenamefont{Naik et~al.}(2009)\citenamefont{Naik, Hanay, Hiebert,
  Feng, and Roukes}}]{Naik2009}
\bibinfo{author}{\bibfnamefont{A.}~\bibnamefont{Naik}},
  \bibinfo{author}{\bibfnamefont{M.}~\bibnamefont{Hanay}},
  \bibinfo{author}{\bibfnamefont{W.}~\bibnamefont{Hiebert}},
  \bibinfo{author}{\bibfnamefont{X.}~\bibnamefont{Feng}}, \bibnamefont{and}
  \bibinfo{author}{\bibfnamefont{M.}~\bibnamefont{Roukes}},
  \bibinfo{journal}{Nat. Nanotechnol.} \textbf{\bibinfo{volume}{4}},
  \bibinfo{pages}{445} (\bibinfo{year}{2009}).

\bibitem[{\citenamefont{Ekinci et~al.}(2004)\citenamefont{Ekinci, Huang, and
  Roukes}}]{Ekinci2004u}
\bibinfo{author}{\bibfnamefont{K.}~\bibnamefont{Ekinci}},
  \bibinfo{author}{\bibfnamefont{X.}~\bibnamefont{Huang}}, \bibnamefont{and}
  \bibinfo{author}{\bibfnamefont{M.}~\bibnamefont{Roukes}},
  \bibinfo{journal}{Appl. Phys. Lett.} \textbf{\bibinfo{volume}{84}},
  \bibinfo{pages}{4469} (\bibinfo{year}{2004}).

\bibitem[{\citenamefont{Lassagne et~al.}(2008)\citenamefont{Lassagne,
  Garcia-Sanchez, Aguasca, and Bachtold}}]{Lassagne2008}
\bibinfo{author}{\bibfnamefont{B.}~\bibnamefont{Lassagne}},
  \bibinfo{author}{\bibfnamefont{D.}~\bibnamefont{Garcia-Sanchez}},
  \bibinfo{author}{\bibfnamefont{A.}~\bibnamefont{Aguasca}}, \bibnamefont{and}
  \bibinfo{author}{\bibfnamefont{A.}~\bibnamefont{Bachtold}},
  \bibinfo{journal}{Nano Lett.} \textbf{\bibinfo{volume}{8}},
  \bibinfo{pages}{3735} (\bibinfo{year}{2008}).

\bibitem[{\citenamefont{Kacem et~al.}(2009)\citenamefont{Kacem, Hentz, Pinto,
  Reig, and Nguyen}}]{Kacem2009}
\bibinfo{author}{\bibfnamefont{N.}~\bibnamefont{Kacem}},
  \bibinfo{author}{\bibfnamefont{S.}~\bibnamefont{Hentz}},
  \bibinfo{author}{\bibfnamefont{D.}~\bibnamefont{Pinto}},
  \bibinfo{author}{\bibfnamefont{B.}~\bibnamefont{Reig}}, \bibnamefont{and}
  \bibinfo{author}{\bibfnamefont{V.}~\bibnamefont{Nguyen}},
  \bibinfo{journal}{Nanotechnology} \textbf{\bibinfo{volume}{20}},
  \bibinfo{pages}{275501} (\bibinfo{year}{2009}).

\bibitem[{\citenamefont{Hiebert}(2012)}]{hiebert2012}
\bibinfo{author}{\bibfnamefont{W.}~\bibnamefont{Hiebert}},
  \bibinfo{journal}{Nat. Nanotechnol.} \textbf{\bibinfo{volume}{7}},
  \bibinfo{pages}{278} (\bibinfo{year}{2012}).

\bibitem[{\citenamefont{Panchal et~al.}(2012)\citenamefont{Panchal, Upadhyay,
  and Harsha}}]{Panchal2012}
\bibinfo{author}{\bibfnamefont{M.~B.} \bibnamefont{Panchal}},
  \bibinfo{author}{\bibfnamefont{S.}~\bibnamefont{Upadhyay}}, \bibnamefont{and}
  \bibinfo{author}{\bibfnamefont{S.}~\bibnamefont{Harsha}},
  \bibinfo{journal}{Nano} \textbf{\bibinfo{volume}{7}},
  \bibinfo{pages}{1250029} (\bibinfo{year}{2012}).

\bibitem[{\citenamefont{Dubi and Di~Ventra}(2011)}]{DiVentra2011}
\bibinfo{author}{\bibfnamefont{Y.}~\bibnamefont{Dubi}} \bibnamefont{and}
  \bibinfo{author}{\bibfnamefont{M.}~\bibnamefont{Di~Ventra}},
  \bibinfo{journal}{Rev. Mod. Phys.} \textbf{\bibinfo{volume}{83}},
  \bibinfo{pages}{131} (\bibinfo{year}{2011}),
  \urlprefix\url{http://link.aps.org/doi/10.1103/RevModPhys.83.131}.

\bibitem[{\citenamefont{Finch et~al.}(2009)\citenamefont{Finch, Garcia-Suarez,
  and Lambert}}]{finch2009}
\bibinfo{author}{\bibfnamefont{C.}~\bibnamefont{Finch}},
  \bibinfo{author}{\bibfnamefont{V.}~\bibnamefont{Garcia-Suarez}},
  \bibnamefont{and} \bibinfo{author}{\bibfnamefont{C.}~\bibnamefont{Lambert}},
  \bibinfo{journal}{Phys. Rev. B} \textbf{\bibinfo{volume}{79}},
  \bibinfo{pages}{033405} (\bibinfo{year}{2009}).

\bibitem[{\citenamefont{Reddy et~al.}(2007)\citenamefont{Reddy, Jang, Segalman,
  and Majumdar}}]{reddy2007}
\bibinfo{author}{\bibfnamefont{P.}~\bibnamefont{Reddy}},
  \bibinfo{author}{\bibfnamefont{S.-Y.} \bibnamefont{Jang}},
  \bibinfo{author}{\bibfnamefont{R.~A.} \bibnamefont{Segalman}},
  \bibnamefont{and} \bibinfo{author}{\bibfnamefont{A.}~\bibnamefont{Majumdar}},
  \bibinfo{journal}{Science} \textbf{\bibinfo{volume}{315}},
  \bibinfo{pages}{1568} (\bibinfo{year}{2007}).

\bibitem[{\citenamefont{Paulsson and Datta}(2003)}]{paulsson2003}
\bibinfo{author}{\bibfnamefont{M.}~\bibnamefont{Paulsson}} \bibnamefont{and}
  \bibinfo{author}{\bibfnamefont{S.}~\bibnamefont{Datta}},
  \bibinfo{journal}{Phys. Rev. B} \textbf{\bibinfo{volume}{67}},
  \bibinfo{pages}{241403} (\bibinfo{year}{2003}).

\bibitem[{\citenamefont{Romero et~al.}(2002)\citenamefont{Romero, Sumanasekera,
  Mahan, and Eklund}}]{Romero2002}
\bibinfo{author}{\bibfnamefont{H.~E.} \bibnamefont{Romero}},
  \bibinfo{author}{\bibfnamefont{G.~U.} \bibnamefont{Sumanasekera}},
  \bibinfo{author}{\bibfnamefont{G.~D.} \bibnamefont{Mahan}}, \bibnamefont{and}
  \bibinfo{author}{\bibfnamefont{P.~C.} \bibnamefont{Eklund}},
  \bibinfo{journal}{Phys. Rev. B} \textbf{\bibinfo{volume}{65}},
  \bibinfo{pages}{205410} (\bibinfo{year}{2002}),
  \urlprefix\url{http://link.aps.org/doi/10.1103/PhysRevB.65.205410}.

\bibitem[{\citenamefont{Koch et~al.}(2004)\citenamefont{Koch, von Oppen, Oreg,
  and Sela}}]{Koch2004}
\bibinfo{author}{\bibfnamefont{J.}~\bibnamefont{Koch}},
  \bibinfo{author}{\bibfnamefont{F.}~\bibnamefont{von Oppen}},
  \bibinfo{author}{\bibfnamefont{Y.}~\bibnamefont{Oreg}}, \bibnamefont{and}
  \bibinfo{author}{\bibfnamefont{E.}~\bibnamefont{Sela}},
  \bibinfo{journal}{Phys. Rev. B} \textbf{\bibinfo{volume}{70}},
  \bibinfo{pages}{195107} (\bibinfo{year}{2004}),
  \urlprefix\url{http://link.aps.org/doi/10.1103/PhysRevB.70.195107}.

\bibitem[{\citenamefont{Segal}(2005)}]{Segal2005}
\bibinfo{author}{\bibfnamefont{D.}~\bibnamefont{Segal}},
  \bibinfo{journal}{Phys. Rev. B} \textbf{\bibinfo{volume}{72}},
  \bibinfo{pages}{165426} (\bibinfo{year}{2005}),
  \urlprefix\url{http://link.aps.org/doi/10.1103/PhysRevB.72.165426}.

\bibitem[{\citenamefont{Krawiec and Jalochowski}(2007)}]{Kraviec2007}
\bibinfo{author}{\bibfnamefont{M.}~\bibnamefont{Krawiec}} \bibnamefont{and}
  \bibinfo{author}{\bibfnamefont{M.}~\bibnamefont{Jalochowski}},
  \bibinfo{journal}{Phys. Status Solidi B} \textbf{\bibinfo{volume}{244}},
  \bibinfo{pages}{2464} (\bibinfo{year}{2007}), ISSN \bibinfo{issn}{1521-3951},
  \urlprefix\url{http://dx.doi.org/10.1002/pssb.200674614}.

\bibitem[{\citenamefont{Remaggi et~al.}(2013)\citenamefont{Remaggi, Ziani,
  Dolcetto, Cavaliere, and Sassetti}}]{Remaggi2013}
\bibinfo{author}{\bibfnamefont{F.}~\bibnamefont{Remaggi}},
  \bibinfo{author}{\bibfnamefont{N.~T.} \bibnamefont{Ziani}},
  \bibinfo{author}{\bibfnamefont{G.}~\bibnamefont{Dolcetto}},
  \bibinfo{author}{\bibfnamefont{F.}~\bibnamefont{Cavaliere}},
  \bibnamefont{and} \bibinfo{author}{\bibfnamefont{M.}~\bibnamefont{Sassetti}},
  \bibinfo{journal}{New J. Phys.} \textbf{\bibinfo{volume}{15}},
  \bibinfo{pages}{083016} (\bibinfo{year}{2013}),
  \urlprefix\url{http://stacks.iop.org/1367-2630/15/i=8/a=083016}.

\bibitem[{\citenamefont{Zhou et~al.}(2015)\citenamefont{Zhou, Thingna,
  H\"{a}nggi, Wang, and Li}}]{Hanggi2015}
\bibinfo{author}{\bibfnamefont{H.}~\bibnamefont{Zhou}},
  \bibinfo{author}{\bibfnamefont{J.}~\bibnamefont{Thingna}},
  \bibinfo{author}{\bibfnamefont{P.}~\bibnamefont{H\"{a}nggi}},
  \bibinfo{author}{\bibfnamefont{J.-S.} \bibnamefont{Wang}}, \bibnamefont{and}
  \bibinfo{author}{\bibfnamefont{B.}~\bibnamefont{Li}}, \bibinfo{journal}{Sci.
  Rep.} \textbf{\bibinfo{volume}{5}} (\bibinfo{year}{2015}).

\bibitem[{\citenamefont{Cr\'epieux et~al.}(2011)\citenamefont{Cr\'epieux,
  \ifmmode~\check{S}\else \v{S}\fi{}imkovic, Cambon, and
  Michelini}}]{Michelini2011}
\bibinfo{author}{\bibfnamefont{A.}~\bibnamefont{Cr\'epieux}},
  \bibinfo{author}{\bibfnamefont{F.}~\bibnamefont{\ifmmode~\check{S}\else
  \v{S}\fi{}imkovic}},
  \bibinfo{author}{\bibfnamefont{B.}~\bibnamefont{Cambon}}, \bibnamefont{and}
  \bibinfo{author}{\bibfnamefont{F.}~\bibnamefont{Michelini}},
  \bibinfo{journal}{Phys. Rev. B} \textbf{\bibinfo{volume}{83}},
  \bibinfo{pages}{153417} (\bibinfo{year}{2011}),
  \urlprefix\url{http://link.aps.org/doi/10.1103/PhysRevB.83.153417}.

\bibitem[{\citenamefont{Chi and Dubi}(2012)}]{Dubi2012}
\bibinfo{author}{\bibfnamefont{F.}~\bibnamefont{Chi}} \bibnamefont{and}
  \bibinfo{author}{\bibfnamefont{Y.}~\bibnamefont{Dubi}}, \bibinfo{journal}{J.
  Phys. Condens. Matter} \textbf{\bibinfo{volume}{24}}, \bibinfo{pages}{145301}
  (\bibinfo{year}{2012}),
  \urlprefix\url{http://stacks.iop.org/0953-8984/24/i=14/a=145301}.

\bibitem[{\citenamefont{Juergens et~al.}(2013)\citenamefont{Juergens, Haupt,
  Moskalets, and Splettstoesser}}]{Juergens2013}
\bibinfo{author}{\bibfnamefont{S.}~\bibnamefont{Juergens}},
  \bibinfo{author}{\bibfnamefont{F.}~\bibnamefont{Haupt}},
  \bibinfo{author}{\bibfnamefont{M.}~\bibnamefont{Moskalets}},
  \bibnamefont{and}
  \bibinfo{author}{\bibfnamefont{J.}~\bibnamefont{Splettstoesser}},
  \bibinfo{journal}{Phys. Rev. B} \textbf{\bibinfo{volume}{87}},
  \bibinfo{pages}{245423} (\bibinfo{year}{2013}),
  \urlprefix\url{http://link.aps.org/doi/10.1103/PhysRevB.87.245423}.

\bibitem[{\citenamefont{Arrachea et~al.}(2007)\citenamefont{Arrachea,
  Moskalets, and Martin-Moreno}}]{Arrachea2007}
\bibinfo{author}{\bibfnamefont{L.}~\bibnamefont{Arrachea}},
  \bibinfo{author}{\bibfnamefont{M.}~\bibnamefont{Moskalets}},
  \bibnamefont{and}
  \bibinfo{author}{\bibfnamefont{L.}~\bibnamefont{Martin-Moreno}},
  \bibinfo{journal}{Phys. Rev. B} \textbf{\bibinfo{volume}{75}},
  \bibinfo{pages}{245420} (\bibinfo{year}{2007}),
  \urlprefix\url{http://link.aps.org/doi/10.1103/PhysRevB.75.245420}.

\bibitem[{\citenamefont{Segal and Nitzan}(2006)}]{Nitzan2006}
\bibinfo{author}{\bibfnamefont{D.}~\bibnamefont{Segal}} \bibnamefont{and}
  \bibinfo{author}{\bibfnamefont{A.}~\bibnamefont{Nitzan}},
  \bibinfo{journal}{Phys. Rev. E} \textbf{\bibinfo{volume}{73}},
  \bibinfo{pages}{026109} (\bibinfo{year}{2006}),
  \urlprefix\url{http://link.aps.org/doi/10.1103/PhysRevE.73.026109}.

\bibitem[{\citenamefont{Kaestner
  et~al.}(2008{\natexlab{a}})\citenamefont{Kaestner, Kashcheyevs, Amakawa,
  Blumenthal, Li, Janssen, Hein, Pierz, Weimann, Siegner et~al.}}]{Kaest2008}
\bibinfo{author}{\bibfnamefont{B.}~\bibnamefont{Kaestner}},
  \bibinfo{author}{\bibfnamefont{V.}~\bibnamefont{Kashcheyevs}},
  \bibinfo{author}{\bibfnamefont{S.}~\bibnamefont{Amakawa}},
  \bibinfo{author}{\bibfnamefont{M.~D.} \bibnamefont{Blumenthal}},
  \bibinfo{author}{\bibfnamefont{L.}~\bibnamefont{Li}},
  \bibinfo{author}{\bibfnamefont{T.~J. B.~M.} \bibnamefont{Janssen}},
  \bibinfo{author}{\bibfnamefont{G.}~\bibnamefont{Hein}},
  \bibinfo{author}{\bibfnamefont{K.}~\bibnamefont{Pierz}},
  \bibinfo{author}{\bibfnamefont{T.}~\bibnamefont{Weimann}},
  \bibinfo{author}{\bibfnamefont{U.}~\bibnamefont{Siegner}},
  \bibnamefont{et~al.}, \bibinfo{journal}{Phys. Rev. B}
  \textbf{\bibinfo{volume}{77}}, \bibinfo{pages}{153301}
  (\bibinfo{year}{2008}{\natexlab{a}}),
  \urlprefix\url{http://link.aps.org/doi/10.1103/PhysRevB.77.153301}.

\bibitem[{\citenamefont{Kaestner
  et~al.}(2008{\natexlab{b}})\citenamefont{Kaestner, Kashcheyevs, Hein, Pierz,
  Siegner, and Schumacher}}]{Kaest2008a}
\bibinfo{author}{\bibfnamefont{B.}~\bibnamefont{Kaestner}},
  \bibinfo{author}{\bibfnamefont{V.}~\bibnamefont{Kashcheyevs}},
  \bibinfo{author}{\bibfnamefont{G.}~\bibnamefont{Hein}},
  \bibinfo{author}{\bibfnamefont{K.}~\bibnamefont{Pierz}},
  \bibinfo{author}{\bibfnamefont{U.}~\bibnamefont{Siegner}}, \bibnamefont{and}
  \bibinfo{author}{\bibfnamefont{H.~W.} \bibnamefont{Schumacher}},
  \bibinfo{journal}{Appl. Phys. Lett.} \textbf{\bibinfo{volume}{92}},
  \bibinfo{eid}{192106} (\bibinfo{year}{2008}{\natexlab{b}}),
  \urlprefix\url{http://scitation.aip.org/content/aip/journal/apl/92/19/10.1063/1.2928231}.

\bibitem[{\citenamefont{Kaestner et~al.}(2009)\citenamefont{Kaestner, Leicht,
  Kashcheyevs, Pierz, Siegner, and Schumacher}}]{Kaest2009}
\bibinfo{author}{\bibfnamefont{B.}~\bibnamefont{Kaestner}},
  \bibinfo{author}{\bibfnamefont{C.}~\bibnamefont{Leicht}},
  \bibinfo{author}{\bibfnamefont{V.}~\bibnamefont{Kashcheyevs}},
  \bibinfo{author}{\bibfnamefont{K.}~\bibnamefont{Pierz}},
  \bibinfo{author}{\bibfnamefont{U.}~\bibnamefont{Siegner}}, \bibnamefont{and}
  \bibinfo{author}{\bibfnamefont{H.~W.} \bibnamefont{Schumacher}},
  \bibinfo{journal}{Appl. Phys. Lett.} \textbf{\bibinfo{volume}{94}},
  \bibinfo{eid}{012106} (\bibinfo{year}{2009}),
  \urlprefix\url{http://scitation.aip.org/content/aip/journal/apl/94/1/10.1063/1.3063128}.

\bibitem[{\citenamefont{Fujiwara et~al.}(2008)\citenamefont{Fujiwara,
  Nishiguchi, and Ono}}]{Fuji2008}
\bibinfo{author}{\bibfnamefont{A.}~\bibnamefont{Fujiwara}},
  \bibinfo{author}{\bibfnamefont{K.}~\bibnamefont{Nishiguchi}},
  \bibnamefont{and} \bibinfo{author}{\bibfnamefont{Y.}~\bibnamefont{Ono}},
  \bibinfo{journal}{Appl. Phys. Lett.} \textbf{\bibinfo{volume}{92}},
  \bibinfo{eid}{042102} (\bibinfo{year}{2008}),
  \urlprefix\url{http://scitation.aip.org/content/aip/journal/apl/92/4/10.1063/1.2837544}.

\bibitem[{\citenamefont{Cavaliere et~al.}(2009)\citenamefont{Cavaliere,
  Governale, and K\"onig}}]{Cavaliere2009}
\bibinfo{author}{\bibfnamefont{F.}~\bibnamefont{Cavaliere}},
  \bibinfo{author}{\bibfnamefont{M.}~\bibnamefont{Governale}},
  \bibnamefont{and} \bibinfo{author}{\bibfnamefont{J.}~\bibnamefont{K\"onig}},
  \bibinfo{journal}{Phys. Rev. Lett.} \textbf{\bibinfo{volume}{103}},
  \bibinfo{pages}{136801} (\bibinfo{year}{2009}),
  \urlprefix\url{http://link.aps.org/doi/10.1103/PhysRevLett.103.136801}.

\bibitem[{\citenamefont{Vavilov et~al.}(2001)\citenamefont{Vavilov, Ambegaokar,
  and Aleiner}}]{Vavi2001}
\bibinfo{author}{\bibfnamefont{M.~G.} \bibnamefont{Vavilov}},
  \bibinfo{author}{\bibfnamefont{V.}~\bibnamefont{Ambegaokar}},
  \bibnamefont{and} \bibinfo{author}{\bibfnamefont{I.~L.}
  \bibnamefont{Aleiner}}, \bibinfo{journal}{Phys. Rev. B}
  \textbf{\bibinfo{volume}{63}}, \bibinfo{pages}{195313}
  (\bibinfo{year}{2001}),
  \urlprefix\url{http://link.aps.org/doi/10.1103/PhysRevB.63.195313}.

\bibitem[{\citenamefont{Moskalets and B\"{u}ttiker}(2002)}]{Mosk2002}
\bibinfo{author}{\bibfnamefont{M.}~\bibnamefont{Moskalets}} \bibnamefont{and}
  \bibinfo{author}{\bibfnamefont{M.}~\bibnamefont{B\"{u}ttiker}},
  \bibinfo{journal}{Phys. Rev. B} \textbf{\bibinfo{volume}{66}},
  \bibinfo{pages}{205320} (\bibinfo{year}{2002}),
  \urlprefix\url{http://link.aps.org/doi/10.1103/PhysRevB.66.205320}.

\bibitem[{\citenamefont{Foa~Torres}(2005)}]{Torres2005}
\bibinfo{author}{\bibfnamefont{L.~E.~F.} \bibnamefont{Foa~Torres}},
  \bibinfo{journal}{Phys. Rev. B} \textbf{\bibinfo{volume}{72}},
  \bibinfo{pages}{245339} (\bibinfo{year}{2005}),
  \urlprefix\url{http://link.aps.org/doi/10.1103/PhysRevB.72.245339}.

\bibitem[{\citenamefont{Foa~Torres et~al.}(2011)\citenamefont{Foa~Torres,
  Calvo, Rocha, and Cuniberti}}]{Torres2011}
\bibinfo{author}{\bibfnamefont{L.~E.~F.} \bibnamefont{Foa~Torres}},
  \bibinfo{author}{\bibfnamefont{H.~L.} \bibnamefont{Calvo}},
  \bibinfo{author}{\bibfnamefont{C.~G.} \bibnamefont{Rocha}}, \bibnamefont{and}
  \bibinfo{author}{\bibfnamefont{G.}~\bibnamefont{Cuniberti}},
  \bibinfo{journal}{Appl. Phys. Lett.} \textbf{\bibinfo{volume}{99}},
  \bibinfo{eid}{092102} (\bibinfo{year}{2011}),
  \urlprefix\url{http://scitation.aip.org/content/aip/journal/apl/99/9/10.1063/1.3630025}.

\bibitem[{\citenamefont{Agarwal and Sen}(2007)}]{Agar2007}
\bibinfo{author}{\bibfnamefont{A.}~\bibnamefont{Agarwal}} \bibnamefont{and}
  \bibinfo{author}{\bibfnamefont{D.}~\bibnamefont{Sen}}, \bibinfo{journal}{J.
  Phys. Condens. Matter} \textbf{\bibinfo{volume}{19}}, \bibinfo{pages}{046205}
  (\bibinfo{year}{2007}),
  \urlprefix\url{http://stacks.iop.org/0953-8984/19/i=4/a=046205}.

\bibitem[{\citenamefont{Ganzhorn and Wernsdorfer}(2012)}]{Ganzhorn2012}
\bibinfo{author}{\bibfnamefont{M.}~\bibnamefont{Ganzhorn}} \bibnamefont{and}
  \bibinfo{author}{\bibfnamefont{W.}~\bibnamefont{Wernsdorfer}},
  \bibinfo{journal}{Phys. Rev. Lett.} \textbf{\bibinfo{volume}{108}},
  \bibinfo{pages}{175502} (\bibinfo{year}{2012}),
  \urlprefix\url{http://link.aps.org/doi/10.1103/PhysRevLett.108.175502}.

\bibitem[{\citenamefont{Brouwer}(1998)}]{Brouwer1998}
\bibinfo{author}{\bibfnamefont{P.~W.} \bibnamefont{Brouwer}},
  \bibinfo{journal}{Phys. Rev. B} \textbf{\bibinfo{volume}{58}},
  \bibinfo{pages}{R10135} (\bibinfo{year}{1998}),
  \urlprefix\url{http://link.aps.org/doi/10.1103/PhysRevB.58.R10135}.

\bibitem[{\citenamefont{Splettstoesser
  et~al.}(2005)\citenamefont{Splettstoesser, Governale, K\"onig, and
  Fazio}}]{Splettstoesser2005}
\bibinfo{author}{\bibfnamefont{J.}~\bibnamefont{Splettstoesser}},
  \bibinfo{author}{\bibfnamefont{M.}~\bibnamefont{Governale}},
  \bibinfo{author}{\bibfnamefont{J.}~\bibnamefont{K\"onig}}, \bibnamefont{and}
  \bibinfo{author}{\bibfnamefont{R.}~\bibnamefont{Fazio}},
  \bibinfo{journal}{Phys. Rev. Lett.} \textbf{\bibinfo{volume}{95}},
  \bibinfo{pages}{246803} (\bibinfo{year}{2005}),
  \urlprefix\url{http://link.aps.org/doi/10.1103/PhysRevLett.95.246803}.

\bibitem[{\citenamefont{Splettstoesser
  et~al.}(2006)\citenamefont{Splettstoesser, Governale, K\"onig, and
  Fazio}}]{Splettstoesser2006}
\bibinfo{author}{\bibfnamefont{J.}~\bibnamefont{Splettstoesser}},
  \bibinfo{author}{\bibfnamefont{M.}~\bibnamefont{Governale}},
  \bibinfo{author}{\bibfnamefont{J.}~\bibnamefont{K\"onig}}, \bibnamefont{and}
  \bibinfo{author}{\bibfnamefont{R.}~\bibnamefont{Fazio}},
  \bibinfo{journal}{Phys. Rev. B} \textbf{\bibinfo{volume}{74}},
  \bibinfo{pages}{085305} (\bibinfo{year}{2006}),
  \urlprefix\url{http://link.aps.org/doi/10.1103/PhysRevB.74.085305}.

\bibitem[{\citenamefont{Riwar and Schmidt}(2009)}]{Riwar2009}
\bibinfo{author}{\bibfnamefont{R.-P.} \bibnamefont{Riwar}} \bibnamefont{and}
  \bibinfo{author}{\bibfnamefont{T.~L.} \bibnamefont{Schmidt}},
  \bibinfo{journal}{Phys. Rev. B} \textbf{\bibinfo{volume}{80}},
  \bibinfo{pages}{125109} (\bibinfo{year}{2009}),
  \urlprefix\url{http://link.aps.org/doi/10.1103/PhysRevB.80.125109}.

\bibitem[{\citenamefont{Tahir and MacKinnon}(2010)}]{Tahir2010}
\bibinfo{author}{\bibfnamefont{M.}~\bibnamefont{Tahir}} \bibnamefont{and}
  \bibinfo{author}{\bibfnamefont{A.}~\bibnamefont{MacKinnon}},
  \bibinfo{journal}{Phys. Rev. B} \textbf{\bibinfo{volume}{81}},
  \bibinfo{pages}{195444} (\bibinfo{year}{2010}),
  \urlprefix\url{http://link.aps.org/doi/10.1103/PhysRevB.81.195444}.

\bibitem[{\citenamefont{Albrecht et~al.}(2012)\citenamefont{Albrecht, Wang,
  M\"uhlbacher, Thoss, and Komnik}}]{Albrecht2012}
\bibinfo{author}{\bibfnamefont{K.~F.} \bibnamefont{Albrecht}},
  \bibinfo{author}{\bibfnamefont{H.}~\bibnamefont{Wang}},
  \bibinfo{author}{\bibfnamefont{L.}~\bibnamefont{M\"uhlbacher}},
  \bibinfo{author}{\bibfnamefont{M.}~\bibnamefont{Thoss}}, \bibnamefont{and}
  \bibinfo{author}{\bibfnamefont{A.}~\bibnamefont{Komnik}},
  \bibinfo{journal}{Phys. Rev. B} \textbf{\bibinfo{volume}{86}},
  \bibinfo{pages}{081412} (\bibinfo{year}{2012}),
  \urlprefix\url{http://link.aps.org/doi/10.1103/PhysRevB.86.081412}.

\bibitem[{\citenamefont{Biggio et~al.}(2014)\citenamefont{Biggio, Cavaliere,
  Storace, and Sassetti}}]{Biggio2014}
\bibinfo{author}{\bibfnamefont{M.}~\bibnamefont{Biggio}},
  \bibinfo{author}{\bibfnamefont{F.}~\bibnamefont{Cavaliere}},
  \bibinfo{author}{\bibfnamefont{M.}~\bibnamefont{Storace}}, \bibnamefont{and}
  \bibinfo{author}{\bibfnamefont{M.}~\bibnamefont{Sassetti}},
  \bibinfo{journal}{Ann. Phys. (Berlin)} \textbf{\bibinfo{volume}{526}},
  \bibinfo{pages}{541} (\bibinfo{year}{2014}), ISSN \bibinfo{issn}{1521-3889},
  \urlprefix\url{http://dx.doi.org/10.1002/andp.201400140}.

\bibitem[{\citenamefont{Wilner et~al.}(2013)\citenamefont{Wilner, Wang, Cohen,
  Thoss, and Rabani}}]{Rabani2013}
\bibinfo{author}{\bibfnamefont{E.~Y.} \bibnamefont{Wilner}},
  \bibinfo{author}{\bibfnamefont{H.}~\bibnamefont{Wang}},
  \bibinfo{author}{\bibfnamefont{G.}~\bibnamefont{Cohen}},
  \bibinfo{author}{\bibfnamefont{M.}~\bibnamefont{Thoss}}, \bibnamefont{and}
  \bibinfo{author}{\bibfnamefont{E.}~\bibnamefont{Rabani}},
  \bibinfo{journal}{Phys. Rev. B} \textbf{\bibinfo{volume}{88}},
  \bibinfo{pages}{045137} (\bibinfo{year}{2013}),
  \urlprefix\url{http://link.aps.org/doi/10.1103/PhysRevB.88.045137}.

\bibitem[{\citenamefont{Mitra et~al.}(2005)\citenamefont{Mitra, Aleiner, and
  Millis}}]{Millis2005}
\bibinfo{author}{\bibfnamefont{A.}~\bibnamefont{Mitra}},
  \bibinfo{author}{\bibfnamefont{I.}~\bibnamefont{Aleiner}}, \bibnamefont{and}
  \bibinfo{author}{\bibfnamefont{A.~J.} \bibnamefont{Millis}},
  \bibinfo{journal}{Phys. Rev. Lett.} \textbf{\bibinfo{volume}{94}},
  \bibinfo{pages}{076404} (\bibinfo{year}{2005}),
  \urlprefix\url{http://link.aps.org/doi/10.1103/PhysRevLett.94.076404}.

\bibitem[{\citenamefont{Koch and von Oppen}(2005)}]{Koch2005}
\bibinfo{author}{\bibfnamefont{J.}~\bibnamefont{Koch}} \bibnamefont{and}
  \bibinfo{author}{\bibfnamefont{F.}~\bibnamefont{von Oppen}},
  \bibinfo{journal}{Phys. Rev. Lett.} \textbf{\bibinfo{volume}{94}},
  \bibinfo{pages}{206804} (\bibinfo{year}{2005}),
  \urlprefix\url{http://link.aps.org/doi/10.1103/PhysRevLett.94.206804}.

\bibitem[{\citenamefont{Koch et~al.}(2006)\citenamefont{Koch, von Oppen, and
  Andreev}}]{Koch2006}
\bibinfo{author}{\bibfnamefont{J.}~\bibnamefont{Koch}},
  \bibinfo{author}{\bibfnamefont{F.}~\bibnamefont{von Oppen}},
  \bibnamefont{and} \bibinfo{author}{\bibfnamefont{A.~V.}
  \bibnamefont{Andreev}}, \bibinfo{journal}{Phys. Rev. B}
  \textbf{\bibinfo{volume}{74}}, \bibinfo{pages}{205438}
  (\bibinfo{year}{2006}),
  \urlprefix\url{http://link.aps.org/doi/10.1103/PhysRevB.74.205438}.

\bibitem[{\citenamefont{Braig and Flensberg}(2003)}]{Braig2003}
\bibinfo{author}{\bibfnamefont{S.}~\bibnamefont{Braig}} \bibnamefont{and}
  \bibinfo{author}{\bibfnamefont{K.}~\bibnamefont{Flensberg}},
  \bibinfo{journal}{Phys. Rev. B} \textbf{\bibinfo{volume}{68}},
  \bibinfo{pages}{205324} (\bibinfo{year}{2003}),
  \urlprefix\url{http://link.aps.org/doi/10.1103/PhysRevB.68.205324}.

\bibitem[{\citenamefont{Piovano et~al.}(2011)\citenamefont{Piovano, Cavaliere,
  Paladino, and Sassetti}}]{Piovano2011}
\bibinfo{author}{\bibfnamefont{G.}~\bibnamefont{Piovano}},
  \bibinfo{author}{\bibfnamefont{F.}~\bibnamefont{Cavaliere}},
  \bibinfo{author}{\bibfnamefont{E.}~\bibnamefont{Paladino}}, \bibnamefont{and}
  \bibinfo{author}{\bibfnamefont{M.}~\bibnamefont{Sassetti}},
  \bibinfo{journal}{Phys. Rev. B} \textbf{\bibinfo{volume}{83}},
  \bibinfo{pages}{245311} (\bibinfo{year}{2011}),
  \urlprefix\url{http://link.aps.org/doi/10.1103/PhysRevB.83.245311}.

\bibitem[{\citenamefont{Cavaliere et~al.}(2010)\citenamefont{Cavaliere,
  Mariani, Leturcq, Stampfer, and Sassetti}}]{Cavaliere2010}
\bibinfo{author}{\bibfnamefont{F.}~\bibnamefont{Cavaliere}},
  \bibinfo{author}{\bibfnamefont{E.}~\bibnamefont{Mariani}},
  \bibinfo{author}{\bibfnamefont{R.}~\bibnamefont{Leturcq}},
  \bibinfo{author}{\bibfnamefont{C.}~\bibnamefont{Stampfer}}, \bibnamefont{and}
  \bibinfo{author}{\bibfnamefont{M.}~\bibnamefont{Sassetti}},
  \bibinfo{journal}{Phys. Rev. B} \textbf{\bibinfo{volume}{81}},
  \bibinfo{pages}{201303} (\bibinfo{year}{2010}),
  \urlprefix\url{http://link.aps.org/doi/10.1103/PhysRevB.81.201303}.

\bibitem[{\citenamefont{Traverso~Ziani
  et~al.}(2011)\citenamefont{Traverso~Ziani, Piovano, Cavaliere, and
  Sassetti}}]{Ziani2011}
\bibinfo{author}{\bibfnamefont{N.}~\bibnamefont{Traverso~Ziani}},
  \bibinfo{author}{\bibfnamefont{G.}~\bibnamefont{Piovano}},
  \bibinfo{author}{\bibfnamefont{F.}~\bibnamefont{Cavaliere}},
  \bibnamefont{and} \bibinfo{author}{\bibfnamefont{M.}~\bibnamefont{Sassetti}},
  \bibinfo{journal}{Phys. Rev. B} \textbf{\bibinfo{volume}{84}},
  \bibinfo{pages}{155423} (\bibinfo{year}{2011}),
  \urlprefix\url{http://link.aps.org/doi/10.1103/PhysRevB.84.155423}.

\bibitem[{\citenamefont{Blanter et~al.}(2004)\citenamefont{Blanter, Usmani, and
  Nazarov}}]{Blanter2004}
\bibinfo{author}{\bibfnamefont{Y.~M.} \bibnamefont{Blanter}},
  \bibinfo{author}{\bibfnamefont{O.}~\bibnamefont{Usmani}}, \bibnamefont{and}
  \bibinfo{author}{\bibfnamefont{Y.~V.} \bibnamefont{Nazarov}},
  \bibinfo{journal}{Phys. Rev. Lett.} \textbf{\bibinfo{volume}{93}},
  \bibinfo{pages}{136802} (\bibinfo{year}{2004}).

\bibitem[{\citenamefont{Blanter et~al.}(2005)\citenamefont{Blanter, Usmani, and
  Nazarov}}]{BlanterErr2005}
\bibinfo{author}{\bibfnamefont{Y.~M.} \bibnamefont{Blanter}},
  \bibinfo{author}{\bibfnamefont{O.}~\bibnamefont{Usmani}}, \bibnamefont{and}
  \bibinfo{author}{\bibfnamefont{Y.~V.} \bibnamefont{Nazarov}},
  \bibinfo{journal}{Phys. Rev. Lett.} \textbf{\bibinfo{volume}{94}},
  \bibinfo{pages}{049904} (\bibinfo{year}{2005}),
  \urlprefix\url{http://link.aps.org/doi/10.1103/PhysRevLett.94.049904}.

\bibitem[{\citenamefont{Armour et~al.}(2004)\citenamefont{Armour, Blencowe, and
  Zhang}}]{Armour2004}
\bibinfo{author}{\bibfnamefont{A.~D.} \bibnamefont{Armour}},
  \bibinfo{author}{\bibfnamefont{M.~P.} \bibnamefont{Blencowe}},
  \bibnamefont{and} \bibinfo{author}{\bibfnamefont{Y.}~\bibnamefont{Zhang}},
  \bibinfo{journal}{Phys. Rev. B} \textbf{\bibinfo{volume}{69}},
  \bibinfo{pages}{125313} (\bibinfo{year}{2004}),
  \urlprefix\url{http://link.aps.org/doi/10.1103/PhysRevB.69.125313}.

\bibitem[{\citenamefont{Clerk and Bennett}(2005)}]{Clerck2005}
\bibinfo{author}{\bibfnamefont{A.~A.} \bibnamefont{Clerk}} \bibnamefont{and}
  \bibinfo{author}{\bibfnamefont{S.}~\bibnamefont{Bennett}},
  \bibinfo{journal}{New J. Phys.} \textbf{\bibinfo{volume}{7}},
  \bibinfo{pages}{238} (\bibinfo{year}{2005}),
  \urlprefix\url{http://stacks.iop.org/1367-2630/7/i=1/a=238}.

\bibitem[{\citenamefont{Boubsi et~al.}(2008)\citenamefont{Boubsi, Usmani, and
  Blanter}}]{Boubsi2008}
\bibinfo{author}{\bibfnamefont{R.~E.} \bibnamefont{Boubsi}},
  \bibinfo{author}{\bibfnamefont{O.}~\bibnamefont{Usmani}}, \bibnamefont{and}
  \bibinfo{author}{\bibfnamefont{Y.~M.} \bibnamefont{Blanter}},
  \bibinfo{journal}{New J. Phys.} \textbf{\bibinfo{volume}{10}},
  \bibinfo{pages}{095011} (\bibinfo{year}{2008}),
  \urlprefix\url{http://stacks.iop.org/1367-2630/10/i=9/a=095011}.

\bibitem[{\citenamefont{Galperin et~al.}(2005)\citenamefont{Galperin, Ratner,
  and Nitzan}}]{Galperin2005}
\bibinfo{author}{\bibfnamefont{M.}~\bibnamefont{Galperin}},
  \bibinfo{author}{\bibfnamefont{M.~A.} \bibnamefont{Ratner}},
  \bibnamefont{and} \bibinfo{author}{\bibfnamefont{A.}~\bibnamefont{Nitzan}},
  \bibinfo{journal}{Nano Lett.} \textbf{\bibinfo{volume}{5}},
  \bibinfo{pages}{125} (\bibinfo{year}{2005}), \bibinfo{note}{pMID: 15792425},
  \eprint{http://dx.doi.org/10.1021/nl048216c},
  \urlprefix\url{http://dx.doi.org/10.1021/nl048216c}.

\bibitem[{\citenamefont{Hussein et~al.}(2010)\citenamefont{Hussein, Metelmann,
  Zedler, and Brandes}}]{Brandes2010}
\bibinfo{author}{\bibfnamefont{R.}~\bibnamefont{Hussein}},
  \bibinfo{author}{\bibfnamefont{A.}~\bibnamefont{Metelmann}},
  \bibinfo{author}{\bibfnamefont{P.}~\bibnamefont{Zedler}}, \bibnamefont{and}
  \bibinfo{author}{\bibfnamefont{T.}~\bibnamefont{Brandes}},
  \bibinfo{journal}{Phys. Rev. B} \textbf{\bibinfo{volume}{82}},
  \bibinfo{pages}{165406} (\bibinfo{year}{2010}),
  \urlprefix\url{http://link.aps.org/doi/10.1103/PhysRevB.82.165406}.

\bibitem[{\citenamefont{Pistolesi et~al.}(2008)\citenamefont{Pistolesi,
  Blanter, and Martin}}]{Pistolesi2008}
\bibinfo{author}{\bibfnamefont{F.}~\bibnamefont{Pistolesi}},
  \bibinfo{author}{\bibfnamefont{Y.~M.} \bibnamefont{Blanter}},
  \bibnamefont{and} \bibinfo{author}{\bibfnamefont{I.}~\bibnamefont{Martin}},
  \bibinfo{journal}{Phys. Rev. B} \textbf{\bibinfo{volume}{78}},
  \bibinfo{pages}{085127} (\bibinfo{year}{2008}),
  \urlprefix\url{http://link.aps.org/doi/10.1103/PhysRevB.78.085127}.

\bibitem[{\citenamefont{Cataudella et~al.}(2011)\citenamefont{Cataudella,
  De~Filippis, and Perroni}}]{Cataudella2011}
\bibinfo{author}{\bibfnamefont{V.}~\bibnamefont{Cataudella}},
  \bibinfo{author}{\bibfnamefont{G.}~\bibnamefont{De~Filippis}},
  \bibnamefont{and} \bibinfo{author}{\bibfnamefont{C.~A.}
  \bibnamefont{Perroni}}, \bibinfo{journal}{Phys. Rev. B}
  \textbf{\bibinfo{volume}{83}}, \bibinfo{pages}{165203}
  (\bibinfo{year}{2011}),
  \urlprefix\url{http://link.aps.org/doi/10.1103/PhysRevB.83.165203}.

\bibitem[{\citenamefont{Mozyrsky et~al.}(2006)\citenamefont{Mozyrsky, Hastings,
  and Martin}}]{Mozyrsky2006}
\bibinfo{author}{\bibfnamefont{D.}~\bibnamefont{Mozyrsky}},
  \bibinfo{author}{\bibfnamefont{M.~B.} \bibnamefont{Hastings}},
  \bibnamefont{and} \bibinfo{author}{\bibfnamefont{I.}~\bibnamefont{Martin}},
  \bibinfo{journal}{Phys. Rev. B} \textbf{\bibinfo{volume}{73}},
  \bibinfo{pages}{035104} (\bibinfo{year}{2006}),
  \urlprefix\url{http://link.aps.org/doi/10.1103/PhysRevB.73.035104}.

\bibitem[{\citenamefont{L\"u et~al.}(2012)\citenamefont{L\"u, Brandbyge,
  Hedeg\aa{}rd, Todorov, and Dundas}}]{Dundas2012}
\bibinfo{author}{\bibfnamefont{J.-T.} \bibnamefont{L\"u}},
  \bibinfo{author}{\bibfnamefont{M.}~\bibnamefont{Brandbyge}},
  \bibinfo{author}{\bibfnamefont{P.}~\bibnamefont{Hedeg\aa{}rd}},
  \bibinfo{author}{\bibfnamefont{T.~N.} \bibnamefont{Todorov}},
  \bibnamefont{and} \bibinfo{author}{\bibfnamefont{D.}~\bibnamefont{Dundas}},
  \bibinfo{journal}{Phys. Rev. B} \textbf{\bibinfo{volume}{85}},
  \bibinfo{pages}{245444} (\bibinfo{year}{2012}),
  \urlprefix\url{http://link.aps.org/doi/10.1103/PhysRevB.85.245444}.

\bibitem[{\citenamefont{Marx and Hutter}(2000)}]{Marx2000}
\bibinfo{author}{\bibfnamefont{D.}~\bibnamefont{Marx}} \bibnamefont{and}
  \bibinfo{author}{\bibfnamefont{J.}~\bibnamefont{Hutter}},
  \bibinfo{journal}{Modern methods and algorithms of quantum chemistry}
  \textbf{\bibinfo{volume}{1}}, \bibinfo{pages}{301} (\bibinfo{year}{2000}).

\bibitem[{\citenamefont{Kartsev et~al.}(2014)\citenamefont{Kartsev, Verdozzi,
  and Stefanucci}}]{Stefanucci2014}
\bibinfo{author}{\bibfnamefont{A.}~\bibnamefont{Kartsev}},
  \bibinfo{author}{\bibfnamefont{C.}~\bibnamefont{Verdozzi}}, \bibnamefont{and}
  \bibinfo{author}{\bibfnamefont{G.}~\bibnamefont{Stefanucci}},
  \bibinfo{journal}{EPJ B} \textbf{\bibinfo{volume}{87}}
  (\bibinfo{year}{2014}),
  \urlprefix\url{http://dx.doi.org/10.1140/epjb/e2013-40905-5}.

\bibitem[{\citenamefont{Li et~al.}(2005)\citenamefont{Li, Tully, Schlegel, and
  Frisch}}]{li2005}
\bibinfo{author}{\bibfnamefont{X.}~\bibnamefont{Li}},
  \bibinfo{author}{\bibfnamefont{J.~C.} \bibnamefont{Tully}},
  \bibinfo{author}{\bibfnamefont{H.~B.} \bibnamefont{Schlegel}},
  \bibnamefont{and} \bibinfo{author}{\bibfnamefont{M.~J.}
  \bibnamefont{Frisch}}, \bibinfo{journal}{J. Chem. Phys.}
  \textbf{\bibinfo{volume}{123}}, \bibinfo{pages}{084106}
  (\bibinfo{year}{2005}).

\bibitem[{\citenamefont{Nocera et~al.}(2011)\citenamefont{Nocera, Perroni,
  Marigliano~Ramaglia, and Cataudella}}]{Nocera2011}
\bibinfo{author}{\bibfnamefont{A.}~\bibnamefont{Nocera}},
  \bibinfo{author}{\bibfnamefont{C.~A.} \bibnamefont{Perroni}},
  \bibinfo{author}{\bibfnamefont{V.}~\bibnamefont{Marigliano~Ramaglia}},
  \bibnamefont{and}
  \bibinfo{author}{\bibfnamefont{V.}~\bibnamefont{Cataudella}},
  \bibinfo{journal}{Phys. Rev. B} \textbf{\bibinfo{volume}{83}},
  \bibinfo{pages}{115420} (\bibinfo{year}{2011}),
  \urlprefix\url{http://link.aps.org/doi/10.1103/PhysRevB.83.115420}.

\bibitem[{\citenamefont{Nocera et~al.}(2012)\citenamefont{Nocera, Perroni,
  Marigliano~Ramaglia, and Cataudella}}]{Nocera2012}
\bibinfo{author}{\bibfnamefont{A.}~\bibnamefont{Nocera}},
  \bibinfo{author}{\bibfnamefont{C.~A.} \bibnamefont{Perroni}},
  \bibinfo{author}{\bibfnamefont{V.}~\bibnamefont{Marigliano~Ramaglia}},
  \bibnamefont{and}
  \bibinfo{author}{\bibfnamefont{V.}~\bibnamefont{Cataudella}},
  \bibinfo{journal}{Phys. Rev. B} \textbf{\bibinfo{volume}{86}},
  \bibinfo{pages}{035420} (\bibinfo{year}{2012}),
  \urlprefix\url{http://link.aps.org/doi/10.1103/PhysRevB.86.035420}.

\bibitem[{\citenamefont{Bode et~al.}(2012)\citenamefont{Bode, Kusminskiy,
  Egger, and von Oppen}}]{Bode2012}
\bibinfo{author}{\bibfnamefont{N.}~\bibnamefont{Bode}},
  \bibinfo{author}{\bibfnamefont{S.~V.} \bibnamefont{Kusminskiy}},
  \bibinfo{author}{\bibfnamefont{R.}~\bibnamefont{Egger}}, \bibnamefont{and}
  \bibinfo{author}{\bibfnamefont{F.}~\bibnamefont{von Oppen}},
  \bibinfo{journal}{Beilstein J. Nanotechnol.} \textbf{\bibinfo{volume}{3}},
  \bibinfo{pages}{144} (\bibinfo{year}{2012}), ISSN \bibinfo{issn}{2190-4286}.

\bibitem[{\citenamefont{Perroni et~al.}(2013)\citenamefont{Perroni, Nocera, and
  Cataudella}}]{Perroni2013}
\bibinfo{author}{\bibfnamefont{C.~A.} \bibnamefont{Perroni}},
  \bibinfo{author}{\bibfnamefont{A.}~\bibnamefont{Nocera}}, \bibnamefont{and}
  \bibinfo{author}{\bibfnamefont{V.}~\bibnamefont{Cataudella}},
  \bibinfo{journal}{EPL (Europhys. Lett.)} \textbf{\bibinfo{volume}{103}},
  \bibinfo{pages}{58001} (\bibinfo{year}{2013}),
  \urlprefix\url{http://stacks.iop.org/0295-5075/103/i=5/a=58001}.

\bibitem[{\citenamefont{Weiss}(2008)}]{Weiss2008}
\bibinfo{author}{\bibfnamefont{U.}~\bibnamefont{Weiss}},
  \emph{\bibinfo{title}{Quantum dissipative systems}}
  (\bibinfo{publisher}{World Scientific}, \bibinfo{year}{2008}),
  \bibinfo{edition}{3rd} ed.

\bibitem[{\citenamefont{Haug and Jauho}(2008)}]{Haug2008}
\bibinfo{author}{\bibfnamefont{H.}~\bibnamefont{Haug}} \bibnamefont{and}
  \bibinfo{author}{\bibfnamefont{A.-P.} \bibnamefont{Jauho}},
  \emph{\bibinfo{title}{Quantum kinetics in transport and optics of
  semiconductors}}, vol. \bibinfo{volume}{123} (\bibinfo{publisher}{Springer,
  Berlin}, \bibinfo{year}{2008}).

\bibitem[{\citenamefont{Perroni et~al.}(2015)\citenamefont{Perroni, Ninno, and
  Cataudella}}]{Perroni2014a}
\bibinfo{author}{\bibfnamefont{C.~A.} \bibnamefont{Perroni}},
  \bibinfo{author}{\bibfnamefont{D.}~\bibnamefont{Ninno}}, \bibnamefont{and}
  \bibinfo{author}{\bibfnamefont{V.}~\bibnamefont{Cataudella}},
  \bibinfo{journal}{New J. Phys.} \textbf{\bibinfo{volume}{17}},
  \bibinfo{pages}{083050} (\bibinfo{year}{2015}).

\bibitem[{\citenamefont{Br\"uggemann et~al.}(2012)\citenamefont{Br\"uggemann,
  Weick, Pistolesi, and von Oppen}}]{VonOppen2012}
\bibinfo{author}{\bibfnamefont{J.}~\bibnamefont{Br\"uggemann}},
  \bibinfo{author}{\bibfnamefont{G.}~\bibnamefont{Weick}},
  \bibinfo{author}{\bibfnamefont{F.}~\bibnamefont{Pistolesi}},
  \bibnamefont{and} \bibinfo{author}{\bibfnamefont{F.}~\bibnamefont{von
  Oppen}}, \bibinfo{journal}{Phys. Rev. B} \textbf{\bibinfo{volume}{85}},
  \bibinfo{pages}{125441} (\bibinfo{year}{2012}),
  \urlprefix\url{http://link.aps.org/doi/10.1103/PhysRevB.85.125441}.

\bibitem[{\citenamefont{Perroni
  et~al.}(2014{\natexlab{a}})\citenamefont{Perroni, Romeo, Nocera, Ramaglia,
  Citro, and Cataudella}}]{Perroni2014b}
\bibinfo{author}{\bibfnamefont{C.~A.} \bibnamefont{Perroni}},
  \bibinfo{author}{\bibfnamefont{F.}~\bibnamefont{Romeo}},
  \bibinfo{author}{\bibfnamefont{A.}~\bibnamefont{Nocera}},
  \bibinfo{author}{\bibfnamefont{V.~M.} \bibnamefont{Ramaglia}},
  \bibinfo{author}{\bibfnamefont{R.}~\bibnamefont{Citro}}, \bibnamefont{and}
  \bibinfo{author}{\bibfnamefont{V.}~\bibnamefont{Cataudella}},
  \bibinfo{journal}{J. Phys. Conden. Matter} \textbf{\bibinfo{volume}{26}},
  \bibinfo{pages}{365301} (\bibinfo{year}{2014}{\natexlab{a}}),
  \urlprefix\url{http://stacks.iop.org/0953-8984/26/i=36/a=365301}.

\bibitem[{\citenamefont{Lu et~al.}(2003)\citenamefont{Lu, Grobis, Khoo, Louie,
  and Crommie}}]{Lu2003}
\bibinfo{author}{\bibfnamefont{X.}~\bibnamefont{Lu}},
  \bibinfo{author}{\bibfnamefont{M.}~\bibnamefont{Grobis}},
  \bibinfo{author}{\bibfnamefont{K.~H.} \bibnamefont{Khoo}},
  \bibinfo{author}{\bibfnamefont{S.~G.} \bibnamefont{Louie}}, \bibnamefont{and}
  \bibinfo{author}{\bibfnamefont{M.~F.} \bibnamefont{Crommie}},
  \bibinfo{journal}{Phys. Rev. Lett.} \textbf{\bibinfo{volume}{90}},
  \bibinfo{pages}{096802} (\bibinfo{year}{2003}),
  \urlprefix\url{http://link.aps.org/doi/10.1103/PhysRevLett.90.096802}.

\bibitem[{\citenamefont{Perroni
  et~al.}(2014{\natexlab{b}})\citenamefont{Perroni, Ninno, and
  Cataudella}}]{Perroni2014}
\bibinfo{author}{\bibfnamefont{C.~A.} \bibnamefont{Perroni}},
  \bibinfo{author}{\bibfnamefont{D.}~\bibnamefont{Ninno}}, \bibnamefont{and}
  \bibinfo{author}{\bibfnamefont{V.}~\bibnamefont{Cataudella}},
  \bibinfo{journal}{Phys. Rev. B} \textbf{\bibinfo{volume}{90}},
  \bibinfo{pages}{125421} (\bibinfo{year}{2014}{\natexlab{b}}),
  \urlprefix\url{http://link.aps.org/doi/10.1103/PhysRevB.90.125421}.

\bibitem[{\citenamefont{Almbladh and Minnhagen}(1978)}]{almb78}
\bibinfo{author}{\bibfnamefont{C.~O.} \bibnamefont{Almbladh}} \bibnamefont{and}
  \bibinfo{author}{\bibfnamefont{P.}~\bibnamefont{Minnhagen}},
  \bibinfo{journal}{Phys. Rev. B} \textbf{\bibinfo{volume}{17}},
  \bibinfo{pages}{929} (\bibinfo{year}{1978}),
  \urlprefix\url{http://link.aps.org/doi/10.1103/PhysRevB.17.929}.

\bibitem[{\citenamefont{Mahan}(2000)}]{mahan2000}
\bibinfo{author}{\bibfnamefont{G.~D.} \bibnamefont{Mahan}},
  \emph{\bibinfo{title}{Many-particle physics}} (\bibinfo{publisher}{Springer
  Science \& Business Media}, \bibinfo{year}{2000}).

\bibitem[{\citenamefont{Panasyuk et~al.}(2012)\citenamefont{Panasyuk, Levin,
  and Yerkes}}]{Pana2012}
\bibinfo{author}{\bibfnamefont{G.~Y.} \bibnamefont{Panasyuk}},
  \bibinfo{author}{\bibfnamefont{G.~A.} \bibnamefont{Levin}}, \bibnamefont{and}
  \bibinfo{author}{\bibfnamefont{K.~L.} \bibnamefont{Yerkes}},
  \bibinfo{journal}{Phys. Rev. E} \textbf{\bibinfo{volume}{86}},
  \bibinfo{pages}{021116} (\bibinfo{year}{2012}),
  \urlprefix\url{http://link.aps.org/doi/10.1103/PhysRevE.86.021116}.

\bibitem[{\citenamefont{Wang}(2007)}]{Wang2007a}
\bibinfo{author}{\bibfnamefont{J.-S.} \bibnamefont{Wang}},
  \bibinfo{journal}{Phys. Rev. Lett.} \textbf{\bibinfo{volume}{99}},
  \bibinfo{pages}{160601} (\bibinfo{year}{2007}),
  \urlprefix\url{http://link.aps.org/doi/10.1103/PhysRevLett.99.160601}.

\bibitem[{\citenamefont{Wang et~al.}(2008)\citenamefont{Wang, Wang, and
  Lü}}]{Wang2008}
\bibinfo{author}{\bibfnamefont{J.-S.} \bibnamefont{Wang}},
  \bibinfo{author}{\bibfnamefont{J.}~\bibnamefont{Wang}}, \bibnamefont{and}
  \bibinfo{author}{\bibfnamefont{J.~T.} \bibnamefont{Lü}},
  \bibinfo{journal}{EPJ B} \textbf{\bibinfo{volume}{62}}, \bibinfo{pages}{381}
  (\bibinfo{year}{2008}), ISSN \bibinfo{issn}{1434-6028},
  \urlprefix\url{http://dx.doi.org/10.1140/epjb/e2008-00195-8}.

\bibitem[{\citenamefont{Yee et~al.}(2011)\citenamefont{Yee, Malen, Majumdar,
  and Segalman}}]{Yee2011}
\bibinfo{author}{\bibfnamefont{S.~K.} \bibnamefont{Yee}},
  \bibinfo{author}{\bibfnamefont{J.~A.} \bibnamefont{Malen}},
  \bibinfo{author}{\bibfnamefont{A.}~\bibnamefont{Majumdar}}, \bibnamefont{and}
  \bibinfo{author}{\bibfnamefont{R.~A.} \bibnamefont{Segalman}},
  \bibinfo{journal}{Nano Lett.} \textbf{\bibinfo{volume}{11}},
  \bibinfo{pages}{4089} (\bibinfo{year}{2011}), \bibinfo{note}{pMID: 21882860}.

\bibitem[{\citenamefont{Kittel}(2004)}]{Kittel2004}
\bibinfo{author}{\bibfnamefont{C.}~\bibnamefont{Kittel}},
  \emph{\bibinfo{title}{Wiley: Introduction to Solid State Physics}}
  (\bibinfo{publisher}{Wiley edition}, \bibinfo{year}{2004}).

\bibitem[{\citenamefont{Liu et~al.}(2010)\citenamefont{Liu, Sun, and
  Xie}}]{Liu2010}
\bibinfo{author}{\bibfnamefont{J.}~\bibnamefont{Liu}},
  \bibinfo{author}{\bibfnamefont{Q.-f.} \bibnamefont{Sun}}, \bibnamefont{and}
  \bibinfo{author}{\bibfnamefont{X.~C.} \bibnamefont{Xie}},
  \bibinfo{journal}{Phys. Rev. B} \textbf{\bibinfo{volume}{81}},
  \bibinfo{pages}{245323} (\bibinfo{year}{2010}),
  \urlprefix\url{http://link.aps.org/doi/10.1103/PhysRevB.81.245323}.

\bibitem[{\citenamefont{Wang et~al.}(2007)\citenamefont{Wang, Carter,
  Lagutchev, Koh, Seong, Cahill, and Dlott}}]{Wang2007}
\bibinfo{author}{\bibfnamefont{Z.}~\bibnamefont{Wang}},
  \bibinfo{author}{\bibfnamefont{J.~A.} \bibnamefont{Carter}},
  \bibinfo{author}{\bibfnamefont{A.}~\bibnamefont{Lagutchev}},
  \bibinfo{author}{\bibfnamefont{Y.~K.} \bibnamefont{Koh}},
  \bibinfo{author}{\bibfnamefont{N.-H.} \bibnamefont{Seong}},
  \bibinfo{author}{\bibfnamefont{D.~G.} \bibnamefont{Cahill}},
  \bibnamefont{and} \bibinfo{author}{\bibfnamefont{D.~D.} \bibnamefont{Dlott}},
  \bibinfo{journal}{Science} \textbf{\bibinfo{volume}{317}},
  \bibinfo{pages}{787} (\bibinfo{year}{2007}).

\bibitem[{\citenamefont{Jezouin et~al.}(2013)\citenamefont{Jezouin, Parmentier,
  Anthore, Gennser, Cavanna, Jin, and Pierre}}]{Jezouin2013}
\bibinfo{author}{\bibfnamefont{S.}~\bibnamefont{Jezouin}},
  \bibinfo{author}{\bibfnamefont{F.~D.} \bibnamefont{Parmentier}},
  \bibinfo{author}{\bibfnamefont{A.}~\bibnamefont{Anthore}},
  \bibinfo{author}{\bibfnamefont{U.}~\bibnamefont{Gennser}},
  \bibinfo{author}{\bibfnamefont{A.}~\bibnamefont{Cavanna}},
  \bibinfo{author}{\bibfnamefont{Y.}~\bibnamefont{Jin}}, \bibnamefont{and}
  \bibinfo{author}{\bibfnamefont{F.}~\bibnamefont{Pierre}},
  \bibinfo{journal}{Science} \textbf{\bibinfo{volume}{342}},
  \bibinfo{pages}{601} (\bibinfo{year}{2013}),
  \eprint{http://www.sciencemag.org/content/342/6158/601.full.pdf},
  \urlprefix\url{http://www.sciencemag.org/content/342/6158/601.abstract}.

\bibitem[{\citenamefont{Meier et~al.}(2014)\citenamefont{Meier, Menges,
  Nirmalraj, H\"olscher, Riel, and Gotsmann}}]{Meier2014}
\bibinfo{author}{\bibfnamefont{T.}~\bibnamefont{Meier}},
  \bibinfo{author}{\bibfnamefont{F.}~\bibnamefont{Menges}},
  \bibinfo{author}{\bibfnamefont{P.}~\bibnamefont{Nirmalraj}},
  \bibinfo{author}{\bibfnamefont{H.}~\bibnamefont{H\"olscher}},
  \bibinfo{author}{\bibfnamefont{H.}~\bibnamefont{Riel}}, \bibnamefont{and}
  \bibinfo{author}{\bibfnamefont{B.}~\bibnamefont{Gotsmann}},
  \bibinfo{journal}{Phys. Rev. Lett.} \textbf{\bibinfo{volume}{113}},
  \bibinfo{pages}{060801} (\bibinfo{year}{2014}),
  \urlprefix\url{http://link.aps.org/doi/10.1103/PhysRevLett.113.060801}.

\bibitem[{\citenamefont{Weick et~al.}(2011)\citenamefont{Weick, von Oppen, and
  Pistolesi}}]{Weick2011}
\bibinfo{author}{\bibfnamefont{G.}~\bibnamefont{Weick}},
  \bibinfo{author}{\bibfnamefont{F.}~\bibnamefont{von Oppen}},
  \bibnamefont{and}
  \bibinfo{author}{\bibfnamefont{F.}~\bibnamefont{Pistolesi}},
  \bibinfo{journal}{Phys. Rev. B} \textbf{\bibinfo{volume}{83}},
  \bibinfo{pages}{035420} (\bibinfo{year}{2011}),
  \urlprefix\url{http://link.aps.org/doi/10.1103/PhysRevB.83.035420}.

\bibitem[{\citenamefont{Nayfeh and Mook}(2008)}]{Nayfeh2008}
\bibinfo{author}{\bibfnamefont{A.~H.} \bibnamefont{Nayfeh}} \bibnamefont{and}
  \bibinfo{author}{\bibfnamefont{D.~T.} \bibnamefont{Mook}},
  \emph{\bibinfo{title}{Nonlinear oscillations}} (\bibinfo{publisher}{John
  Wiley \& Sons}, \bibinfo{year}{2008}).

\bibitem[{\citenamefont{Meerwaldt et~al.}(2012)\citenamefont{Meerwaldt,
  Labadze, Schneider, Taspinar, Blanter, van~der Zant, and
  Steele}}]{Meerwaldt2012}
\bibinfo{author}{\bibfnamefont{H.~B.} \bibnamefont{Meerwaldt}},
  \bibinfo{author}{\bibfnamefont{G.}~\bibnamefont{Labadze}},
  \bibinfo{author}{\bibfnamefont{B.~H.} \bibnamefont{Schneider}},
  \bibinfo{author}{\bibfnamefont{A.}~\bibnamefont{Taspinar}},
  \bibinfo{author}{\bibfnamefont{Y.~M.} \bibnamefont{Blanter}},
  \bibinfo{author}{\bibfnamefont{H.~S.~J.} \bibnamefont{van~der Zant}},
  \bibnamefont{and} \bibinfo{author}{\bibfnamefont{G.~A.}
  \bibnamefont{Steele}}, \bibinfo{journal}{Phys. Rev. B}
  \textbf{\bibinfo{volume}{86}}, \bibinfo{pages}{115454}
  (\bibinfo{year}{2012}),
  \urlprefix\url{http://link.aps.org/doi/10.1103/PhysRevB.86.115454}.

\bibitem[{\citenamefont{Thouless}(1983)}]{Thouless1983}
\bibinfo{author}{\bibfnamefont{D.~J.} \bibnamefont{Thouless}},
  \bibinfo{journal}{Phys. Rev. B} \textbf{\bibinfo{volume}{27}},
  \bibinfo{pages}{6083} (\bibinfo{year}{1983}),
  \urlprefix\url{http://link.aps.org/doi/10.1103/PhysRevB.27.6083}.

\bibitem[{\citenamefont{Altshuler and Glazman}(1999)}]{Altshuler1999}
\bibinfo{author}{\bibfnamefont{B.~L.} \bibnamefont{Altshuler}}
  \bibnamefont{and} \bibinfo{author}{\bibfnamefont{L.~I.}
  \bibnamefont{Glazman}}, \bibinfo{journal}{Science}
  \textbf{\bibinfo{volume}{283}}, \bibinfo{pages}{1864} (\bibinfo{year}{1999}),
  \urlprefix\url{http://www.sciencemag.org/content/283/5409/1864.short}.

\bibitem[{\citenamefont{Buitelaar et~al.}(2008)\citenamefont{Buitelaar,
  Kashcheyevs, Leek, Talyanskii, Smith, Anderson, Jones, Wei, and
  Cobden}}]{Buitelaar2008}
\bibinfo{author}{\bibfnamefont{M.~R.} \bibnamefont{Buitelaar}},
  \bibinfo{author}{\bibfnamefont{V.}~\bibnamefont{Kashcheyevs}},
  \bibinfo{author}{\bibfnamefont{P.~J.} \bibnamefont{Leek}},
  \bibinfo{author}{\bibfnamefont{V.~I.} \bibnamefont{Talyanskii}},
  \bibinfo{author}{\bibfnamefont{C.~G.} \bibnamefont{Smith}},
  \bibinfo{author}{\bibfnamefont{D.}~\bibnamefont{Anderson}},
  \bibinfo{author}{\bibfnamefont{G.~A.~C.} \bibnamefont{Jones}},
  \bibinfo{author}{\bibfnamefont{J.}~\bibnamefont{Wei}}, \bibnamefont{and}
  \bibinfo{author}{\bibfnamefont{D.~H.} \bibnamefont{Cobden}},
  \bibinfo{journal}{Phys. Rev. Lett.} \textbf{\bibinfo{volume}{101}},
  \bibinfo{pages}{126803} (\bibinfo{year}{2008}),
  \urlprefix\url{http://link.aps.org/doi/10.1103/PhysRevLett.101.126803}.

\bibitem[{\citenamefont{Wei et~al.}(2001)\citenamefont{Wei, Wang, Guo, and
  Roland}}]{Wei2001}
\bibinfo{author}{\bibfnamefont{Y.}~\bibnamefont{Wei}},
  \bibinfo{author}{\bibfnamefont{J.}~\bibnamefont{Wang}},
  \bibinfo{author}{\bibfnamefont{H.}~\bibnamefont{Guo}}, \bibnamefont{and}
  \bibinfo{author}{\bibfnamefont{C.}~\bibnamefont{Roland}},
  \bibinfo{journal}{Phys. Rev. B} \textbf{\bibinfo{volume}{64}},
  \bibinfo{pages}{115321} (\bibinfo{year}{2001}),
  \urlprefix\url{http://link.aps.org/doi/10.1103/PhysRevB.64.115321}.

\bibitem[{\citenamefont{Pe\~na Aza et~al.}(2013)\citenamefont{Pe\~na Aza,
  Scorrano, and Gorelik}}]{Gorelik2013}
\bibinfo{author}{\bibfnamefont{M.~E.} \bibnamefont{Pe\~na Aza}},
  \bibinfo{author}{\bibfnamefont{A.}~\bibnamefont{Scorrano}}, \bibnamefont{and}
  \bibinfo{author}{\bibfnamefont{L.~Y.} \bibnamefont{Gorelik}},
  \bibinfo{journal}{Phys. Rev. B} \textbf{\bibinfo{volume}{88}},
  \bibinfo{pages}{035412} (\bibinfo{year}{2013}),
  \urlprefix\url{http://link.aps.org/doi/10.1103/PhysRevB.88.035412}.

\bibitem[{\citenamefont{Romeo and Citro}(2009)}]{Citro2009}
\bibinfo{author}{\bibfnamefont{F.}~\bibnamefont{Romeo}} \bibnamefont{and}
  \bibinfo{author}{\bibfnamefont{R.}~\bibnamefont{Citro}},
  \bibinfo{journal}{Phys. Rev. B} \textbf{\bibinfo{volume}{80}},
  \bibinfo{pages}{235328} (\bibinfo{year}{2009}),
  \urlprefix\url{http://link.aps.org/doi/10.1103/PhysRevB.80.235328}.

\bibitem[{\citenamefont{Nocera et~al.}(2013)\citenamefont{Nocera, Perroni,
  Ramaglia, Cantele, and Cataudella}}]{Nocera2013}
\bibinfo{author}{\bibfnamefont{A.}~\bibnamefont{Nocera}},
  \bibinfo{author}{\bibfnamefont{C.~A.} \bibnamefont{Perroni}},
  \bibinfo{author}{\bibfnamefont{V.~M.} \bibnamefont{Ramaglia}},
  \bibinfo{author}{\bibfnamefont{G.}~\bibnamefont{Cantele}}, \bibnamefont{and}
  \bibinfo{author}{\bibfnamefont{V.}~\bibnamefont{Cataudella}},
  \bibinfo{journal}{Phys. Rev. B} \textbf{\bibinfo{volume}{87}},
  \bibinfo{pages}{155435} (\bibinfo{year}{2013}),
  \urlprefix\url{http://link.aps.org/doi/10.1103/PhysRevB.87.155435}.

\bibitem[{\citenamefont{Weick et~al.}(2010)\citenamefont{Weick, Pistolesi,
  Mariani, and von Oppen}}]{Weick2010}
\bibinfo{author}{\bibfnamefont{G.}~\bibnamefont{Weick}},
  \bibinfo{author}{\bibfnamefont{F.}~\bibnamefont{Pistolesi}},
  \bibinfo{author}{\bibfnamefont{E.}~\bibnamefont{Mariani}}, \bibnamefont{and}
  \bibinfo{author}{\bibfnamefont{F.}~\bibnamefont{von Oppen}},
  \bibinfo{journal}{Phys. Rev. B} \textbf{\bibinfo{volume}{81}},
  \bibinfo{pages}{121409} (\bibinfo{year}{2010}),
  \urlprefix\url{http://link.aps.org/doi/10.1103/PhysRevB.81.121409}.

\bibitem[{\citenamefont{Li et~al.}(2012)\citenamefont{Li, Ren, Wang, Zhang,
  H\"anggi, and Li}}]{Li2012}
\bibinfo{author}{\bibfnamefont{N.}~\bibnamefont{Li}},
  \bibinfo{author}{\bibfnamefont{J.}~\bibnamefont{Ren}},
  \bibinfo{author}{\bibfnamefont{L.}~\bibnamefont{Wang}},
  \bibinfo{author}{\bibfnamefont{G.}~\bibnamefont{Zhang}},
  \bibinfo{author}{\bibfnamefont{P.}~\bibnamefont{H\"anggi}}, \bibnamefont{and}
  \bibinfo{author}{\bibfnamefont{B.}~\bibnamefont{Li}}, \bibinfo{journal}{Rev.
  Mod. Phys.} \textbf{\bibinfo{volume}{84}}, \bibinfo{pages}{1045}
  (\bibinfo{year}{2012}),
  \urlprefix\url{http://link.aps.org/doi/10.1103/RevModPhys.84.1045}.

\end{thebibliography}

\end{document}